\documentclass[12pt,preprint]{aastex}

\usepackage{psfig}
\usepackage{amsmath}
\usepackage{amssymb}
\usepackage{graphicx}
\usepackage{appendix}
\usepackage{color}

\definecolor{red}{RGB}{255,0,0}

\newcommand{\lSect}[1]{{\label{sec:#1}}}
\newcommand{\lFig}[1]{{\label{fig:#1}}}
\newcommand{\lEq}[1]{{\label{eq:#1}}}
\newcommand{\lTab}[1]{{\label{tab:#1}}}
\newcommand{\Msun}{\mbox{M$_\odot$}}
\newcommand{\Lsun}{\mbox{L$_\odot$}}

\newcommand{\dmb}{\ensuremath{\Delta M_{15}(B)}}
\newcommand{\Nifs}{\ensuremath{^{56}{\rm Ni}}}
\newcommand{\kms}{\ensuremath{{\rm km~s}^{-1}}}

\def\gtaprx {\lower .1ex\hbox{\rlap{\raise .6ex\hbox{\hskip .3ex
	{\ifmmode{\scriptscriptstyle >}\else
		{$\scriptscriptstyle >$}\fi}}}
	\kern -.4ex{\ifmmode{\scriptscriptstyle \sim}\else
		{$\scriptscriptstyle\sim$}\fi}}}
\def\ltaprx {\lower .1ex\hbox{\rlap{\raise .6ex\hbox{\hskip .3ex
	{\ifmmode{\scriptscriptstyle <}\else
		{$\scriptscriptstyle <$}\fi}}}
	\kern -.4ex{\ifmmode{\scriptscriptstyle \sim}\else
		{$\scriptscriptstyle\sim$}\fi}}}
\newcommand{\FIGFF}[2]{{\ref{fig:#2}{#1}}}

\newcommand{\FIG}[2]{{Fig.~\FIGFF{#1}{#2}}}
\newcommand{\Fig}[1]{{\FIG{}{#1}}}

\newcommand{\Sectff}[1]{{\ref{sec:#1}}}
\newcommand{\Sect}[1]{{\S~\Sectff{#1}}}

\newcommand{\Eqref}[1]{{\ref{eq:#1}}}
\newcommand{\Eqff}[1]{{(\Eqref{#1})}}

\newcommand{\EQ}[1]{{Equation~\Eqff{#1}}}

\newcommand{\Eq}[1]{{eq.~\Eqff{#1}}}

\newcommand{\Tab}[1]{{Table \ref{tab:#1}}}

\begin{document}


\title{Sub-Chandrasekhar Mass Models For Supernovae}

\author{S. E. Woosley\altaffilmark{1} and Daniel Kasen\altaffilmark{2,3}}

\altaffiltext{1}{Department of Astronomy and Astrophysics, University
  of California, Santa Cruz, CA 95064; woosley@ucolick.org}
\altaffiltext{2}{Departments of Physics and Astronomy, University of
  California, Berkeley, CA 94720} \altaffiltext{3}{Nuclear Science
  Division, Lawrence Berkeley National Laboratory, Berkeley, CA}
  
\begin{abstract} 
For carbon-oxygen white dwarfs accreting hydrogen or helium at rates
in the range $\sim1$ - 10 $\times 10^{-8}$ \Msun \ y$^{-1}$, a variety
of explosive outcomes is possible well before the star reaches the
Chandrasekhar mass. These outcomes are surveyed for a range of white
dwarf masses (0.7 - 1.1 \Msun), accretion rates ($1 - 7 \times
10^{-8}$ \Msun \ y$^{-1}$), and initial white dwarf temperatures (0.01
and 1 \Lsun). The results are particularly sensitive to the convection
that goes on during the last few minutes before the explosion. Unless
this convection maintains a shallow temperature gradient, and unless
the density is sufficiently high, the accreted helium does not
detonate.  Below a critical helium ignition density, which we estimate
to be $5 - 10 \times 10^5$ g cm$^{-3}$, either helium novae or helium
deflagrations result. The hydrodynamics, nucleosynthesis, light
curves, and spectra of a representative sample of detonating and
deflagrating models are explored. Some can be quite faint indeed,
powered at peak for a few days by the decay of $^{48}$Cr and
$^{48}$V. Only the hottest, most massive white dwarfs considered with
the smallest helium layers, show reasonable agreement with the light
curves and spectra of common Type Ia supernovae. For the other models,
especially those involving lighter white dwarfs, the helium shell mass
exceeds 0.05 \Msun \ and the mass of the $^{56}$Ni that is synthesized
exceeds 0.01 \Msun. These explosions do not look like ordinary Type Ia
supernovae, or any other frequently observed transient.
\end{abstract}

\section{INTRODUCTION}
\lSect{intro}

The possibility that explosive burning initiated in the thick helium
shell of an accreting carbon-oxygen white dwarf (CO-dwarf) might lead
to a supernova-like transient has been explored extensively in the
literature.  In the early 80's, it was realized that, for a range of
accretion rates around 10$^{-8}$ \Msun \ y$^{-1}$, a thick helium
layer would accumulate and ignite explosively
\citep{Taa80a,Taa80b,Nom80,Nom82a,Nom82b,Woo80}. In some cases, the
helium runaway ignited as a single detonation producing a faint
supernova and leaving behind an intact white dwarf. In others, the
helium detonation led to a secondary explosion of the CO core as well,
and the star was completely disrupted
\citep{Nom80,Woo80,Nom82b,Liv90}. Detailed studies of the energetics,
nucleosynthesis and light curves of these explosions were carried out
\citep[e.g.,][]{Woo86,Woo94}. A critical density for the detonation of
helium in this context was noted, $1 - 2 \times 10^6$ g cm$^{-3}$
\citep{Nom82b}, as was the possibility of intrinsically faint Type Ia
supernovae. The key role of such supernovae in synthesizing $^{44}$Ca
(as its radioactive progenitor $^{44}$Ti) was also discovered
\citep{Woo94}.

Exploratory multi-dimensional studies of this model were carried out
in the 90's \citep{Liv90b,Liv91,Liv95,Arn97,Ben97,Liv97,Gar99}, in
both two and three dimensions. A critical issue was whether helium
detonation, if it occurred, would lead to carbon detonation and the
explosion of the whole star. The general consensus was that it would,
but there were a variety of caveats. First, it was much easier to
ignite a carbon detonation if the helium detonation did not occur
right at the CO-helium interface, but at some altitude above it. A
mixed layer between CO and helium also favored detonation.  The helium
layers considered were also all rather thick, at least 0.1 \Msun. If
``direct drive'' did not work, a carbon detonation might still occur
by the focusing, within the CO-core, of compressional waves caused by
helium detonation, providing that the zoning in the calculation was
fine enough.

More recently, multi-dimensional studies
\citep{Fin07,Fin10,Sim10,Kromer_2010} have explored combined helium and
carbon detonations in some depth. Among the assumptions and
conclusions are: 1) helium ignition will occur as a detonation for
almost any value of helium shell mass, including very small ones down
to 0.0035 \Msun \ \citep{Bil07,Fin10}; 2) helium detonation will
invariably cause detonation of the carbon-oxygen core \citep{Fin10};
and 3) for low mass helium shells, detonation of carbon in a range of
core masses will give a variety of supernova light curves with a
width-luminosity relation not unlike that observed for all but the
brightest Type Ia supernovae \citep{Sim10,Kromer_2010}. \citet{Bil07}
also refocused attention on the sub-Chandrasekhar mass models, because
of the possibility that they might make an observable class of
sub-luminous supernovae, ``point-one-a'''s. They associated this
phenomenon with a particular class of binary system, AM Canum
Venaticorum binaries.

These conclusions have far reaching implications and motivate a
careful, independent evaluation of the problem. While definitive
studies will need to be, and now can be done in three-dimensions, this
paper explores, in 1D, the physics and parameter space that such
calculations will need to examine. Light curves and spectra for a
representative sample of our models are calculated and compared with
observations. As we shall see, there is still considerable uncertainty
regarding the outcome of sub-Chandrasekhar mass models, and more calculations
and observations will be needed to determine what really happens.

We begin by discussing the problem set up - the relevant nuclear
physics and how accretion is simulated in the code. 

\section{Physics and Procedures}

\subsection{Nuclear Physics}
\lSect{nuclear}

Because of advances in computer technology and reaction data bases, it
is now possible to employ more realistic nuclear physics in models for
supernovae than was feasible in the 1990's. For this study, we used
the ``adaptive reaction network'' described in \citet{Rau02}. This
network was used for both energy generation and nucleosynthesis up to
the point where a helium detonation was well underway. After that,
energy generation was calculated using the usual 19 isotope network
\citep{Woo94}, but nucleosynthesis continued to be followed using the
larger network. One calculation that used the adaptive network
throughout showed no appreciable differences with the (cheaper)
calculation using the small network for energy generation.

The adaptive network begins with a basic set of 49 nuclei: n, $^1$H,
$^4$He, $^{12-14}$C, $^{14-15}$N, $^{15-18}$O, $^{18-19}$F,
$^{20-22}$Ne, $^{21-24}$Na, $^{23-26}$Mg, $^{25-28}$Al, $^{27-32}$Si,
$^{30-33}$P, and $^{31-36}$S. As the calculation proceeds, nuclei are
added or subtracted according to rules regarding minimum abundances
and projected flows. By the time that the core was freezing out from
nuclear statistical equilibrium following the supernova, the reaction
network had grown to over 500 nuclei and included species as heavy as
$^{96}$Ru. Nuclei were retained in the network if they had mass
fractions greater than 10$^{-23}$. Nuclei were added when they were
potential products of any reaction on any nucleus in the network with
mass fraction greater than 10$^{-10}$. In addition to assuring no
important nucleus was overlooked, this also allowed the optimal use of
CPU and did not waste storage and time on nuclei with trivial
abundances. In principle, the network could also shrink with time, but
in practice this did not occur.  The necessary reaction rates were taken
from a variety of sources. See \citet{Woo07} for a summary.  All
reaction rates were appropriately corrected for electron screening.

Using the large reaction network for energy generation in the
hydrodynamical calculation is important because ignition is influenced
by heating from the reaction sequence
$^{14}$N(e$^-$,$\nu$)$^{14}$C($\alpha,\gamma)^{18}$O
\citep{Has86}. Since the electron capture is endoergic, this
sequence has a threshold density of $5.78 \times 10^5$ g cm$^{-1}$ and
occurs at an interesting rate only for helium densities above about
10$^6$ g cm$^{-3}$. Such densities occur at the base of the helium,
shell for all but the lowest mass CO-dwarfs and the highest accretion
rates considered in this paper. As the burning progresses and the
temperature rises, the $3 \alpha$ reaction takes over as the dominant
source of energy generation, but electron capture is very important,
both for altering the neutron-excess of the material (which affects
nucleosynthesis) and for determining the location where the runaway
ignites. In particular, the depletion of $^{14}$N by electron capture
at the base of the accreted helium layer can act to shift the ignition
point outwards for helium shells igniting over $2 \times 10^6$ g
cm$^{-3}$. The rate for $^{14}$N(e$^-$,$\nu$)$^{14}$C here is taken
from \citet{Has86}, as implemented in machine usable form by
Martinez-Pinedo (private communication). The triple-alpha rate is from
\citet{Cau88} as adapted for high density by \citet{Nom85} and the
$^{14}$C($\alpha,\gamma)^{18}$O rate is also from \citet{Cau88} as modified by
\citet{Buc96} and \citet{Woo07}.

Also important to the ignition and the possibility of detonation is
the reaction sequence $^{12}$C(p,$\gamma)^{13}$N($\alpha$,p)$^{16}$O,
in which the proton plays a catalytic role in accelerating the
relatively inefficient $^{12}$C($\alpha,\gamma)^{16}$O
\citep{She09}. This reaction proceeds on carbon that is either created
by pre-explosive helium burning or dredged up from the CO-core. The
abundance of free protons can be significant because the composition
of the fuel, almost entirely $^4$He and $^{14}$N, lacks any
appreciable neutron excess. The relevant reaction rates are well known
\citep{Cau88}.  We found that this reaction sequence increased the
energy generation rate during the runaway, for temperatures near
10$^9$ K, by a factor of several. Since detonation involves a race
between the burning rate and a sound wave crossing a small region, a
change of a factor of three can have very significant consequences.

\subsection{Accretion}
\lSect{accretion}

All calculations in this paper were carried out using the Kepler
one-dimensional, implicit hydrodynamics code \citep{Wea78,Woo02}. The
grid in Kepler is Lagrangian, so a proper treatment of accretion is
challenging \citep{Woo94}. Accretion is simulated here using a surface
boundary pressure that gradually increases at a rate corresponding to
the weight of accreted matter until a designated mass is reached. A
new Lagrangian shell is then added with this mass, a temperature equal
to that of the current outer zone, and a density equal to one-half
that of the current outer zone. For sufficiently fine zoning, the new
zone adjusts promptly to hydrostatic equilibrium. However, the newly
accreted surface layers of the star are not in thermal equilibrium, so some
memory of the the entropy of the accreted matter is retained. In all
cases studied here, the entropy of the accreted material was in the
range $S/N_A k$ = 3 to 4, increasing radially outwards. A decreasing
entropy gradient would have caused convection. The entropy here is
mostly due to the ions.  Two runs, with identical zoning, where
the density of the accreted shell was increased by a factor of two
(Model 10B) had a critical helium shell mass that differed by 4\%. As
expected, the calculation with the higher accretion density (lower
entropy) ignited later and had the larger mass.

As a consequence of this memory of the accretion entropy, our
results are somewhat zoning dependent. With finer surface zoning, the
(thinner) outer zone of the star has a higher entropy when in
hydrostatic equilibrium. Since the zoning in the present study is
finer than in \citet{Woo94}, the entropies of the accreted shells are
higher and, consequently, the critical masses for runaway are smaller,
for the same accretion rate and CO-dwarf mass. Smaller zones also
mean a shorter time between zone addition and less time for
cooling. This too contributes to reducing the critical mass. For
example, in the 1994 study, a 0.7 \Msun \ CO core accreting at $5
\times 10^{-8}$ \Msun \ y$^{-1}$ accumulated 0.13 \Msun \ before
running away. In the present study, it accretes 0.12. Previously, a 0.9
\Msun \ core accreting at $3.5 \times 10^{-8}$ \Msun \ y$^{-1}$
accreted 0.18 \Msun \ of helium. The same core here, accreting at $3
\times 10^{-8}$ and $4 \times 10^{-8}$ \Msun \ y$^{-1}$, runs away
after accreting only 0.13 and 0.11 \Msun \ respectively.

While this variation is troublesome, the magnitude of the effect is
not large compared with that expected if other uncertain parameters of
the problem are varied, such as the metallicity (which affects
ignition by $^{14}$N(e$^-$,$\nu$)$^{14}$C($\alpha,\gamma)^{18}$O), the
CO-core temperature (i.e., the luminosity) of the accreting CO dwarf, and
time varying accretion rates. The actual entropy of the accreted
material will depend on difficult-to-model radiative efficiencies as
the matter accumulates through a shock onto the white dwarf.  As a
result, the correspondence between accretion rate and critical mass
derived here is only approximate. To compensate, we examined a broad
range of models.  These models might be more appropriately categorized
by their CO-dwarf mass, helium shell mass, and location of the
ignition than by their accretion rate. At the present time though,
there is no compelling reason to rule any of them out.

\subsection{Initial Models}

Initial models were constructed as described in \citet{Woo94}, except
that finer zoning was used, especially near the center of the white
dwarf where a detonation might occur, and in the accreting helium
shell.  We also considered heavier mass CO-cores, higher accretion
rates, and hotter white dwarfs than \citet{Woo94}. These changes were
motivated by a desire to examine models with smaller helium shell
masses at the time of explosion. It will turn out that the higher mass
helium shells give supernovae that look less like the Type Ia
supernovae (SN Ia) that have been observed so far. We also wanted to
compare results with \citet{Bil07} and \citet{Sim10}, who considered
heavier CO-cores.

CO-dwarfs with masses of 0.7, 0.8, 0.9, 1.0 and 1.1 \Msun \ were
allowed to cool and relax into thermal equilibrium with a luminosity
of either 0.01 \Lsun \ or 1.0 \Lsun. Cooling to 0.01 \Lsun \ takes
about 200 My, while cooling to 1 \Lsun, in the absence of accretion
only takes about 10 My \citep{Ren10}, comparable to the accretion
time. White dwarfs with the higher luminosity had typical temperatures
in the outer CO-core just beneath the accreting helium of 7 - $8
\times 10^7$ K during the accretion. The lower luminosity dwarfs were
much cooler with central temperatures $\sim 1 - 2 \times 10^7$ at the
onset and the heat generated by accretion flowed more readily into
them. The actual value of temperature in these models depended much
more on the individual mass, accretion rate, location, and stage of
accretion.

Coulomb corrections were included the equation of state. The initial
composition of the CO-dwarfs was 49.5\% $^{12}$C, 49.5\% $^{16}$O, and
1\% $^{22}$Ne. The composition of the accreted material was 99\%
$^4$He and 1\% $^{14}$N, i.e., approximately what is expected for
solar metallicity. No elements heavier than $^{22}$Ne were
included. Once the white dwarf had relaxed to the desired luminosity,
set here by a central boundary condition, accretion was initiated at a
specified rate and nuclear burning turned on. During the accretion and
ignition phases and the early stages of helium detonation, the full
reaction network (\Sect{nuclear}) was coupled to the hydrodynamic
calculation. Time-dependent mixing-length convection was included,
though no convection occurred until helium burning ignited.  The
convective speed was calculated according to local (zonal) gradients,
but the convective speed was not allowed to exceed 20\% sonic.

The combinations of CO-dwarf mass, accretion rate and white dwarf
luminosity studied are summarized in Tables 1 and 2.  Models are named
according to the mass of the CO-core, times 10, a letter indicating
relative accretion rate (``A'' or ``AA'' being the highest accretion
rate considered for a given mass), and, in some cases, an ``H''
(\Tab{models2}) to indicate that the CO-substrate was ``hot'', i.e.,
that the initial luminosity of the accreting white dwarf was 1 \Lsun,
not 0.01 \Lsun. A ``1'' at the end of the model name indicates that
detonation of the CO-core was suppressed and only the helium layer
exploded.

Helium was added as described in \Sect{accretion}. For most of the
runs, the mass of each helium zone was $1 - 2 \times 10^{30}$ g,
though in some studies, especially those with smaller critical helium
shell masses, finer, variable mass zones were used. The number of
helium zones added is indicated for each model in Tables 1 and 2. As
the helium shell grew by accretion, the density and temperature at its
base increased due to gravitational compression and, eventually,
nuclear burning. The helium layer was close to isothermal, but often
with a small inverted temperature gradient, that is higher
temperatures occurred closer to the surface. The highest temperature
was always many zones below the surface though. For example, for a 1.0
\Msun \ CO-core accreting at $5 \times 10^{-8}$ \Msun \ y$^{-1}$
(Model 10B), the temperature at the base of the helium shell when
helium burning first became rapid enough to power convection was $8.55
\times 10^7$ K and the density there was $1.90 \times 10^6$ g
cm$^{-3}$. However the maximum temperature and the base of the
convective region was situated 0.026 \Msun \ farther out where the
temperature and density were $9.66 \times 10^7$ K and $1.35 \times
10^6$ g cm$^{-3}$.  As the runaway progressed, the convective region
grew and its base moved inwards to 0.020 \Msun \ above the interface
(0.0618 \Msun \ beneath the surface) as the temperature rose and
density declined (\Fig{moda1aign}).

The ignition characteristics of other CO-core masses and accretion
rates included in this study are given in Tables 1 and 2.  
M$_{\rm acc}$ is the mass of helium accreted prior to runaway and 
M$_{\rm ign}$ is the mass, measured inwards from the surface, where the
runaway develops. If M$_{\rm ign}$ equals M$_{\rm acc}$, the runaway
develops at the CO-core helium interface. The two densities given,
$\rho_{\rm ign}$ and $\rho_{\rm run}$, are the values at the base of
the accreted helium shell and the location of the runaway. Both
densities are evaluated at the time when the maximum power in the
convective shell reached 10$^{47}$ erg s$^{-1}$ (\Sect{ignition}) for
those models that detonated. For the models that did not detonate
(those indicated with a ``d'' in the table), the densities were
evaluated when the energy generation was a maximum, typically for
luminosities between 10$^{46}$ erg s$^{-1}$ and 10$^{47}$ erg
s$^{-1}$. Additional information concerning ignition conditions is
given in \Tab{ignpoint}, which gives the temperature and density at
the base of the helium convection zone when the maximum convective
luminosity equals the values indicated.

\section{Ignition, Convection and Outcomes}
\lSect{channels}
\subsection{The Freezing Out of Convection}
\lSect{ignition}

Convective energy transport and mixing were followed in Kepler using
time-dependent mixing length theory \citep{Wea78,Woo88}. While this
should be adequate during ignition and the early stages of the
runaway, mixing length theory certainly breaks down when the burning
time scale becomes shorter than the convective turnover time.

Consider again Model 10B. As \Fig{moda1aign} shows, as the runaway
progresses, the convection zone grows outwards from the ignition
region, raising the temperature and lowering the density so as to
maintain an approximately adiabatic profile. As the temperature at the
base of the convective shell rises, the luminosity gradually
increases. It is only after about 15 years of convection that the
runaway finally culminates in explosion. During this interval there is
ample time for many convective turnovers and an adiabatic temperature
profile is maintained until close to the end.

Characteristics of the convection can be estimated using mixing length
arguments \citep[e.g.,][]{Woo04}. The convective speed is approximately
\begin{equation}
v_{\rm rms} \approx \ \left(\frac{4 \, G \, \delta_{\rm P} \, L}{3 \, c_{\rm P}
\, T}\right)^{1/3}.
\lEq{speed}
\end{equation}
where L is a typical luminosity in the convective shell and
\begin{equation}
\begin{split}
\delta_{\rm P} \ &= \ - \left(\frac {\partial \, ln \, \rho}{\partial \, ln
\, T}\right) \\
&= \ - \frac{T}{\rho} \ \frac{\Delta \rho}{\Delta T}\\
&= \ \frac{T}{\rho} \ \left(\frac{\partial \, {\rm P}}{\partial T}\right)_{\rho} 
\left(\frac{\partial \, {\rm P}} {\partial \rho}\right)_T^{-1}\\
&\approx \ 0.20 \ \left(\frac{T_8}{2.5}\right)
\left(\frac{1}{\rho_6}\right).
\lEq{dp}
\end{split}
\end{equation}
Here $T_8$ is the temperature in units of 10$^8$ K, $\rho_6$, the
density in units of 10$^6$ g cm$^{-1}$, and $c_{\rm P}$, the specific
heat at constant pressure. The partial derivatives of the pressure can
be evaluated using the Helmholtz equation of state \citep{Tim00a}. The
specific heat varies slowly with temperature and density in the region
of interest and has a value $7.2 \times 10^7$ erg g$^{-1}$ K$^{-1}$
for $T_8$ = 2.5 and $\rho_6$ = 1.

The luminosity within the convective shell is not actually a constant,
but builds in the energy generating region at the shell's base and
declines in the outer regions due to heating and expansion. It has a
maximum value though, which is achieved a short distance above the
energy generating region, and that will be treated here as
representative. When $L_{\rm max} \sim 10^{45}$ erg, the temperature
and density at the base of the convective shell for Model 10B are $2.10
\times 10^8$ K and $1.34 \times 10^6$ g cm$^{-3}$
respectively. \EQ{speed} then gives a convective speed of 90 km
s$^{-1}$, or about 3\% sonic.  The pressure scale height is 340 km, so
the time for matter to rise a scale height is about 4 s.  The nuclear
time scale at this point in Model 10B, the time for the helium to
runaway to high temperature if convection were to be abruptly turned
off, is 4.1 s, but with convection left on the time scale is
considerably longer. The time for a sound wave to go half way round
the star at the radius of the convective shell, $3.8 \times 10^8$ cm,
is 4.6 s.  Since both these time scales are still comparable to the
time convection requires to cool the burning region, this might be
regarded as the ``last good mixing-length convection model''.  The
explosion is still developing in an approximately spherically
symmetric manner and convection is able to carry away the heat
generated by the helium burning reactions.

As the temperature continues to rise however, the luminosity and
speeds in the convection zone increase while the nuclear time scale
decreases rapidly. Twenty two seconds later (if convection is left
on), the maximum luminosity in the convective shell reaches 10$^{46}$
erg s$^{-1}$ and the temperature at its base is $2.44 \times 10^8$
K. The nuclear time scale, in the absence of convection, has now
shrunk to 0.713 s while the convective speed has only grown to 210 km
s$^{-1}$, so the convective turnover time is still a second or so
(depending upon the choice of the mixing length). Thus the nuclear
time scale has become shorter than the convective time scale. This
critical temperature, about $2.5 - 3.0 \times 10^8$ K for Model 10B,
plays the same role in the sub-Chandrasekhar mass models as the better
known number, $7 \times 10^8$ K, does in the break down of convection
in the standard carbon-deflagration model \citep{Woo04}. It will take
similar multi-dimensional calculations \citep{Zin09} to follow the
subsequent evolution. Certainly by this point though, communication
around the star has been lost and the runaway(s) will proceed
according to local conditions established at the last good mixing
length convection model (see \Sect{multipoint}).

Using mixing length theory beyond this point is clearly perilous, but
continuing the calculation to a luminosity of 5, 10, and 50 $\times
10^{46}$ erg s$^{-1}$ with convection still on, the temperature at the
base of the convection zone rises to $2.76 \times 10^8$, $2.96 \times
10^8$, and $3.68 \times 10^8$ respectively. The corresponding nuclear
time scales drop to 0.253, 0.149, and 0.046 s. Meanwhile, the
convective speed only increases modestly, roughly as $L^{1/3}$. So by
the time these latter temperatures are reached, the runaway has
certainly become localized to a small fraction of a scale height.

Different outcomes result if the six models with temperature gradients
given in \Fig{moda1aign} are allowed to evolve with convection turned
off. The first three models, one with the ignition conditions (when
convection first started to occur) and the two with luminosities 0.1
and $1 \times 10^{46}$ erg s$^{-1}$ give rise to deflagrations. The
temperature gradient steepens until finally a small region runs away
on a time longer than the sound crossing time of the region. The
density goes down while the pressure remains constant. A thin
shell instability develops with an inverted density. The other
three models with luminosities 5, 10 and 50 $\times 10^{46}$ erg
s$^{-1}$ generate outwards moving detonation waves. This result is
analogous to that obtained by \citet{Woo90} for the usual
carbon-deflagration (Chandrasekhar-mass) scenario. There, if
convection was halted at a base temperature of $8.5 \times 10^8$ K a
carbon-burning-initiated detonation occurred, but if convection were
halted a bit earlier at $8.0 \times 10^8$ K, a deflagration
occurred. For the carbon burning case, it is now thought that the
deflagration is favored \citep{Zin09}. What happens in the helium
burning case?

\subsection{Conditions for Helium Detonation}
\lSect{hedetcond}

Shallow temperature gradients are necessary to initiate a detonation
by the Zeldovich mechanism; steep ones inhibit it
\citep{Zel70,Bli87,Woo90}. Convection, so long as it operates
efficiently, keeps the temperature gradient adiabatic. At each point
on that adiabatic temperature profile there is a nuclear time
scale. If there exist regions that are separated by a distance such
that the ratio of that distance to the difference in nuclear time
scale implies a phase velocity that is supersonic, a detonation can
form. On the other hand, if the phase velocity is subsonic, a
deflagration results. The extreme non-linearity of the burning rate
tends to sharpen temperature gradients and only efficient mixing will
suppress the tendency to give birth to deflagrations.

There are three ways of estimating the necessary conditions for helium
detonation, each with its own strengths and shortcomings.

First, is the analytic approach.  Consider the range of peak
temperatures and densities where convection is expected to freeze out
in most of our models: $2.0 \ltaprx T_8 \ltaprx 3.5$ and $0.5 \ltaprx
\rho_6 \ltaprx5$. \Tab{ignpoint} and \Fig{detconds} show the
temperature and density at the base of the convective shell that is
running away for the models defined in Tables 1 and 2. The conditions
are given at two times that approximately define the limits of mixing
length convection - luminosity equals 10$^{46}$ and 10$^{47}$ erg
s$^{-1}$.

It is useful to define a nuclear time scale, $\tau_{\rm run}$, equal
to the time required to run away to high temperature, say over $1.2
\times 10^9$ K, starting from the given conditions. This time scale is
evaluated here empirically using the models and includes the
complications of energy generation by
$^{12}$C(p,$\gamma)^{13}$N($\alpha,$p)$^{16}$O, as well as the local
thermodynamic evolution of the model.  That evolution is isobaric
until quite close to the end, when the pressure starts to accumulate
in those models that detonate.  To an accuracy of 50\%
in the range of temperature and density given above, the time scale to
run away is given by
\begin{equation}
\tau_{\rm run} \ \approx \ 3.4 \times 10^{-4} \,  \exp \, (20/T_8)
\,\rho_6^{-2.3} \ \ {\rm s}.
\lEq{taurun}
\end{equation}
An exponential was found to fit the actual values of $\tau_{\rm run}$
significantly better than a power law, possibly reflecting the
temperature sensitivity of the helium burning reaction rate.  A
{\sl necessary} condition for detonation is that the absolute value of the
temperature gradient approximately satisfy
\begin{equation}
c_s \ \frac{d T}{d r} \ \frac{d \tau_{\rm run}}{d T} \ \ltaprx \ 1. 
\lEq{zeld}
\end{equation}
For the conditions of interest, the sound speed, $c_s$, varies from
2300 to 3200 km s$^{-1}$ and an appropriate temperature gradient is
the adiabatic one,
\begin{align}
\left( \frac{d T}{d r} \right)_{\rm ad} \ &= \left( 1 \, - \,
\frac{1}{\Gamma_2}\right) \ \frac{T}{P} \, \frac{d P}{d r}\\ 
&= \left( 1 \, - \, \frac{1}{\Gamma_2}\right) \, \frac{T}{H},
\lEq{adiabat}
\end{align}
where $H = (1/P \ dP/dr)^{-1}$, is the pressure scale height.
Typically H $\approx 300$ km, and, with $c_s = 3000$ km s$^{-1}$, $T
\approx 3 \times 10^8$ K, and $\Gamma_2 = 3/2$, one finds $dT/dr
\approx$ 3 K cm$^{-1}$. This value for the gradient agrees very well
with that found in the models that detonated at the times and
conditions given in \Tab{ignpoint}. \EQ{zeld} then gives, as a
condition for detonation,
\begin{equation}
\rho_6 \ \gtaprx \ \left( \frac{0.0607}{T_8^2} \, \exp \, (20/T_8) \right)^{1/2.3}.
\lEq{dtdr}
\end{equation}
This equation, plotted as the topmost solid line in \Fig{detconds},
delineates those regions that are likely to detonate from those that
are not. {\sl Based solely upon this constraint, none of the models
  calculated in this paper would detonate.} However, several
approximations were made in deriving this condition, and \Eq{zeld},
itself, is an approximation that depends upon the dimensionality of the
problem and equation of state. It also neglects gas dynamic effects. A
region close to burning on a sonic time scale will not stay in
pressure equilibrium with its surroundings. Higher pressure will
shorten the local time scale and expansion of the burning region will
alter the temperature gradient in its vicinity.  As a result,
detonation can happen at a lower density than \Eq{dtdr}
suggests. 

A more accurate estimate of the conditions for detonation {\sl in a
  plane} comes from the models themselves. Empirically, all of the
models with convective powers of 10$^{47}$ erg s$^{-1}$ detonated
along with a substantial fraction of those with 10$^{46}$ erg
s$^{-1}$. The dashed line in \Fig{detconds}, which is the density
given by \Eq{dtdr} divided by 4, is arbitrary, but is a more accurate
representation of the model results.

The story does not end here, though, because the detonation probably
does not originate in spherical shell (which locally resembles a
plane) as these one-dimensional models necessarily assume. A more
realistic description is probably detonation starting from a point or
set of points (\Sect{multipoint}). To examine the conditions required
for detonation starting from a point, spheres of helium of constant
density and size were constructed using an approach similar to that
employed by \citet{Nie97} to study carbon detonation.  Each sphere,
about 100 km in radius, was zoned into spherical shells, each 1 km
thick. A temperature gradient was imposed such that $dT/dr = 3$ K
cm$^{-1}$ and three choices of central temperature were explored, 2.5,
3.0, and $3.5 \times 10^8$ K. The spheres were evolved with the Kepler
code {\sl using the large nuclear reaction network} and therefore
including carbon burning by
$^{12}$C(p,$\gamma)^{13}$N($\alpha,$p)$^{16}$O. The density was varied
and the critical value of density required to develop a successful
detonation was determined. For example, at $T_8$ = 3.5, a 100 km
sphere with constant density $1.5 \times 10^6$ g cm$^{-3}$ detonated
while one with density $1.0 \times 10^6$ g cm$^{-3}$ did not. The
resulting points are plotted with error bars as triangles in
\Fig{detconds}.

The constraints from the numerical detonation experiments agree
reasonably well with the curve given by \Eq{dtdr}, although they lie
below it, as expected. The correct answer is obviously dependent upon
the actual geometry of the runaway. Is it locally more ``point-like''
or more like a plane? For now, detonations below 10$^6$ g cm$^{-3}$
should be treated with caution \citep[see also][]{Nom82b}, until the
defining multi-dimensional calculations of ignition are done.

Also shown in \Fig{detconds}, as the lower solid line, is the
condition $\tau_{\rm run}$ = 2 s, hence 
\begin{equation}
\rho_{\rm defl,6} \ \approx \ \left( 1.68 \times 10^{-4} \,
\exp(20/T_8)\right)^{1/2.3}.
\end{equation}
Very roughly, models below this line will be able to transport their
nuclear energy by convection throughout the explosion and not
experience a violent, hydrodynamic event. Between this line and the
detonation line, deflagrations occur. Given the uncertainties
discussed here, deflagration could happen in anywhere from none to
most of our models. Since we are restricted here to one-dimensional
simulations, except for a few models with ``d'' in \Tab{models1} and
\Tab{models2}, all our simulations experienced helium
detonation. Those few that did not were either helium novae or
deflagrations.

\subsection{Helium Deflagration}
\lSect{defl}
 
If convection fails to set up the proper conditions for detonation,
but energy is generated at a much faster rate than convection can
carry, a deflagration will develop (\Fig{detconds}). The temperature
rapidly rises in the ignition region, becoming almost discontinuous
and localized to one or a few zones, but the pressure does not change
greatly. Consequently the density goes down dramatically. This is the
starting point for a deflagration.

In the carbon deflagration model for Chandrasekhar-mass explosions, a
thin flame forms at this point and that flame is moved around by the
Rayleigh-Taylor instability and turbulence. Here the situation is
different, both because the runaway is happening on a shell rather
than near the center of a sphere and because a helium burning
``flame'' is thicker and more subject to disruption by turbulence.
The properties of “laminar” conductive flames in helium are well
determined, \citep{Tim00}.  Typical flame speeds and widths for
densities $10^6 - 10^7$ g cm$^{-3}$ are 10$^3$ - 10$^5$ cm s$^{-1}$
and 100 - 1 cm, respectively. However, the flames here are born in a
medium that is both turbulent on small scales and characterized by
rapid bulk flows on large scales due to the convection. Convective
speeds are $\sim$100 km s$^{-1}$ (\Sect{ignition}) and the pressure
scale height is a few hundred km. This convection drives turbulence
through shear instabilities on a smaller scale. Judging from the
analogous case in carbon deflagration, roughly one order of magnitude
in length and velocity scale separate the bulk flow from turbulence
that might be described as isotropic with a Kolmogorov spectrum
\citep{Zin05,Roe07}. Thus, very approximately, we expect the
turbulence due to convection at the time the runaway ignites to have a
characteristic speed $\sim10$ km s$^{-1}$ on an integral scale of
about 30 km. Kolmogorov scaling ($v \propto l^{1/3}$) down to the
flame width (e.g., 100 cm at 10$^6$ g cm$^{-3}$) then gives a
turbulent speed of $\sim 3 \times 10^4$ cm s$^{-1}$, significantly
greater than the laminar speed at that density, but still very
subsonic. Unlike carbon deflagration, helium deflagration at densities
below about 10$^7$ g cm$^{-3}$ commences in the distributed burning
regime \citep{Tim00,Shen_2010} where turbulent transport dominates
conduction \citep[e.g.,][]{Asp08}.

However, the rate of helium burning is not determined by the speed of
a flame but by the large scale flows in which the burning is embedded
\citep{Dam40}. The ash produced by the helium burning has a lower
density than the cold helium fuel and floats. Initially, regions of
hot ash will form with sizes and geometries determined by the chaotic
temperature conditions in the convective flow. These regions will grow
as a consequence of the turbulence generated by convection and, later,
the turbulence generated by their own rise. For a density contrast
near unity and a gravitational acceleration, $g = GM/r^2 \sim 10^9$ cm
s$^{-2}$, hot ash can already reach float speeds of 100 km s$^{-1}$,
in a few hundredths of a second.  After that, the Rayleigh-Taylor
instability, governs the overall progress of the burning while
turbulence governs the flame speed on small scales. The problem at
that point has clearly become multi-dimensional. We shall attempt to
model such events here using a ``Sharp-Wheeler'' description
\citep{Nie97} of the burning, but clearly any one-dimensional
treatment is very approximate (\Sect{multidefl}).
 
Two models were explored. One, Model 8DEFL, was based upon Model
8C. Detonation was suppressed by taking the starting conditions to be
when the luminosity equaled 10$^{45}$ erg s$^{-1}$ rather than
10$^{47}$ erg s$^{-1}$.  The other, Model 10DEFL, was a recomputation
of Model 10D with finer zoning. Because of the finer zoning the
runaway ignited at a somewhat reduced critical mass and density
(\Sect{accretion}; \Tab{models1}). When the nuclear time step had
declined to less than 1 ms in the models, one-dimensional deflagration
physics was turned on in Kepler. The clock was zeroed and a flame was
propagated according to the prescription
\begin{equation}
v_{\rm flame} \ = \ {\rm Max} \, \left( A \, g_{\rm eff} \, t, v_{\rm conv}\right).
\lEq{paramdefl}
\end{equation}
Here $g_{eff}$ is the local acceleration due to gravity, $G M(r)/r^2$,
times the Atwood number. The Atwood number measures the density
contrast between fuel and ash and is close to unity for helium burning
around 10$^6$ g cm$^{-3}$.  The constant A is approximately 0.12 $\pm$
a factor of two \citep{Gli88}. Here, motivated by a desire to have
substantial burning in the deflagration, we chose A = 0.2. Convection
prior to the runaway with speed $v_{\rm conv} \approx 100$ km s$^{-1}$
sets a lower bound to the flame speed. This velocity is measured in
the co-moving Lagrangian frame and the rate at which mass is burned is
\begin{equation}
\dot M \ = \ 4 \pi r^2 \rho_{\rm fuel} v_{\rm flame}.
\end{equation}
For the large gravitational acceleration typical here, $g_{\rm eff}
\sim 10^9$ cm s$^{-2}$, floatation dominates over convection after
only 0.01 s. Since the local speed of sound is about 3000 km s$^{-1}$,
this equation cannot be used beyond about 1 s, but the explosion was
over by then in both models studied. Larger values of ``A'' result in
a delayed transition to detonation which is probably not
physical. This is not necessarily to say that no transition to
detonation occurs as the flame moves down the density gradient near
the surface, only that ascribing this to floating bubbles moving at
supersonic speeds is not correct.

\section{Model Results}

\subsection{Helium Detonation}
\lSect{hedet}

If carbon detonation is suppressed, either by coarse central zoning or
limiting the propagation of the helium detonation into the carbon,
then only the helium layer explodes and very little additional matter,
if any, is ejected. These models are indicated by a ``1'' at the end
of the job name in Tables 1 and 2 and the final remnant masses and
kinetic energies are given in \Tab{explosions}. The kinetic energies
range from a few times 10$^{49}$ erg to a few $\times 10^{50}$ erg,
depending on the mass of material that burns, its initial binding
energy, and the nucleosynthetic products. Typical values for the
kinetic energy are $\sim 10^{51} \ {\rm M_{Eject}}$, if M$_{\rm
  Eject}$ is measured in solar masses. This corresponds to an energy
per mass of $q \sim 5 \times 10^{17}$ erg g$^{-1}$ indicating that not
all of the helium burns, and implies a typical ejection speed, $(2 \,
q)^{1/2} \approx$ 10,000 km s$^{-1}$. Of course, the peak speed will
be higher than the mean.  The lowest energy explosions, Models 7A1,
7B1, 8A1, 8HBC1, 9A1, 10A1, 10HC1, 10HCD1, and 11HD1, make almost no
$^{56}$Ni and will thus be very faint. In fact, all but Model 10HCD1
make very little of anything above $^{44}$Ti (\Tab{radio}). Some will
be very faint indeed. They will not be ``point Ia'' or even ``point
zero Ia'' supernovae (\Sect{defllite}).

The rest of the models, 7E1, 8D1, 8E1, 8HC, 9C1, 9E1, 10B1, 10E1,
11E1, and 11F1, all but one derived from the cooler accreting white
dwarf, make appreciable $^{56}$Ni and will make potentially detectable
transients with varying properties (\Sect{fullstarlite}). Typical
abundances and speeds are like those in the right hand panels of
\Fig{moda1aign}, \Fig{p7d1a} and \Fig{1p1a1}, though the speeds in the
latter two figures have been boosted by the explosion of the CO core.

\subsection{Carbon Detonation}
\lSect{cdet}

\subsubsection{Double-Detonation Models}

Most of the models studied here were of the double detonation variety
where the helium not only detonated, but touched off a secondary
detonation in the CO-core. These were either of the edge-lit variety
(``a'' in Tables 1 and 2), or detonation ignited by compression at the
center of the CO-core (\Sect{cdetcomp}; ``b'' in the
tables). Qualitatively, the results were the same. In edge-lit
detonations, the whole star detonated before any appreciable expansion
of the CO-core occurred. In the compressional detonation, most of the
core was only slightly compressed before the CO-core
detonated. Detonation of the CO-core assured the complete detonation
of any unburned portion of the helium shell on the way out. This
degeneracy in results will probably not hold for multi-dimensional
simulations of the explosion.

The kinematic results of all the explosions are summarized in
\Tab{explosions} and the nucleosynthesis will be discussed in
\Sect{nucleo}. As expected, the higher mass CO-cores produce more
$^{56}$Ni, both because there is more mass, to burn and because burning
goes farther at the higher density. Typical results, as exemplified by
Models 7D, 9C, 11C and 11E, are shown in \Fig{p7d1a} and
\Fig{1p1a1}. In the higher mass models, most of the core burns to
$^{56}$Ni, leaving only a narrow range of mass and velocity where
intermediate mass elements are produced. In the lower mass cores, more
intermediate mass elements are made with a broad range of velocities,
but the supernova will be faint.  Light curves and spectra for these
models are discussed in \Sect{fullstarlite}.

\subsubsection{Condition for Compressional Detonation of the CO-Core}
\lSect{cdetcomp}

For sufficiently fine central zoning, all models in which helium
detonated also detonated their CO-core. Is this reasonable?

Convergent spherical and cylindrical shocks have been studied
extensively because of their wide application to such disparate topics
as bombs, laser pellet implosions, sonoluminescence, and diamond
synthesis. The solution depends on the dimensionality of the problem,
adiabatic exponent of the gas, and density gradient, but is typically
given by a ``scaling law of the second kind'' \citep{Zel66},
\begin{equation}
R  = R_0 \left(1- \frac{t}{to} \right)^\alpha,
\lEq{shock}
\end{equation}
 where $R_0$ is the radius at t = 0 and $t_o$ is the time when the
 shock reaches the origin. Typical values of $\alpha$ are 0.7 for
 spherical geometry and 0.85 for cylindrical geometry \citep[][and
   references therein]{You86}. As the shock converges to a central
 point or line, constant density is a reasonable approximation.  For
 $\alpha < 1$, this implies a velocity that increases as the shock
 converges to the center. As the shock velocity increases so does the
 temperature behind it.  Given the singularity implied in \Eq{shock},
 the temperature achieved at the center of a calculation with perfect
 spherical symmetry (e.g., a one-dimensional calculation in radial
 coordinates) will approach an arbitrarily high value depending on how
 finely one zones the center of the star.

If the region compressed is large enough that its burning time scale
is less than its sound crossing time, a detonation can result. The
necessary resolution implies very fine zoning in a Lagrangian stellar
evolution code such as Kepler. For models with a thick helium shell,
above about 0.1 \Msun, a resolution at the center of the star of about
100 km, or about 10$^{-4}$ \Msun \ sufficed, but for models like 10HC
and 11HD, a central zone smaller than 50 km was required, or about
10$^{-5}$\Msun.  The CO core was ``kicked'' by the expansion of the
helium layer so as to give speeds behind the inwards moving shock in
the range 500 km s$^{-1}$ (low mass helium shell explosions) to 1000
km s$^{-1}$ (high mass shells). Small time steps were taken so as to
suppress the damping of such weak shocks by the implicit hydrodynamics
of the Kepler code. Successful detonation required a temperature of
about $1.6 \times 10^9$ K in the inner zone, and this in turn required
that the compressional speeds be boosted to over 3000 km s$^{-1}$ by
the focusing at the stellar center.

While the concentration of $\sim10^{45}$ erg into 10$^{-5}$ \Msun \ of
the mass does not seem unreasonable, the timing is
critical. The sound crossing time for the ignition region is about 10
ms. Compressional waves from a substantial fraction of the star must
arrive simultaneously to that tolerance. This happens naturally for
ignition with spherical or cylindrical symmetry, but may be more
difficult in the case of asymmetric, asynchronous ignition
(\Sect{multipoint}).

\subsection{Helium Deflagration}
\lSect{hedefl}

Typical results for a helium deflagration are shown for Model 10DEFL
in \Fig{deflg1c} and \Tab{models1}.  Model 10DEFL ejected 0.077 \Msun
\ at speeds up to 6500 km s$^{-1}$, but most of this was unburned
helium. Only $9 \times 10^{-5}$ \Msun \ of $^{56}$Ni was synthesized
and the major radioactivities were instead $^{44}$Ti (0.0025 \Msun),
$^{48}$Cr (0.0063 \Msun), and $^{52}$Fe (0.0016 \Msun). Because of the
$^{48}$Cr and $^{52}$Fe, these will be brighter events than classical
novae, but because of the lack of $^{56}$Ni, they will be much fainter
than a SN Ia. They are indeed similar to the theoretical predictions
of \citet{Bil07}, but are a result of deflagration, not detonation.
The $^{44}$Ti synthesis is also very large and nucleosynthetically
interesting. It would not take many of these explosions to make the
solar abundance of $^{44}$Ca and yet they are faint and may have escaped
discovery (see \Sect{lcurve}).

Model 8DEFL was quite similar in outcome. A total mass of 0.141 \Msun
\ was ejected of which 0.114 \Msun \ was $^4$He. About half of the
rest (0.015 \Msun) was $^{40}$Ca and the masses of radioactivities were
$^{44}$Ti, 0.0035 \Msun; $^{48}$Cr, 0.0037 \Msun; $^{52}$Fe, $4.5
\times 10^{-4}$ \Msun; and $^{56}$Ni, $6.3 \times 10^{-5}$ \Msun. The
total kinetic energy was $3 \times 10^{49}$ erg with most of the
material moving at about 6,000 km s$^{-1}$ (helium) and less (the
heavier elements). Very little silicon and sulfur were produced and
what was made was in a thin shell. Both the mass and location of the
intermediate mass elements would change in a multi-dimensional
simulation.

Deflagrations are uniquely able to reach a high burning temperature
and yet burn only a small fraction of their mass. This is because the
matter ahead of the burning front has time to expand, ultimately
putting out the flame. The light curves and spectra of these models
are discussed in \Sect{defllite}.

\subsection{Helium Novae}
\lSect{nova}

Helium novae occur when a massive CO-dwarf accretes a low mass layer
of helium that runs away and transports its energy by convection
\citep{Kat03}. They may have been observed \citep{Ash03}.  Models in
Tables 1 and 2 with a ``d'' that were not deflagrations are able to
transport the energy developed in the runaway by convection without
exceeding a power of 10$^{47}$ erg s$^{-1}$. These models were not
followed through mass ejection which is likely to occur over an
extended period, but they were followed until the density at the base
of the convection region had declined more than an order of magnitude
and the energy generation had declined substantially.

The net binding energy of a gram of helium in the pre-explosive star
varies from $1.3 \times 10^{17}$ erg g$^{-1}$ for CO-cores of 0.7
\Msun \ to $3.9 \times 10^{17}$ erg g$^{-1}$ for CO-cores of 1.1 \Msun
\ with thick helium shells. The variation mostly reflects the higher
gravitational potential at the edge of more massive CO-cores (which
also have a smaller radius).  These numbers are to be compared with
the energy that is released by helium burning, $1.33 \times 10^{18}$
erg g$^{-1}$ if the helium burns to silicon, and $1.51 \times 10^{18}$
erg g$^{-1}$ if it burns to nickel. As with classical novae, only a
small amount of fuel must burn to eject the accreted layer. The mass
is probably ejected as a wind over a long period of time during which
the luminosity may not greatly exceed the Eddington value
\citep{Ibe91,Yoo04}.

Typically the composition of these nova-like explosions is enhanced
in elements up to calcium, but not beyond. For the less energetic
events, those with the lowest ignition densities, magnesium was an
abundant product.

\section{Multi-Dimensional Aspects of Ignition and Burning}
\lSect{multid}

While the models in this paper are one dimensional, the physical
conditions and time scales obtained are relevant to predicting the
multi-dimensional behavior and defining the parameter space for future
work. In this section, we discuss the possibility that the helium
runaway ignites neither as a spherically symmetric shell (the 1D
limit), nor at a single point, but as something in between, either an
extended, but bounded region, or a collection of widely separated points
ignited at different times. The necessary conditions for a helium
detonation to smoothly transition into an inwards carbon detonation at
the CO-core interface are also examined.

\subsection{Multi-Point Ignition?}
\lSect{multipoint}

Convective energy transport in most of our calculations was halted
when the maximum power being transported anywhere in the convective
shell passed 10$^{47}$ erg s$^{-1}$ (\Sect{ignition}). This value
assured detonation in most models. However, this assumption in a
one-dimensional model implies a degree of coherency in the burning
around the star that is clearly unrealistic. Consider, for example,
Model 9C. If convection is halted at 10$^{47}$ erg s$^{-1}$,
a detonation ensues 0.16 s later. Even if convection is
(unrealistically) left on, reaching values of 10$^{50}$ erg s$^{-1}$,
the time until runaway is only extended to 0.44 s. Yet the time for
a sound wave ($c_s$ = 2520 km s$^{-1}$) to travel half-way around the
star (ignition radius equals 3950 km) is 4.9 s. Even a detonation
front moving at 12,000 km s$^{-1}$ would take about a second to reach
the opposite pole.  During that time, the entire helium shell would
have already detonated spontaneously. Starting at 10$^{46}$ erg
s$^{-1}$ only helps a little. Runaway occurs after 0.95 s if
convection is halted and 4.3 s if it is not. None of these
one-dimensional simulations is a realistic starting model for a
multi-dimensional study of detonation initiated at a single point
\citep[e.g.,][]{Fin10}.

To find a realistic starting model, one must consider the evolution at
an earlier time when communication among various regions burning
around the star first starts to break down. Our arguments here
resemble those used by \citet{Woo82} and \citet{Spi02} to explain the
spreading of nuclear burning on neutron stars experiencing Type I
x-ray bursts. Unlike \citet{Spi02} however, we consider the non-rotating
case. Also, there is no flame at this stage and hence no large change in
pressure scale height, just a small temperature and pressure contrast
between burning regions.

Communication among regions burning at different locations on a
gravitational equipotential (to good approximation here, a sphere)
occurs by the {\sl baroclinic instability}. A convective cell with a
higher temperature at its base than surrounding cells has a higher
entropy. Associated with this higher entropy is a larger scale height
which moves the center of mass of the convective region slightly
outwards, reducing its weight. Compared with cooler regions farther
away, this expansion causes lateral pressure imbalances. At the same
gravitational equipotential, the pressure in the hotter convective
cell at all locations above the burning shell is higher than its
surroundings.  As a result, matter spreads outwards from the high
entropy cell. Because of the increased scale height, the substrate
below, upon which the burning rests, be it helium or carbon and
oxygen, also feels reduced weight and rises, bringing up material from
beneath to replace what is lost due to the spreading above. If the
runaway occurs at the CO-He interface, this instability causes mixing.

To illustrate this idea with some approximate numbers, consider the
helium burning shell in Model 9C at a time 40 s before the runaway
when the maximum convective power is $\sim$10$^{45}$ erg s$^{-1}$. The
temperature, density, pressure, and radius at the base of the
convection zone are $2.021 \times 10^8$ K, $1.429 \times 10^6$ g
cm$^{-1}$, $5.288 \times 10^{22}$ dyne cm$^{-2}$, and 3955.7 km. A
short time later, when the temperature at the base has increased 5\%,
the density, pressure and radius of this Lagrangian mass point have
changed to $1.416 \times 10^6$ g cm$^{-3}$, $5.263 \times 10^{22}$
dyne cm$^{-2}$, and 3958.1 km respectively. At constant Lagrangian
coordinates the pressure has gone down, but the pressure at this
latter radius, 3958.1 km, in the starting model (at 10$^{45}$ erg
s$^{-1}$) was $5.259 \times 10^{22}$ dyne cm$^{-2}$. That is, at a
given radius the pressure inside the hot cell has gone up and there is
now a pressure differential along a line of constant radius connecting
the cold and hot models of $4 \times 10^{19}$ dyne cm$^{-2}$.  The
difference is even larger, $1.6 \times 10^{20}$ dyne cm$^{-2}$, one
pressure scale height above the base of the convective
region. Examination at other times in the run, 1000 s and 10 s before
runaway gives similar numbers for the pressure difference because the
temperature at the base of the helium convection zone is not varying
greatly. Since the differential pressure is, for small changes in
temperature, proportional to the size of the temperature fluctuation,
we conclude that fluctuations in temperature at the base of the helium
burning region, $T_{\rm He}$ give rise to lateral pressure imbalances,
in Model 9C, of order $10^{19} \, (100 \, \delta T_{\rm He})/T_{\rm
  He}$ dyne cm$^{-2}$, where $\delta T_{\rm He}$ is the variation in
the temperature at the base of the helium convective shell. In three
dimensions, the higher pressure in the hotter region would lead to
outflows, probably greatest about one pressure scale height above the
base. The loss of matter from the convective cell would reduce the
pressure upon its base, allowing inflow and upwelling to occur.

Given sufficient time, this mixing keeps regions of the convective
shell that are on gravitational equipotentials around the star burning
at the same temperature. During the last day before the explosion,
however, this thermostat breaks down. A pressure difference
of 10$^{19}$ dyne cm$^{-2}$ operating over a distance, $\Delta x$, of
12,000 km, the largest distance separating two points on the spherical
shell, at a density of $1.4 \times 10^6$ g cm$^{-3}$, gives an
acceleration, $a = \rho^{-1} \, dP/dx \sim 10^4$ cm s$^{-2}$. A lower
bound to the time to spread 12,000 km, given this acceleration, is
$\sqrt{2 \Delta x/a} \sim 500$ s. This assumes free streaming, as in a
wind flowing over the surface. Diffusion in the convecting fluid could
take much longer. 

Past this point, when the first significant global variations in
temperature begin to persist, the surface of the star fractures into a
large number of smaller regions whose temperatures evolve
independently and can potentially develop large contrasts. The typical
scale of such regions is probably a pressure scale height, $\sim$300
km, so there could be hundreds of such cells. How many finally run
away within one second of another depends on the size of fluctuations
that are imprinted from an earlier era and is very difficult to
estimate without a full multi-dimensional simulation. There is no
compelling reason at this stage though to assume that there is only
one or that the distribution has any special symmetry.

Multiple ignitions would vary in starting time as well as location, so
``multi-point asynchronous ignition'' might best describe the
situation.  Numerical studies in 2D of the closely related problem of
ignition in classical novae suggest that single-point ignition is
unlikely \citep{Gla07,Cas10}. On the other hand, some observations of
x-ray bursts are consistent with ignition at a single point
\citep{Bha06}.  Recent work on sub-Chandrasekhar mass models for
supernovae has frequently assumed single-point ignition in three
dimensions \citep[e.g.][]{Sim10} or multi-point synchronous ignition
in two-dimensions \citep{Fin07}. Given the inherent cylindrical
symmetry of a 2D calculation where ignition occurs on tori, it would
be best to repeat these calculations in three dimensions. Spherical
(1D) ignition, single point (2D or 3D) ignition, and toroidal ignition
all have a preferred axis of symmetry. Multi-point asynchronous
ignition does not and may not so easily lead to detonation of the CO
core.  \citet{Gar99} studied five point asynchronous ignition in 3D
and found no compressional detonation of the CO core. However, they
did see strong collisions among the various helium detonations that
might have ignited a carbon detonation. These calculations need to be
repeated with higher resolution and lower mass helium, shells.

Our models show that the inward shocks must converge in a region
smaller than 100 km for carbon detonation to occur. For the lower
helium shell masses of the 10H series, about 50 km, was required. It
seems reasonable that this strong focusing effect will be lost if the
ignition is asymmetric by much more than 100 km, e.g., elliptical with
a minor axis of 100 km and a major axis of 300 km. Since the pressure
scale height is about 300 km, irregularities in ignition on this scale
seem reasonable.

These arguments also suggest a novel mixing mechanism that might be
applicable to a broad range of astrophysical problems. In the present
context, it suggests that, just as in classical novae, that there
probably is appreciable mixing between the CO dwarf and the helium
shell prior to runaway, at least for those explosions that ignite at
the interface. This mixture of carbon and helium is easier to detonate
than either alone \citep{She09}.

Unless the number of ignition points becomes very large though, most
of the burning during the runaway will occur in a helium layer that,
except at a few points, has not reached the conditions appropriate for
a runaway. The best starting model for a 3D simulation then is a model
from Table 1 or 2 for the density at the CO interface indicated, but
with a temperature that is cool enough everywhere on the star except
the ignition point(s) to assure that a spontaneous runaway does not
happen in several seconds.

\subsection{Ignition at Altitude}
\lSect{altitude}

It is generally agreed that a helium detonation ignited at a sharp
CO-He interface will not successfully propagate into the CO-core
\citep[e.g.,][]{Liv90b}, while a detonation ignited well above the
interface will, if the density is high enough
\citep[e.g.,][]{Arn97,Ben97,Gar99}. Many of our models, those labeled
with an ``a'' in \Tab{models1} and \Tab{models2}, ignited their helium
sufficiently high above the interface that an inwards as well as
outwards detonation was generated. In those cases, the helium
detonation transitioned successfully into an inwards moving carbon
detonation. Generally, those models had thicker helium shells and
colder white dwarf accretors than the others that ignited carbon
detonation by focused, star-wide compression. However, detonations are
stronger and more robust for nearly planar 1D geometry than for the
spherically divergent 3D geometry that actually characterizes ignition
at a single point. An unanswered question is how high above the
interface helium detonation must begin in order to propagate directly
into the carbon. \citet{Gar99} studied this issue numerically, but for
a limited choice of (high) densities around $5 \times 10^6$ g
cm$^{-3}$. Their results, admittedly with crude zoning, suggested that
below $4 \times 10^6$ g cm$^{-3}$, ignition had to occur more than 100
km above the CO interface in order for the helium detonation to
propagate into the carbon. Detonation thus became unlikely below this
density. What physics sets this critical density and how robust is the
constraint?  Once again, this is a question that 3D simulation will
probably answer soon, but we can make some estimates.

Assume that the helium detonation propagates into the carbon when its
momentum is sufficiently large to ignite a ``critical mass'' of
carbon. Then the necessary critical mass of helium must approximately
equal or exceed the critical mass of carbon at the same
density. Critical masses for {\sl carbon} detonation have been
determined numerically by \citet{Roe07b}, who found a sensitive
dependence of the size on density. For a density of 10$^{7}$ g
cm$^{-3}$, a length scale of 10 km suffices to detonate carbon, but
for $3 \times 10^6$ g cm$^{-3}$ the necessary size climbs to 100 km,
i.e, a substantial fraction of a pressure scale height in the present
models. This is consistent with \citet{Gar99}.  At 10$^6$ g cm$^{-3}$,
it became impossible to detonate the carbon. Many of our ``double
detonation'' models have densities at the CO interface in the range 1
- 3 $\times 10^6$ g cm$^{-3}$. The CO detonations in those cases must
be regarded as questionable.  However, helium detonation is more
energetic than carbon detonation and may be facilitated if there is a
mixed region of carbon and helium rather than a sharp
discontinuity. So this conclusion could be overly pessimistic. Also if
``edge-lit'' carbon detonation fails, compression of the core or
collisions of helium detonations \citep{Gar99} could still lead to
carbon detonation.

\subsection{Deflagration}
\lSect{multidefl}

If detonation does not occur, then a deflagration of some sort will
ensue (\Sect{hedefl}). For those models noted with a ``d'' in
\Tab{models1} and \Tab{models2}, convection is able to maintain a
nearly adiabatic gradient throughout the runaway and the result
resembles an ordinary, though powerful classical nova. All of the
other models will develop a strong density inversion and grossly
superadiabatic temperature gradients. While a flame in the traditional
sense of a nearly discontinuous composition change propagated by
conduction (or turbulence) and burning may not develop, the burning
rate is greatly in excess of what ordinary convection can carry, yet
the speeds are subsonic.

Like the detonation discussed in \Sect{multipoint}, we expect the
deflagration to be ignited at one or several points scattered around
the star. It will not happen simultaneously everywhere on a spherical
shell. The propagation of this sort of burning has been discussed by
\citet{Fry82} and \citet{Spi02}. The larger scale height associated
with the burning region leads to lateral spreading and mixing of the
helium layer. For a {\sl rotating} white dwarf, \citet{Spi02} gives an
approximate expression for the burning speed around the star,
\begin{equation}
v_{\rm flame} \  \approx \ \frac{\sqrt{g L}}{\tau_{\rm nuc} f}
\end{equation}
where $L$ is the pressure scale height, $g$, the gravitational
acceleration, $\tau_{\rm nuc}$, some estimate of the burning time
scale and $f = 2 \, \Omega {\rm Cos}(\theta)$. Here $\Omega$ is the
angular rotation rate of the white dwarf and $\pi/2 - \theta$ is the
latitude.  Typically the burning time, once a luminosity $\sim10^{46}$
erg s$^{-1}$ is reached, is about a second. Taking a value of $f \sim
1$ rad s$^{-1}$, i.e., moderate rotation, but not near break up, the
flame speed is $\sqrt{g L}$, or about 1000 km s$^{-1}$, and the time
to go half way round the star, about 10 s.  A similar value is
obtained from scaling arguments for a non-rotating star. The time for
buoyant material to float a scale height is about $\sqrt{L/g}$ and
during that time the burning might be expected to also mix with a
region of size $\sim L$. This time and length scale gives a speed of
approximately $\sqrt{g L}$.

In our one-dimensional models we used a Sharp-Wheeler scaling law for
the flame speed (\Sect{hedefl}), $v_{\rm flame} = 0.2 \, g_{\rm eff} \, t$.
  Except for the uncertain number out front, this is also $\sqrt{g L}$
  as can be seen by substituting $t = \sqrt{L/g}$ in the Sharp-Wheeler
  expression.

\section{Nucleosynthesis}
\lSect{nucleo}

\subsection{Nucleosynthesis in the Models}
\lSect{modelnucleo}

The nucleosynthesis of all models is summarized in Tables 5 - 10.
\Tab{radio} gives the abundances of the most interesting
radioactivities. \Tab{himddyield} and \Tab{lomddyield} give the
``production factors'' for major species in those models where both
the helium shell and carbon-oxygen core detonate. For these stars that
completely explode, the production factor is defined as the ratio of
the overall mass fraction of the given species in the exploded star
compared with its mass fraction in the sun. The actual mass produced
as a given species is thus this factor times the mass of the whole
star times the mass fraction in the sun as given by \citet{Lod03}.
Since the production factors for iron are in the range 200 to 700,
production factors less than 10 are considered unimportant and are not
given. The stars in Tables 8 and 10 leave a white dwarf
remnant. Nevertheless, so long as we confine our attention to species
that exist only in the ejecta, the definition of production factor is
similar - the average mass fraction in the ejecta {\sl plus remnant
  white dwarf} compared to the solar value. This is certainly true of
all the spcies we are interested in, namely those heavier than
neon. So the mass ejected is again the mass of the star (white dwarf
plus helium shell; Table 1) times the yield factor times the solar
mass fraction. For progenitors of solar metallicity, the yield factor
is thus the mass produced as a given species divided by the mass of
that species in the star before the explosion. Because the tables
start at silicon, all species given were produced in the
explosion. Since no $r$-process occurs here and because no initial
abundances heavier than $^{22}$Ne were included in the calculation,
nucleosynthesis above the iron group is not very well represented. An
exception is a number of ``p-process'' isotopes up to about A = 80
that are produced by alpha capture in the detonation of the helium
shell.

The yields of trace isotopes above mass 40, except for $^{56}$Ni, are
also quite uncertain since they depend on uncertain reaction rates for
radioactive targets (\Tab{prog}). The relevant cross sections have not
been measured in the laboratory, but come from theoretical estimates
which may be particularly uncertain for self-conjugate (Z = N)
nuclei. In many cases, the yields also depend upon the metallicity of
the star and accreted material. Solar values have been assumed here,
but that need not always be the case. These uncertainties may underlie
the large yields of nickel isotopes in all models and of $^{48}$Ti in
the helium detonations and deflagrations (\Tab{hedetyield} and
\Tab{deflyield}). These yields could easily be off by a factor of two
or more.

In general, the nucleosynthesis in those models where both helium and
carbon detonate resemble closely that reported previously in
\citet{Woo94}. Substantial amounts of $^{43,44}$Ca, $^{46,47,48}$Ti,
$^{51}$V, $^{50,52,53}$Cr, $^{55}$Mn, $^{56,57}$Fe,
$^{58,60,61,62}$Ni, $^{63,65}$Cu, and $^{64}$Zn are made. Especially
large are the productions of $^{44}$Ca, $^{48}$Ti and $^{52}$Cr made
as $^{44}$Ti, $^{48}$Cr and $^{52}$Fe respectively. The iron group is
also produced in substantial abundances, including of course $^{56}$Fe
made as $^{56}$Ni. Given the appreciable uncertainty in the cross
sections discussed above, the overall yield pattern is reasonably
close to solar, and not markedly inferior to that made in
Chandrasekhar-mass explosions \citep{Iwa99}.  In fact, the reduced
production of $^{58}$Ni and production instead of $^{44}$Ca,
$^{47}$Ti, $^{63,65}$Cu and $^{64}$Zn are significant
improvements. Cobalt and manganese are somewhat deficient however. The
species $^{44}$Ca is particularly interesting because it is not made
in the Chandrasekhar mass models and is inadequately produced in
core-collapse supernovae \citep{Tim96}. It is also inadequately
produced for the higher mass double detonations here
(\Tab{himddyield}), but is made sufficiently in the lower mass ones
(\Tab{lomddyield}). Overall though, it is difficult to make both a
solar ratio for $^{44}$Ca/$^{56}$Fe and sufficient $^{56}$Ni to give a
bright Type Ia supernova (though Model 10B may be an exception).

The rest of the models, where the carbon does not detonate, produce
too little $^{56}$Ni to be typical SN Ia. Still, their nucleosynthesis
is interesting.  The single (helium) detonation models produce many of
the same species as the double detonations nut production of the iron
group is greatly diminished (\Tab{hedetyield}).  The largest yields
are for $\alpha$-particle nuclei above $^{40}$Ca and their
neighbors. After decay (\Tab{prog}) these make $^{43,44}$Ca,
$^{46,47,48}$Ti, $^{51}$V, $^{52}$Cr, $^{60,61,62}$Ni, $^{64}$Zn, and
interesting amounts of $^{68}$Zn $^{74}$Se, and $^{78}$Kr. While the
$^{44}$Ca yields are now sufficiently high compared with $^{56}$Fe,
depending on the specific model and uncertain reaction rates, they are
still low compared with other trace species like $^{48}$Ti and
$^{61}$Ni. 

Some of the most interesting and most uncertain nucleosynthesis
happens in the helium deflagration models (\Tab{deflyield}).  Lines
from stable $^{40}$Ca would be very prominent in the spectrum of this
sort of explosion.  In some models, $^{44}$Ca is the largest
nucleosynthetic production. Lacking in $^{56}$Ni and having low
abundances of other short-lived radioactivities (\Tab{radio}), these
would be very faint supernovae and might happen frequently. It is
important when comparing the resulting light curves and observations
to follow the radioactive decays. Thus while $^{44}$Ca is an abundant
nucleosynthetic product, it is Ti that will be abundant in the
spectrum. The large production of $^{48}$Ti reflects synthesis as
$^{48}$Cr. This nucleus decays rapidly to $^{48}$V, but the half-life
of $^{48}$Va is 16 days, so vanadium lines should be prominent in the
peak light spectrum. The nucleus $^{45}$Sc is made as $^{45}$Ti, but
this decays to $^{45}$Sc in a few hours, so the large production of
$^{45}$Sc should give prominent lines of the element scandium.
Analysis of the network flows shows that $^{45}$Ti is made by a
variety of reactions during explosive helium burning. Chief among them
are $^{41}$Ca($\alpha,\gamma)^{45}$Ti,
$^{43}$Ca(p,$\gamma)^{44}$Sc(p,$\gamma)^{45}$Ti, and
$^{44}$Ti(n,$\gamma)^{45}$Ti. The $^{45}$Ti is destroyed by
$^{45}$Ti(p,$\gamma)^{46}$V. Its sysnthesis thus depends on uncertain
nuclear physics like the $^{44}$Sc and $^{44,45}$Ti cross sections, as
well as the neutron excess.

The large production of $^{44}$Ti (half-life 60 years) also has
implications for the late time supernova light curve. The large
productions indicated could keep the supernova shining at
$\sim10^{38}$ erg s$^{-1}$ for a century or more \citep{Tim96}.

None of the three classes of models discussed - double detonation,
helium detonation and helium deflagration - are presently mutually
exclusive.  SN Ia could possibly come from big dwarfs detonating their
cores (if the light curves and spectra were acceptable) and $^{44}$Ti
from helium deflagrations. For now, there is no nucleosynthetic reason
to exclude any of the sub-Chandrasekhar mass models; indeed there is
some advantage to having their contribution.

\subsection{Comparison with Previous Results}
\lSect{MPAcompare}

As noted above, the nucleosynthesis calculated here, for those models
where the helium detonates, agrees well with previous one-dimensional
studies of this class of model. It differs appreciably, however, with
that found in more recent multi-dimensional studies by \citet{Fin10}
and \citet{Kromer_2010}. These simulations found far less production
of $^{56}$Ni and production instead of intermediate mass
radioactivities like $^{44}$Ti, $^{48}$Cr, and $^{52}$Fe. In fact, the
nucleosynthesis for their {\sl detonations} resembles that coming from
our {\sl deflagrations}. An interesting comparison is Model 1 of
\citep{Fin10} and our Model 8B. Both began as 0.8 \Msun \ CO-cores
(0.81 for the Fink model) that accreted 0.13 - 0.14 \Msun \ of helium
and exploded. The mass of the two stars when they exploded was nearly
identical, 0.936 \Msun \ for their Model 1, 0.941 \Msun \ for our Model
8B, yet detonation of the helium shell in their Model 1 made $8.4 \times
10^{-4}$ \Msun \ of $^{56}$Ni, while Model 8B made 0.046 \Msun \ of
$^{56}$Ni. Why are the two results so different?

There are two reasons. One is the inherent hydrodynamical difference
between detonation in a spherical shell (our case) and a sliding
detonation that moves around the star at the He-CO-core interface
(theirs). In the latter, which should be more realistic, pressure is
lost from behind the detonation because the material has an open
direction for expansion. This reduces the pressure and temperature in
the detonation front and the helium burning reactions do not proceed
as far.

A larger effect, however, probably arises from the difference in
density at the base of the helium shell in the two initial models:
$3.7 \times 10^5$ g cm$^{-3}$ in their Model 1 and $1.55 \times 10^6$
g cm$^{-3}$ in our Model 8B (the ignition density in \Tab{models1} is
less for Model 8B because ignition actually occurs further out in the
star). Detonation at the larger density gives higher temperatures and
heavier elements.  But then why are the densities so different? The
assumptions underlying Model 8B have been presented in this paper.

Model 1 of \citet{Fin10}, on the other hand, is derived from a
construction provided to them by \citet{Bil07}. In that construction,
the helium envelope was assumed to be completely convective, i.e.,
adiabatic, with a temperature at its base (over $6 \times 10^8$ K)
adequate to cause helium to burn on a hydrodynamic time scale. As we
have discussed in \Sect{ignition}, the more correct assumption is that
convection freezes out when the nuclear time scale is comparable to
the convective turnover time.  Since convection at this point is still
very subsonic, a lower temperature is implied.  This makes physical
sense and has been demonstrated to be the correct assumption for
carbon ignition in Chandrasekhar mass models for SN Ia several times
\citep[e.g.][]{Zin09}. Consequently our temperatures at the base of
helium convective shells are in the range $2 - 3 \times 10^8$ K, not
$6 \times 10^8$ K. As the runaway proceeds, $6 \times 10^8$ K is
eventually reached in our models, but only in a small region. In
multi-dimensional simulations, the region will probably be smaller
still, on just one side of the star.  This high temperature is not a
characteristic value at the base of a spherically symmetric, adiabatic
envelope.  A lower entropy in the helium shell means that, with a
similar gravitational potential at the base, it would have a higher
density.  The difference probably accounts for the discrepancy in
density in the two models.

The best result will be achieved when our present pre-supernova models
are exploded in three-dimensions. That can be done. For now though, we
believe our nucleosynthesis, though one-dimensional, is closer to
correct. Table 3 of \citet{Fin10} shows a rapid rise in the fraction
of helium that burns to $^{56}$Ni as the density approaches 10$^{6}$ g
cm$^{-3}$ even in the multi-dimensional sliding-detonation models. We
find it difficult to detonate helium in models where the density is as
low as $3.7 \times 10^5$ g cm$^{-1}$.

\section{Light Curves and Spectra}
\lSect{lcurve}

We calculated synthetic light curves and spectra of a representative
subset of our models using the time dependent radiation transport code
SEDONA \citep{Kasen_MC} .  The code parameters were similar to those
described in previous transport calculations for SNe~Ia
\citep[e.g.,][]{Kasen_2009}.  The radioactive decay chain of
\Nifs\ was included as well as those of $^{52}$Fe and $^{48}$Cr, along
with a gamma-ray transport scheme to determine the energy deposition.
Local thermodynamics equilibrium (LTE) was assumed to determine the
ionization and excitation state of the ejecta.  While often a good
approximation for SNe~Ia in the early phases, LTE breaks down at later
times when nebular line emission and non-thermal ionization effects
become significant.

The light curves of the models show considerable diversity in
brightnesses and duration as illustrated in
Figure~\ref{fig:vband_lc}.  The fundamental observable properties of
the entire set of model light curves and spectra are summarized in
Figures~\ref{fig:mb_dm15}-\ref{fig:mb_vp}.  We divide our discussion
into those models in which the entire star (both shell and core)
explode, and those in which only the helium shell is disrupted.

\subsection{Light Curves and Spectra - Full Star Explosions}
\lSect{fullstarlite}

Double detonation explosions of sub-Chandrasekhar white dwarfs were
investigated some time ago as possible models for normal SNe~Ia
\citep{Woo94,Liv95} but the idea eventually fell out of favor when
calculations of model spectra did not match those of observed events
\citep{Hoeflich_1996,Nugent_1997}.  In particular, the spectral
features of intermediate mass elements (IMEs) were too weak in the
models compared to observations, and the continuum was too blue.  More
recent, multi-dimensional calculations confirmed the weakness of IME
features, but suggested that the models were in fact too red compared
to observations, especially after maximum light
\citep{Kromer_2010}. In either case, the spectral peculiarities are
due to the outer shell of radioactive material produced in the
detonation of the helium layer.  If the helium layer is omitted from
the model altogether, the light curves and spectra of detonated bare
sub-Chandrasekhar CO white dwarfs actually agree quite well with
observations \citep{Sim10}.  The mass of burned material in the outer
shell is therefore crucial to the predicted observables.

We find that the models in our set which assumed a ``cold" white dwarf
(i.e., one with initial luminosity of 0.01~\Lsun) generally have
maximum light spectra that do not resemble observed normal SNe~Ia.
Figure~\ref{fig:spec_compare} shows as a typical example the maximum
light spectra of Model 9C.  The IME absorption features (Si~II, S~II,
Ca~II) in the model are too weak to match the normal Type~Ia SN2003du,
and overall the model more closely resembles the spectroscopically
peculiar SN~1991T (although the model lacks the two prominent Fe~III
lines near 4300~\AA\ and 5000~\AA\ seen in SN~1991T).  At later times
($\sim 7$ days past peak) weak IME lines do begin to appear and the
spectrum starts to look more normal (\Fig{9C_series}).  In this
sense, the early spectral evolution of these models
resembles the class of SN~1999aa-like supernovae \citep{Li01}, which
appear to be intermediate between the normal and spectroscopically
peculiar SNe~Ia.  At yet later times ($\sim 2$ weeks after peak) the
line blanketing from iron group elements in the outer layers becomes
strong, leading to a very red spectrum, in contrast to what is
typically observed in either the normal or the SN~1999aa-like SNe~Ia.
A similar spectral evolution characterizes most of the ``cold" models
we calculated, except those with the smallest accreted shell masses.
The observational counterpart to the bulk of these models is unclear.

The weakness of the IME features in these models is due to several
effects.  First, the radioactive heating from the shell serves to
further ionize Si~II, S~II and Ca~II, thereby reducing their opacity.
Second, the absorption features are diluted because a portion of the
continuum luminosity is being generated in the radioactive shell above
the IME line forming region.  A third 
reason why the IME features are weak in these models has to do with
the dynamical effects of the helium shell on the velocity of ejected
IMEs.  Regardless of whether the outer layers burn to heavier elements
or not, the mass in the shell serves to decelerate the detonated core,
thereby limiting the maximum velocity of the IME ejected in the CO
detonation. This effect confines the IME to rather narrow shell of
velocity range 11,000-13,000~\kms.  This is in contrast to observed
SNe~Ia (as well as standard delayed-detonation models) in which the
IME velocities typically span a range $9,000-15,000~\kms$.  The
thinness of the IME layer in the present models restricts the
geometrical area over which this material covers the photosphere,
reducing the strength of the features.  Thus, even if we turn off (by
hand) the radioactivity in the outer shell, the IME features still
remain too weak to reproduce observations (\Fig{9C_series}).  In
principle, Raleigh-Taylor instabilities at the ejecta-shell interface
could broaden the radial distribution of IMEs and increase the depth
of the absorption features, an effect not captured in these 1D
simulations.  In addition, the
structure of the ejecta can be different and aspherical in multi-dimensional
models.  \cite{Fin10} show that when the detonation of the helium
shell occurs at a point, IMEs are ejected at significantly higher
velocities on one side of the ejecta while they are confined
to a narrow velocity range on the other.

In order to systematically explore the effects of the outer helium
shell on the observables, we compare, in Figure~\ref{fig:10_series},
the spectra of four models which had an identical core mass of
$1~\Msun$, but which varied in the amount of material which had
accreted into the helium shell at the time of detonation.  Of these
models, 10HC had the smallest accreted mass: $M_{\rm acc} =
0.0445~\Msun$ within which was produced only $8 \times 10^{-5}~\Msun$
of \Nifs.  The maximum light spectrum of this model shows moderately
strong IME lines which are in reasonable agreement with those observed in
normal SNe~Ia (\Fig{10HC_spec}).  Model 10A, with a slightly larger
helium shell ($M_{\rm acc} = 0.064~\Msun$) also shows some IME
absorption features, albeit weaker than what is typically observed.
Models 10B and 10D, which had yet larger accreted masses, do not show
any IME absorptions at all.

We conclude that the spectra of some double detonation models do in fact
resemble normal SNe~Ia, but only if the accreted mass is small, $M_{\rm
  acc} \la 0.05~\Msun$, and only if a small amount of this is burned to
radioactive material.  In the present model set, this was only
regularly achieved for models in which the white dwarf was ``hot"
i.e., had an initial luminosity of $1~\Lsun$.  \Fig{hot_spectra} shows
the spectra of a four such ``hot" models which varied in the initial
white dwarf mass and consequently the amount of \Nifs\ produced in the
core detonation. All of these models have normal looking spectra. In
the case of the ``cold" white dwarfs ($L = 0.01~\Lsun$), on the other
hand, the star accretes a larger mass before the helium shell
explodes, and the maximum light spectrum generally appears peculiar,
as seen for Model 9C (Fig.~\ref{fig:10_series} and
\ref{fig:9C_series}).

The presence of the outer shell of burned material also impacts the
light curves of the models.  Figure~\ref{fig:LC_9C} shows two
separate light curve calculations for Model 9C, a typical full star
explosion with a ``cold" white dwarf.  In the first calculation, we
included the outer shell of material, while in the second, that shell
was removed after explosion -- i.e., once the ejecta had reached the
homologous expansion phase.  The calculations illustrate two main
effects of the outer shell on the observables: (1) At and before peak,
the decay of \Nifs\ and other radioactive isotopes in the shell leads
to additional heating at the surface, causing the light curves to be
slightly brighter and bluer at peak.  (2) A week or two after peak,
the iron group material in the shell begins to cool to a singly
ionized state, which increases the line opacity in the blue wavelength
bands.  Fluorescence in these lines redistributes flux from shorter to
longer wavelengths \citep{Pinto_Eastman_2000, Kasen_2006}, causing the
colors to become progressively redder after peak,  
In the multi-dimensional
models studied by \cite{Kromer_2010}, iron group elements are ejected
at significantly higher velocities on one side of the ejecta, causing these line 
blanketing effects to set in even earlier (for certain viewing angles).  
In our models, the outer shell results in a
faster decline in the U- and B-band light curves after peak, leading
to a poor (although not terrible) fit to the observed light curve of
the normal Type~Ia SN~2003du \citep{Stanishev_2007}.  
The overall fit
to observations is significantly improved when the outer shell is
removed.

The light curves of other full star explosions show similar behavior
to that described above, but vary in brightness depending on the
\Nifs\ mass produced in the explosion.  The B-band peak magnitudes
range from -18.3 to -19.6 and the B-band decline rates from \dmb =
0.7-2.5~mag.  Figure~\ref{fig:mb_dm15} shows that the set of models
obeys a rather tight width-luminosity relation, however the offset and
slope of models with ``cold" white dwarfs are not in good agreement
with recent observed values for the relation \citep{Folatelli_2010}.
Systematically, the models decline too rapidly for a given brightness,
a consequence of the presence of the burned helium shell.  The models
with ``hot" white dwarfs are in somewhat better agreement with the
normalization of the observed WLR, but have too steep a slope, with
the decline rate changing by only a small amount (0.2~mag) for a $\sim
1$~mag change in peak B-band magnitude.

\subsection{Light Curves and Spectra - Helium Shell Deflagration 
and Detonation Models}
\lSect{defllite}

The models in which only the helium shell explodes (leaving the CO
core intact) produce fainter and more rapidly evolving transients due
to the smaller ejecta masses.  The observables of helium shell
detonations (``.Ia" explosions) have been studied previously by
\citet{Shen_2010}.  We explore here a wider range of models which show 
diversity in both their peak magnitudes and spectroscopic properties.
 The helium shell deflagrations, which involve lower explosion
energies than the detonations, have distinct light curves and spectra
and are explored for the first time here.

For models assuming initially ``cold" white dwarfs, the helium shell
detonations produce moderate quantities of \Nifs, which powers the
light curves.  Figure~\ref{fig:mb_tpeak} shows that within this set of
models, the B-band peak magnitudes vary from -15 to 
-18.3.  The B-band rise times range from 4 to 9 days, and correlate
tightly with the brightness.  This is because the more massive shells
both have a longer diffusion time and typically produce a larger mass
of radioactive isotopes.  The decline rate of the B-band light curves
varies from rapid ($\dmb = 2$~mag) to extremely rapid ($\dmb =
7$~mag).  These fast decline rates reflect not only the low ejecta
mass, but also the strong evolution of the colors to the red as the
ejecta cool over time.  The photospheric velocities of the helium
shell detonations are always around 7,000-10,000~\kms\ at peak, and
the spectral features are broad, with absorptions extending to $\sim
15,000$~\kms\ (Figure~\ref{fig:shell_spec}).  
The spectra of models which 
assumed an initially ``cold" white dwarf generally
lack strong IME absorptions of Si~II and S~II, and instead show broad
features of titanium and iron group elements, similar to those studied
in \citet{Shen_2010}.

The models which assumed an initially ``hot" white dwarf generally
produce less \Nifs\ in the shell explosion and greater abundance of
IMEs.  In this case, radioactive $^{48}$Cr becomes the significant
source of heating.  \Fig{8HBC1} shows the light curves and spectra of
Model 8HBC1, which ejected 0.097~\Msun\ of ejecta, but only $1.5
\times 10^{-4}$~\Msun\ of \Nifs, along with $9 \times
10^{-4}$~\Msun\ of $^{48}$Cr.  An additional $2.5\times 10^{-4}$ of
$^{52}$Fe was produced, but given the short decay time of this isotope
and its daughter, $^{52}$Mn, it does not contribute significantly to
powering the observed transient.  The B-band light curve of 8HBC1
peaks in only 3 days at a remarkably dim magnitude of -13.5. The
maximum light spectrum shows significant absorptions due to silicon and
sulfur, and
overall resembles the spectra of observed subluminous SNe~Ia such as
SN~1991bg.  This is in contrast to the shell explosions of the ``cold"
white dwarf models, which lacked these features.

The two helium shell deflagration models calculated here (8DEFL and
10DEFL) show dramatically different properties than the helium
detonations.  Since the total mass of \Nifs\ is negligible, the light
curves are powered primarily by the decay of $^{48}$Cr.  Both models
are extremely dim (B-band peak magnitudes of -14.5 and -15) though in
principle deflagration of more or less massive shells could produce
brighter or dimmer events as well.  Unfortunately, we were unable to
obtain stable numerical models of thick helium shell deflagration
using our simple one-dimensional treatment of the burning.  Burning in
the deflagration models is mostly incomplete and a significant
percentage of helium remains.  The explosion energy per unit ejected
mass is thus low, which leads to a longer rise time and slower decline
rate relative to a detonation model of the same brightness.  The
photospheric velocities in the deflagration models are very low ($\sim
4000~\kms$ at peak) and the spectra are characterized by numerous
narrow absorption features (Figure~\ref{fig:shell_spec}).  Many of the
lines are from species not commonly seen in SNe~Ia, including features
from Sc~II.  Helium lines are not seen in the synthetic spectra,
however it is well known that these lines are extremely sensitive to
non-thermal excitation effects due to radioactive decay products.
Thus, non-LTE calculations will be needed to predict the actual
strength of the helium lines.

Recent observations have revealed a diverse class of dim, brief
transients which have occasionally been linked to helium shell
explosions.  These observed events show a variety of properties.  In
some cases, for example, the observed photospheric velocities are
relatively high and the light curve declines very rapidly (e.g.,
SN~2002bj \cite{Dovi_2010}, SN~2005E \cite{Perets_2010}, SN2010X
\cite{Kasl10}).  These properties generally resemble the helium
detonation models, however a detailed case-by-case spectral comparison
is needed before a firm link can be drawn.  Most of the models do
predict a a significant Ca~II IR triplet feature, which is observed in
many events, while only a subset of the models predict the Si~II and
S~II features that are occasionally seen.  On the other hand, many of
the observed events show absorption features of carbon and oxygen,
which are not seen in our models, While the 1D model do not ejecta a
significant amount of C/O, it is possible that in more realistic
multi-D calculation some C/O may be dredged up from the core and mixed
into the ejected shell material.
  
In other observed peculiar SNe, the spectral features have very low
velocities which are inconsistent with the helium detonations (e.g.,
SN~2008ha \cite{Foley_2009}).  It is interesting to suggest that these
low velocity events may represent helium shell deflagrations. More
detailed model calculations will be needed to clarify the situation.
In particular, the 1-D models do not properly capture the large scale
mixing that is characteristic of realistic 3-D deflagration burning.
  
\section{Conclusions}
\lSect{conclude}

We have studied one-dimensional models for the explosion of
sub-Chandrasekhar mass white dwarfs accreting helium in a binary
system. The models included white dwarfs in the mass range 0.7 to 1.1
\Msun \ that, at the onset of accretion, had luminosities of
either 0.01 or 1 solar luminosities, models denoted here as ``cold''
or ``hot''. The accretion was followed, as well as the convective
phase leading up to the explosion. Depending upon the relative
magnitudes of the nuclear, convective, and sonic time scales at the
time when energy generation reaches a maximum (\Sect{channels}), four
possible outcomes were found and explored. 1) A runaway in which
convection carries the energy from start to end, an outcome that
resembles classical novae, though with much greater mass ejected. 2)
Helium deflagration, in which rapid subsonic burning creates a density
inversion at the base of the convective helium shell leading to rapid
mixing and burning, also ejecting only the helium shell and a bit of
dredged up carbon. 3) Helium shell detonation that leaves behind a
hot, but intact CO dwarf. 4) Detonation of the helium shell and of the
CO core, either by compression of the latter or direct propagation of
the helium detonation into the core. Representative examples of all
these outcomes can be found in \Tab{models1} and \Tab{models2}. The
nucleosynthesis of all models was calculated and the light
curves and spectra of many. Possible multi-dimensional modifications
of our results were discussed in \Sect{multid}.

We find, in agreement with \citet{Bil07} and \citet{She09}, a minimum
helium shell mass for detonation of M$_{\rm min} \approx 0.025$ \Msun
\ (\Tab{models2}) for a 1.1 \Msun\ white dwarf with a high crustal
temperature. This is substantially less than found in earlier studies
\citep{Woo94} because of the use of finer zoning during the accretion,
which alters the entropy of the accreted material (\Sect{accretion};
up to 40\% change in M$_{\rm min}$); the inclusion, in the energy
generation rate, of new reaction sequences for burning helium and
carbon (\Sect{nuclear}; a factor of two decrease in M$_{\rm min}$);
and the inclusion, in the study, of hotter, higher mass CO-cores (an
additional factor of two decrease in M$_{\rm min}$).  Even for hot
white dwarfs however, this minimum mass increases to 0.10 \Msun \ for
a white dwarf of 0.7 \Msun\ (\Fig{mcrit}), and for cooler white dwarfs
with initial luminosities at the onset of accretion near 1\% solar,
these minimum critical masses are substantially larger
(\Fig{detconds}).

We confirm \citep{She09} that the reaction sequence
$^{12}$C(p,$\gamma)^{13}$N($\alpha,$p)$^{16}$O plays a critical role
in accelerating the burning during the runaway, facilitating helium
detonation. Including this reaction sequence had the effect of
reducing the critical density required for helium detonation in a
spherical model from about $1 \times 10^6$ g cm$^{-3}$ to about $5
\times 10^5$ g cm$^{-3}$ and this allowed lower mass helium shells to
detonate.  However, a study of helium detonation initiated at a point,
rather than in a plane suggested that, depending upon the geometry of
ignition, the required density might be $1 \times 10^6$ g cm$^{-3}$
even when this reaction sequence is included (\Sect{hedetcond}).

At the lower densities in the smallest helium shells,
helium detonation often produced intermediate mass elements,
especially $^{40}$Ca, and not $^{56}$Ni. For those models where the
CO-core detonated, this resulted in models whose spectra and light
curves more closely resembled common SN Ia \citep{Sim10}.  We also
found carbon mass fractions from pre-explosive burning over 10\% in
the region where the detonation occurs.

Explosions with helium density so low that the nuclear time scale
never becomes shorter than the convective time scale (\Fig{detconds})
lead to helium novae, not supernovae. The subsequent evolution of
these systems after ignition and their final nucleosynthesis was not
followed in detail, but their composition will be rich in calcium and
other lighter elements and, of course, devoid of hydrogen. Given the
large nuclear energy content of the helium fuel, $\sim10^{50}$ erg,
these novae could, in principle, shine at the Eddington luminosity for
a very long time. However, a more likely outcome is the ejection of
the accreted shell over a shorter time with only partial burning of
the helium. This sets the stage for recurrence.

The outcome of the other models depended critically upon how
multi-dimensional aspects of the problem were approximated
(\Sect{multid}), especially how convection was treated the last
moments of the runaway (\Sect{ignition}) and the location of and
temperature gradients in the vicinity of the ignition point(s)
(\Sect{hedetcond}).  Here, these conditions were determined by the
maximum power allowed to develop in the convection zone before
convection was turned off. Arguments are given (\Sect{ignition}) that
this critical luminosity lies between approximately 10$^{46}$ and
10$^{47}$, corresponding to a temperature at the base in the range
$2.2 - 3.7 \times 10^8$ K (\Fig{detconds}; \Tab{ignpoint},
\Sect{ignition}). For these conditions the nuclear burning time scale
becomes shorter than the convective transport time scale. Once
convection is halted, the temperature gradient becomes increasingly
superadiabatic and the burning, localized to a small fraction of the
pressure scale height.  Depending upon the temperature gradient when
convection freezes out, the helium may burn as a deflagration or a
detonation.

Lacking clear indication of the correct choice, we explored both
possibilities, though most were of the helium detonation variety. In
all those models where helium detonated, the carbon core also
detonated, though very fine zoning, less than 50 km, was sometimes
required to make this happen in the models with low mass helium
shells.  By imposing spherical or cylindrical symmetry on the
ignition, helium detonation will always lead to the detonation of the
carbon-oxygen white dwarf in a calculation with sufficient resolution
and low numerical damping. However, we also argued that the ignition
may be multi-point, asymmetric and asynchronous
(\Sect{multipoint}). Earlier exploration of this sort of ignition in
3D models suggests that CO detonation by compression is not robust
when cylindrical symmetry is lost \citep{Gar99}. The degree of
asymmetry required to cause defocusing may not be large, perhaps no
more than a few hundred km (\Sect{multipoint}).

A substantial fraction of our models, those labeled with an ``a'' in
\Tab{models1} and \Tab{models2}, ignited their CO-cores more directly
by a helium detonation that crossed smoothly into the carbon. However,
the assumed geometry was again critical. We confirm previous studies
showing that helium detonation ignited at the helium-CO interface does
not yield CO-detonation. Some altitude is required. Based upon
calculations of critical masses of carbon and oxygen by \citet{Roe07}
and prior studies by \citet{Gar99}, we estimate that ``edge-lit''
carbon detonation may only happen for densities at the interface above
about $3 \times 10^6$ g cm$^{-3}$. At lower densities the radius of
carbon required to sustain a detonation exceeds 100 km, which is a
rough upper bound on the altitude of the helium detonation.  By this
criterion, many of the models considered here would not detonate their
carbon at the edge as indicated. Compressional ignition at the center
would still occur for these one-dimensional models though, and this
constraint is conservative. Helium detonations are stronger than
carbon detonations and are already well formed before they reach the
carbon. Three-dimensional simulations of this problem are feasible and
should be carried out.

Nevertheless, we focused on models that detonated their CO cores
because they are the only kind of sub-Chandrasekhar mass model capable
of producing a supernova resembling common Type Ia events. Most of the
models in this category had helium shell masses $\sim$0.1 \Msun \ and
made substantial $^{56}$Ni in those shells, which they ejected with
high velocity. Generally, the spectrum of these models did not
resemble those of ordinary Type Ia supernovae. In some models,
however, especially the hotter and heavier white dwarfs, the helium
shell mass was smaller (\Fig{mcrit}).  However, while
sub-Chandrasekhar mass models may constitute some important fraction
of common SN Ia, there are a number of potential objections to them as
a general solution to the ``SN Ia problem''. Some can be addressed
with further computation and may be resolved shortly. Others might be
addressed with further observation.

1) The models that resemble closely SN Ia, for example 9HC, 10HC
(\Fig{hot_spectra}), and 11HD, occupy a relatively narrow niche of
parameter space (\Fig{mcrit}). For low luminosity white dwarfs, none are
found, and for high luminosity white dwarfs, the necessary range of
accretion rates is quite narrow. Why, for example, does
a 1.0 \Msun \ white dwarf have L = \Lsun \ and not L = 0.01 \Lsun
\ and $\dot {\rm M} = 4 \times 10^{-8}$ \Msun \ y$^{-1}$ and not 3 or
$5 \times 10^{-8}$ \Msun \ y$^{-1}$?  

One possibility is that the explosion occurs in a binary system with a
gradually declining accretion rate so that repeated nova-like
outbursts precede a single helium detonation \citep{Bil07}. The
nova-like outbursts could have left the outer layers of the white
dwarf in a heated state and our ``hot white dwarf models'' might then
be appropriate. The last flash would have a helium shell mass just
over the minimum value for helium detonation and could trigger a CO
detonation that destroyed the system. But then the accretion rate must
always decline just enough to produce a detonation with a minimum mass
and not so much as to give a thick helium shell. Also, the mass of the
white dwarf must increase during the presupernova evolution to 1.1
\Msun \ to explain typical SN Ia luminosities. Nova-like outbursts at
the high accretion rate will actually shrink the white dwarf
mass. Perhaps this can all be made to work, but is it so much harder
to grow the mass to 1.38 \Msun \ (and make a Chandrasekhar-mass
explosion), rather than 1.1 \Msun?

2) Despite numerous papers on the subject, neither the detonation of
the helium nor of the carbon has been conclusively demonstrated,
especially for densities less than 10$^6$ g cm$^{-3}$
(\Sect{hedetcond}). Helium detonation can be averted if shallow
temperature gradients are not maintained by efficient convection at
late times.  Even if the helium detonates, carbon detonation is not
assured for the general case of asymmetric, asynchronous, multi-point
ignition (\Sect{multid}). The existence of supernovae like SN~2010X
\cite{Kasl10}) is also suggestive that, at least occasionally,
helium detonation or deflagration is realized without detonating the
CO core.

3) Even the models that closely resemble SN Ia differ in subtle ways.
The spectral features due to intermediate mass elements, while present
in the models, are not as deep as seen in many normal SNe~Ia.  In
addition, the minimum velocity of IME in the models is typically no
lower than $11,000~\kms$, whereas in observed SNe~Ia the IMEs are
generally observed to extend to lower velocities, $v \approx
9,000$~\kms.  We caution, however, that ours are 1-D models and both
the nucleosynthesis and degree of radial mixing of intermediate mass
elements may differ in multi-D models.

4) Past models may have erred in taking too low a density for the
helium shell when it detonated. This density came from assuming a)
that the criterion for transitioning to a hydrodynamical simulation
was the nuclear time scale equal the sonic time scale when it should
be the nuclear time scale equals the convective transport time
scale. Also the conditions at the runaway point were adopted for the
whole helium shell, not just the small locus of ignition. Many of our
models also ignite well above the CO core - helium interface so there
is denser helium beneath.

If the CO-core does not detonate all the time, there could be a
diverse set of fainter transients awaiting discovery. These could
range in peak brightness all the way from bright radioactive powered
novae ($\sim 10^{39}$ erg s$^{-1}$) to subluminous Type Ia supernovae
(\Fig{vband_lc}). The faintest events would be powered by just a trace
  of $^{48}$Cr \citep[e.g., \Fig{8HBC1} and Models 7B1, 9A1, 10HC1,
    and 10HCD1; see also][]{Bil07}. Each leaves behind a hot white
  dwarf that might potentially be detectable for thousands of years
  \citep{Woo86}. For lower accretion rates and cooler white dwarfs,
  larger masses of $^{56}$Ni and smaller masses of $^{48}$Cr and
  $^{52}$Fe are produced resulting in a brighter supernova - though
  still much fainter than an ordinary SN Ia. If these transients
  exist, they await discovery.

A novel outcome explored here is helium {\sl deflagration}. If the
temperature gradient at the base of the helium shell is too steep at
the time of the final runaway to produce a supersonic phase velocity
for the burning rate and form a detonation, the burning will continue
to steepen that gradient until a strong density inversion
develops. The rise of buoyant burning material leads to mixing that
accelerates the burning rate. In about 10 s, the burning could sweep
around the star, but no supersonic speeds are developed anywhere. We
modeled this sort of explosion (\Sect{defl} using a Sharp-Wheeler
model for the advancement of the Rayleigh-Taylor instability. This is
a gross approximation, and the value used for the uncertain constant
``A'' in \Eq{paramdefl}, which determines the degree of 
burning, was on the high side of its expected range.  Any realistic
model will have to be computed in multi-dimensions (though see
\Sect{multidefl}. However, some generic features may survive. First,
far less helium burns than in a detonation so the supernova is less
energetic; the speeds are much slower. Second, for a given density,
burning in a deflagration occurs at a lower temperature than in a
detonation. For the two models studied, substantial amounts of
$^{44}$Ti and $^{48}$Cr were produced and little $^{56}$Ni. These
transients would be quite faint. Detonation in the same models gave
large energies and large $^{56}$Ni mass fractions.

As noted in previous studies, the nucleosynthesis from the
sub-Chandrasekhar mass models is acceptable, perhaps even superior to
that reported for standard Chandrasekhar mass models. Large
overproductions of $^{58}$Ni are avoided in the sub-Chandrasekhar mass
models, and important contributions to $^{44}$Ca, $^{47}$Ti,
$^{63,65}$Cu and $^{64}$Zn are made (\Sect{nucleo}; Tables 5 -
10). Especially important is the species $^{44}$Ca, made as
$^{44}$Ti. The ratio of $^{44}$Ca/$^{56}$Fe is subsolar in models
where the entire star detonates and the CO-core exceeds one solar
mass. But lighter models, especially the helium deflagrations can
produce large ratios of $^{44}$Ca to iron. Since $^{44}$Ca is
inadequately produced in both massive stars and Chandrasekhar-mass SN
Ia, the present abundance of $^{44}$Ca in the sun thus a strong
argument that sub-Chandrasekhar mass explosions have occurred. It
further suggests an important component in which either the CO-core is
lighter than 1 \Msun \ or does not detonate. These would not be
ordinary SN Ia. A vexing uncertainty in these models, however, are the
poorly determined reaction rates for unstable nuclei heavier than
calcium with Z $\approx$ N. While we did not carry out a sensitivity
study, we expect factor of two or more variations in key abundances
could occur as a consequence of uncertain rates
\citep[e.g.][]{Hof10}.  With present choices for these rates the large
production of $^{48}$Ti and $^{61}$Ni (as $^{48}$Cr and $^{61}$Zn)
often preclude the production of anything else.

We explored the spectra and light curves of a representative subset of
our models and found, in agreement with \citet{Sim10}, that a wide
range of luminosities are possible for the SN Ia resulting from
sub-Chandrasekhar mass explosions that detonate their CO cores, even
very brilliant ones if the requisite accretion rate and CO-dwarf mass
are realized. These luminosities obey, roughly, a width luminosity
relation, however the slope and normalization do not agree well with
recent observations.  In general, the models decline more rapidly in
the B-band than do the observations. We differ fundamentally with
\citet{Sim10} and \citet{Kromer_2010} regarding the nucleosynthesis
expected from helium detonation in these models (\Sect{MPAcompare}).

The observational counterparts of the high helium shell mass,
double-detonation models remain unclear. These did not closely
resemble those of common SN Ia.  In all cases, the presence of the
outer shell led to a peculiar, fairly featureless spectrum around
maximum light.  This problem is due in part to the opacity of
iron-group element in the shell, as has been emphasized in previous
studies.  We find in addition that the dynamical effects of the shell
are important -- the mass of the shell decelerates the detonated core,
thereby restricting the velocity spread of the IME ejected in the CO
detonation to a narrow shell $\sim 11,000-13,000$~\kms.  This reduces
the strength and blueshift of the IME absorption features and is
largely responsible for the poor fit to observed SNe~Ia.  In order to
reproduce the spectra of normal SNe Ia, the mass in the helium shell
should be small, $\la 0.05$ \Msun\ and should be relatively free of
radioactive isotopes.  It is possible that some explosions with larger
helium shells could be be associated with the spectroscopically
peculiar class of SN~1999aa or SN~19991T-like events.  Or perhaps such
low accretion rates are not frequently sustained long enough to cause
an explosion in nature.

Models in which only the shell explodes produce relatively dim and
fast transients.  Within this class, we found a wide range of peak
brightness (peak B-band magnitudes ranging from -13.5 to -18.5) and
diverse spectroscopic properties.  In the models which assumed an
initially ``cold" white dwarf, the helium shell detonations generally
produced transients powered by \Nifs\ decay.  The spectra were
dominated by titanium and iron group lines and lacked IME features,
similar to the ".Ia" explosions studied by \cite{Shen_2010}.  The
``hot" white dwarf models, on the other hand, generally produced lower
abundances of \Nifs\ and were powered mainly by $^{48}$Cr.  Their
spectra showed strong IME features, similar to low-luminosity SNe~Ia
like SN~1991bg.  The models in which the helium shell was assumed to
explode by deflagration, rather than detonation, had distinct
observational properties altogether.  The velocities of line
absorptions in these models were extremely low ($\sim 4000$~\kms).
Calcium lines were strong, and some of the features were due to
unusual species such as Sc~II. Spectral lines from (radioactive)
titanium and vanadium might also be searched for. Only two such shell
deflagrations were modeled here, each with peak B-band magnitudes
around $-15$, but in principle a wider range of luminosities and
durations are possible, depending on the mass of the shell at the time
of explosion.  We suggest that the scenario may be relevant for
explaining the class of dim transients showing low velocities in their
spectra \citep[e.g., SN~2008ha][]{Foley_2009}.

Fortunately, it is feasible to do all these problems in three
dimensions. The convective ignition can be studied using a low Mach
number code \citep[e.g.][]{Non10}, and several groups are developing
the capabilities to study multi-dimensional deflagrations and
detonations. Three dimensional light curves and spectra can also be
computed \citep{Kas07}. Given the forthcoming all sky surveys that
will discover many new optical transients, and the probability that
phenomena similar to those explored here should exist at some level,
we expect this to be a fertile field for further exploration.

\acknowledgements

This research has been supported by the DOE SciDAC Program under
contract DE-FC02-06ER41438; the National Science Foundation (AST
0909129) and the NASA Theory Program (NNX09AK36G). We acknowledge
useful discussions with Ron Taam on the treatment of accretion in
Lagrangian codes, and with Fritz R\"opke, Stuart Sim, and Michael
Fink, helping to elucidate the differences between our models and
theirs. Rob Hoffman and Alex Heger helped develop the reaction network
and Kepler code. Discussions with Gary Glatzmaier and Mike Zingale
helped to elucidate the nature of the baroclinic
instability. Correspondence with Ken Shen and Lars Bildsten helped to
explore important aspects of accretion, helium detonation, and of
their published works. The anonymous referee was extraordinarily
helpful in helping us to clarify and, in some cases, substantially
revise, our major conclusions.

\newpage


\clearpage

\begin{figure}
\begin{center}
\includegraphics[width=0.475\textwidth]{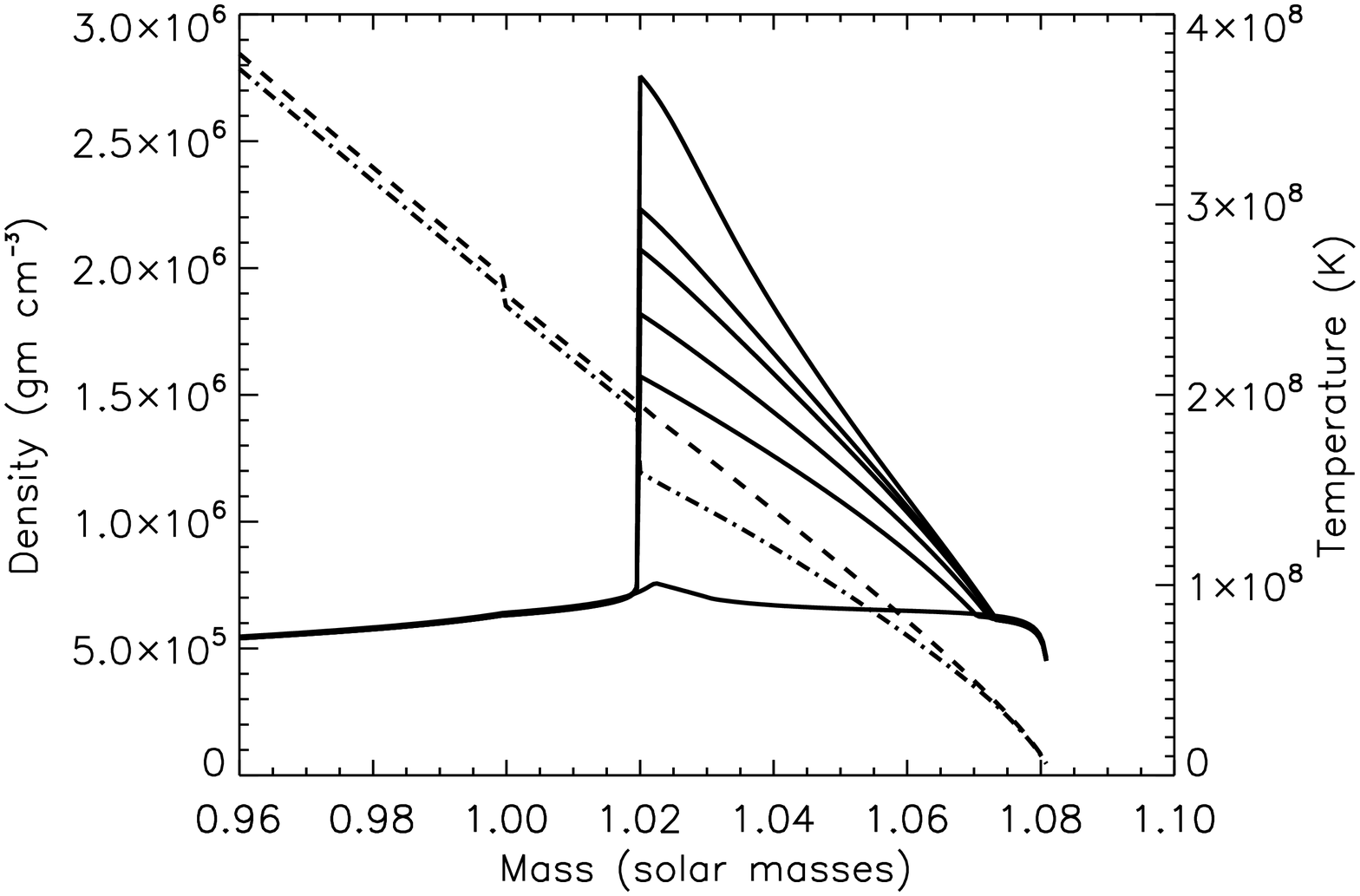}
\hfill
\includegraphics[width=0.475\textwidth]{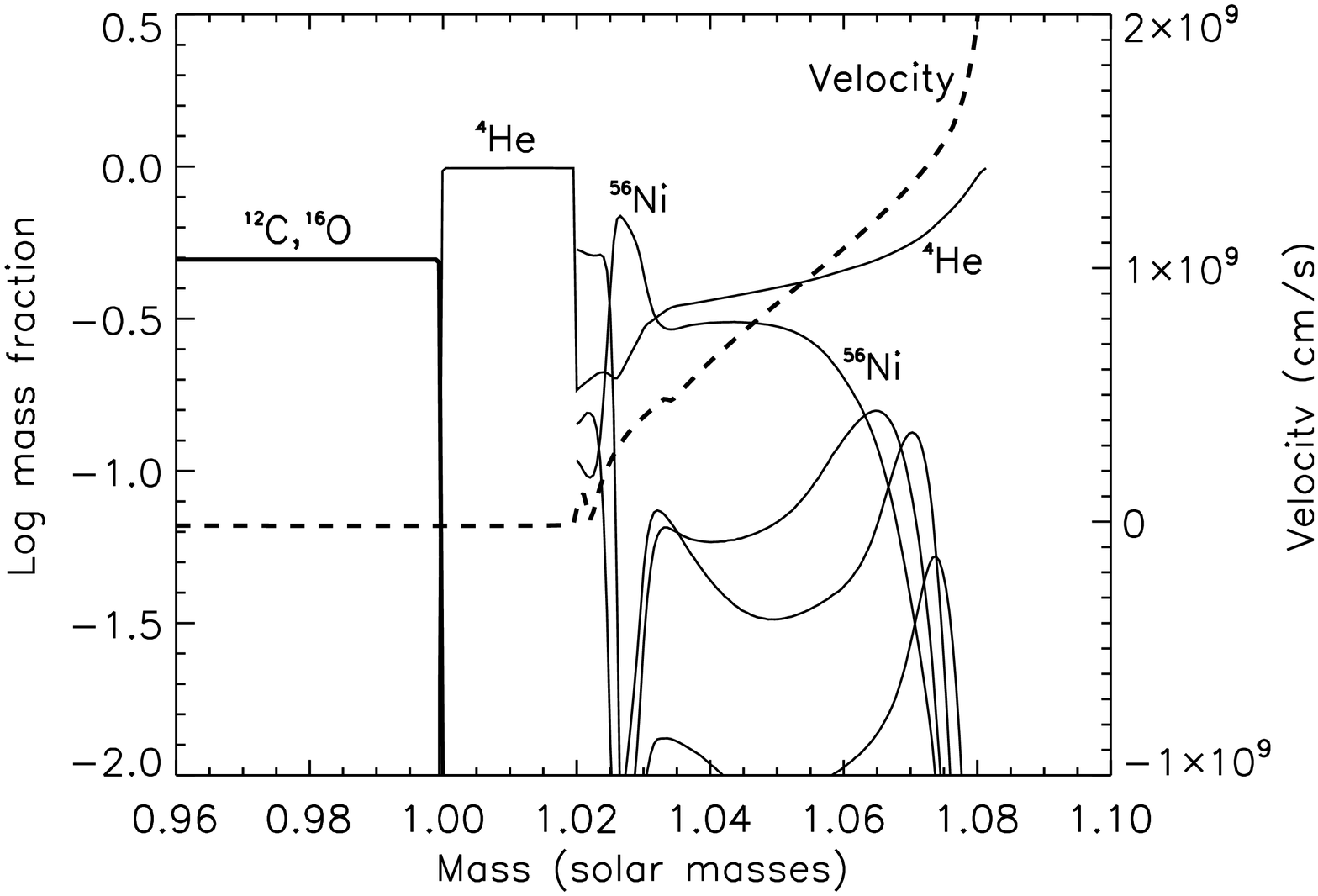}
\caption{The left frame shows the history of the temperature and
  density in Model 10B as the helium runaway develops. Solid lines,
  from bottom to top, give the temperature profile a) when nuclear
  burning first becomes adequate to cause convection in the accreted
  layer; b) the maximum convective luminosity in the shell equals
  10$^{45}$ erg s$^{-1}$; c) 10$^{46}$ erg s$^{-1}$; d) $5 \times
  10^{46}$ erg s$^{-1}$; e) 10$^{47}$ erg s$^{-1}$, and f) $5 \times
  10^{47}$ erg s$^{-1}$. The dashed lines show the density at first
  ignition (point a) and at L = $5 \times 10^{47}$ erg s$^{-1}$ (point
  f). When convection is left on until point f and then turned off, an
  outwards propagating detonation wave develops quickly, but no
  inwards detonation occurs, despite the substantial separation
  between the ignition point and helium shell - CO-core interface at
  1.0 \Msun.  The right hand frame shows some of the abundant isotopes
  after the helium detonation is over and their velocities. The
  unlabeled solid curves are from top to bottom, $^{52}$Fe, $^{48}$Cr,
  and $^{44}$Ti. In this Model 10B1 there was no detonation induced in
  the CO-white dwarf.  \lFig{moda1aign}}
\end{center}
\end{figure}

\begin{figure}
\begin{center}
\includegraphics[width=1.0\textwidth]{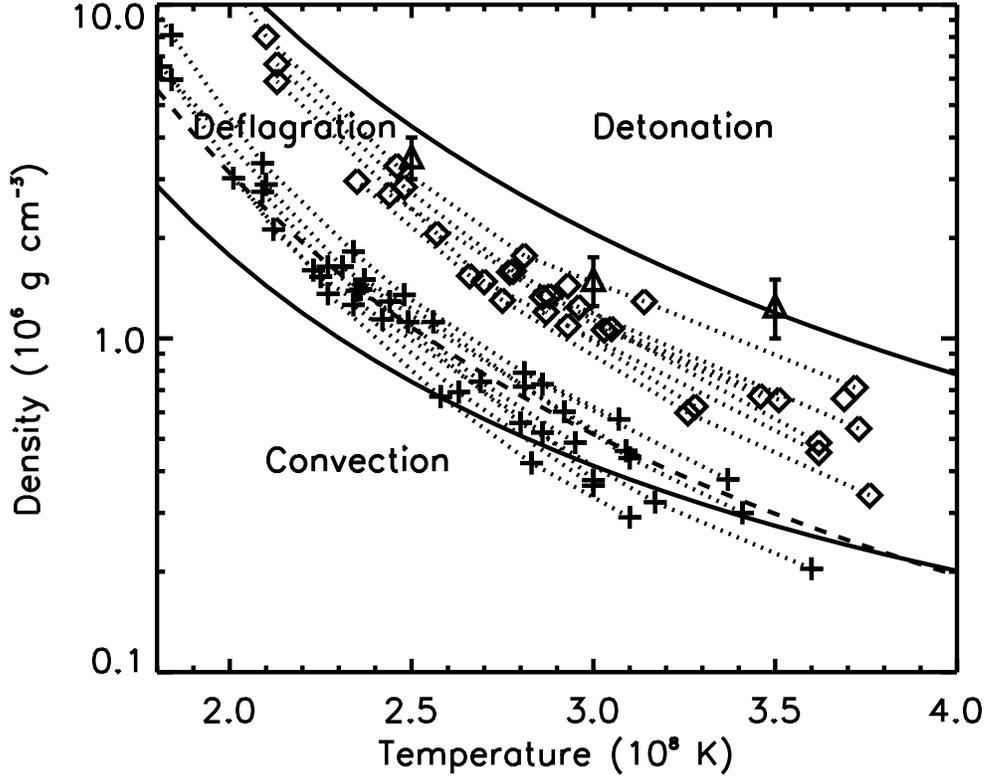}
\caption{Temperatures and densities at the base of the helium burning
  shell when the peak luminosity in that shell reaches 10$^{46}$ erg
  s$^{-1}$ (``+'' symbols) and 10$^{47}$ erg s$^{-1}$ (diamond
  symbols).  Dotted lines connect the ignition conditions in
  \Tab{ignpoint} for series of models based upon the same CO white
  dwarf mass and temperature, but with different accretion rates. The
  upper solid line is the analytic condition for detonation
  (\Eq{dtdr}) and the bottom solid line is the condition $\tau_{\rm
    run}$ = 2 s, a nominal condition for the first break down of
  convection. In between these two lines a deflagration can occur, a
  subsonic runaway that is transported neither by convection or
  detonation. In practice, all of the 10$^{47}$ erg s$^{-1}$ points
  detonated and most of the 10$^{46}$ erg s$^{-1}$ points, showing
  that planar detonation is easier than suggested by
  \Eq{dtdr}. Dividing the top line by an arbitrary factor of 4 gives
  the dashed line that is more consistent with the model
  results. Triangles with error bars at T$_8$ = 2.5, 3.0, and 3.5 show
  the location of the detonation transition as determined by
  fine-zoned hydrodynamical studies where the ignition occurred at a
  point rather than a plane. These data points are more consistent
  with the solid line given by \Eq{dtdr} rather than the dashed one
  and suggest that a) no detonation occurs for a density less than
  10$^{6}$ g cm$^{-3}$, and b) there could be an appreciable range of
  models that will deflagrate rather than detonate.  \lFig{detconds}}
\end{center}
\end{figure}

\begin{figure}
\begin{center}
\includegraphics[width=0.475\textwidth]{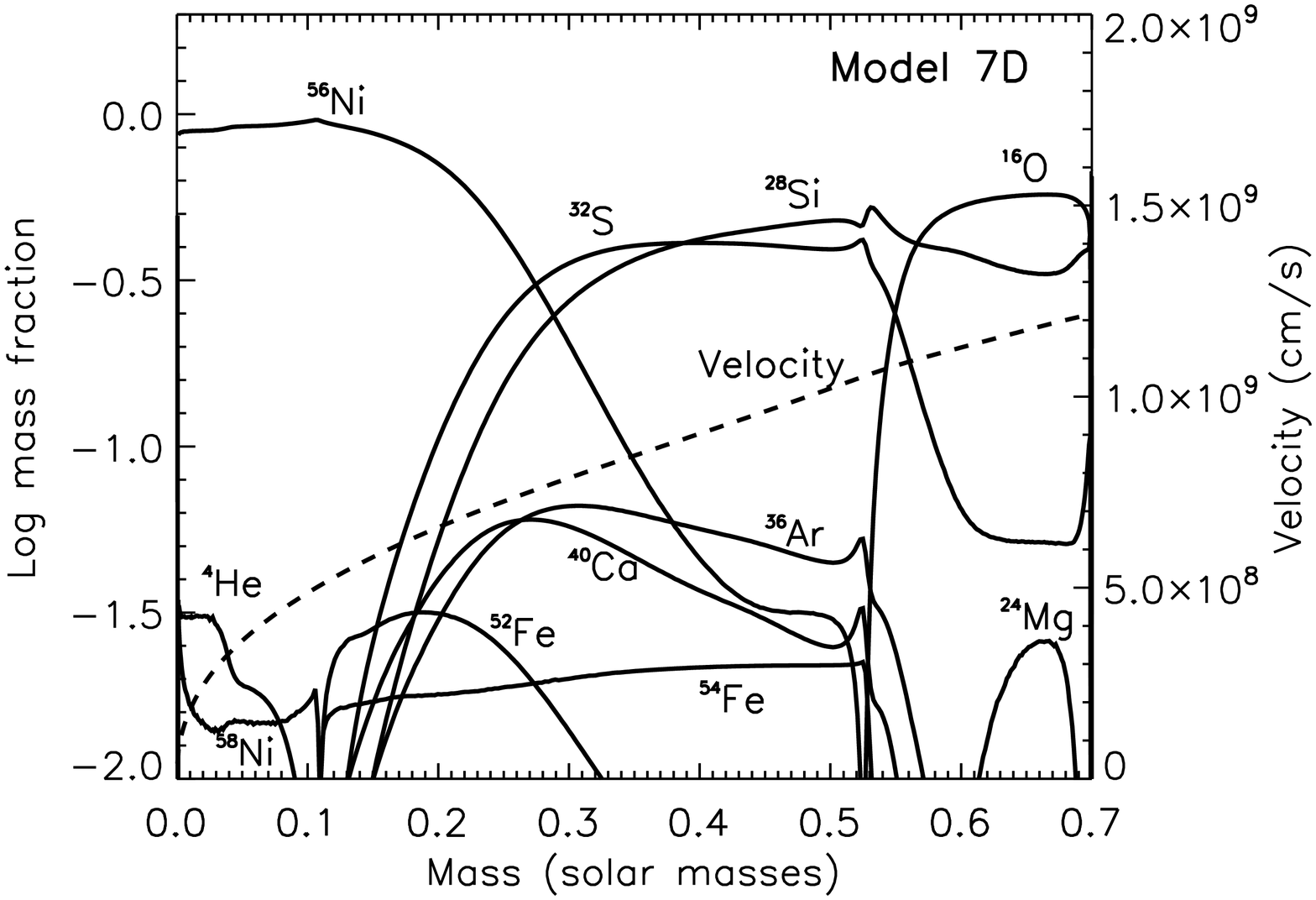}
\hfill
\includegraphics[width=0.475\textwidth]{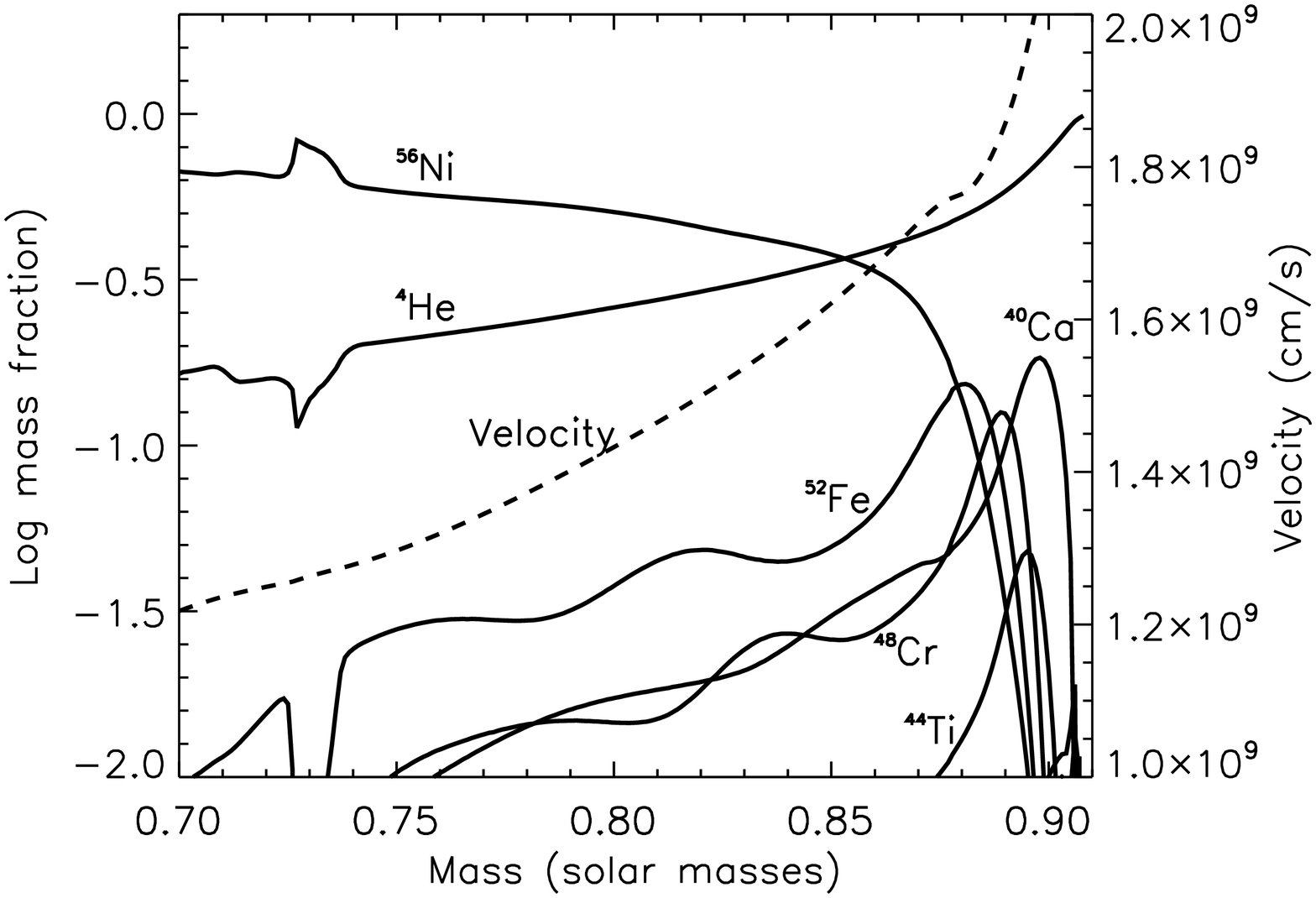}
\vskip 24pt
\includegraphics[width=0.475\textwidth]{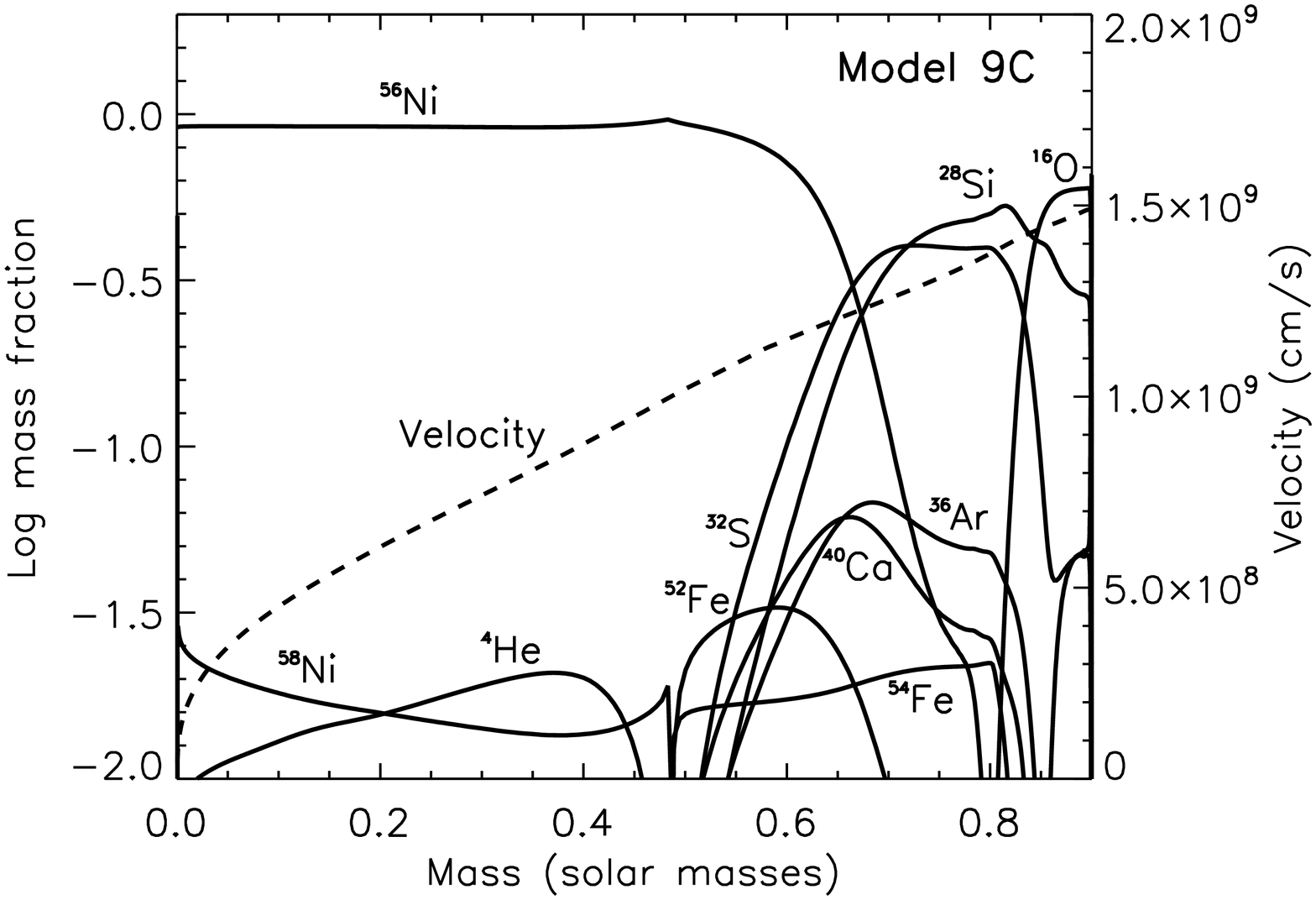}
\hfill
\includegraphics[width=0.475\textwidth]{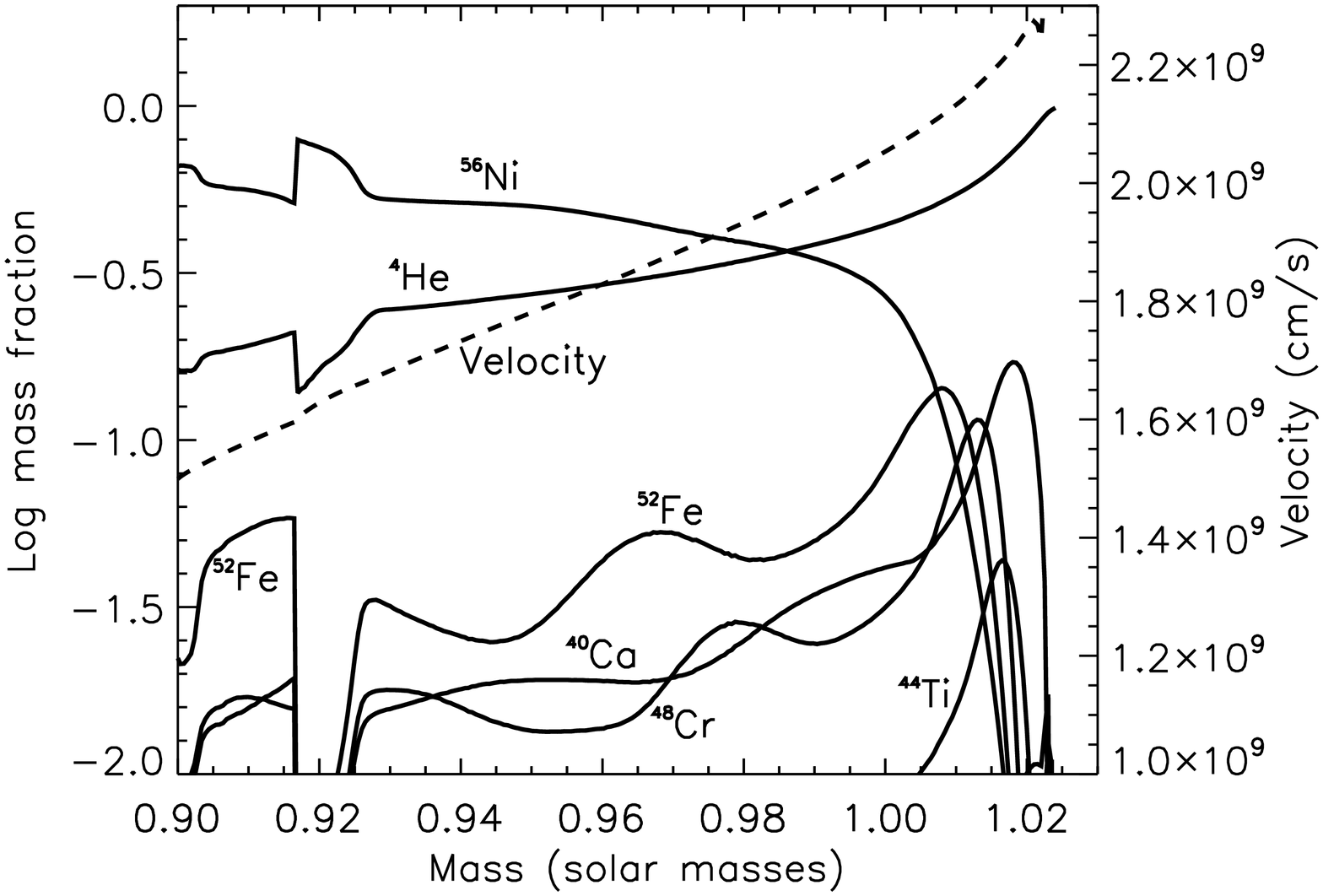}
\caption{Final abundances and velocity in Models 7D and 9C showing a
  range of nickel and intermediate mass synthesis. In each case the
  left frame shows the log of the mass fraction for species produced
  within the original CO-core and the right frame shows details of the
  helium shell detonation. The total mass for Model 7D was 0.907 \Msun
  \ and ignition occurred at 0.725 \Msun. Ignition resulted in both
  outwards and inwards moving detonations in the helium layer so that
  the whole star exploded. The total mass of Model 9C was 1.02 \Msun
  \ and ignition occurred at 0.918 \Msun. The prompt part of the
  explosion consisted of an outwards moving helium detonation but
  compression resulted, with some delay, in a secondary, centrally
  ignited detonation of the CO core so that, in the end the whole star
  exploded. Because of its larger total mass and higher density, less
  intermediate mass elements are produced in Model 9C.\lFig{p7d1a}}
\end{center}
\end{figure}

\begin{figure}
\begin{center}
\includegraphics[width=0.475\textwidth]{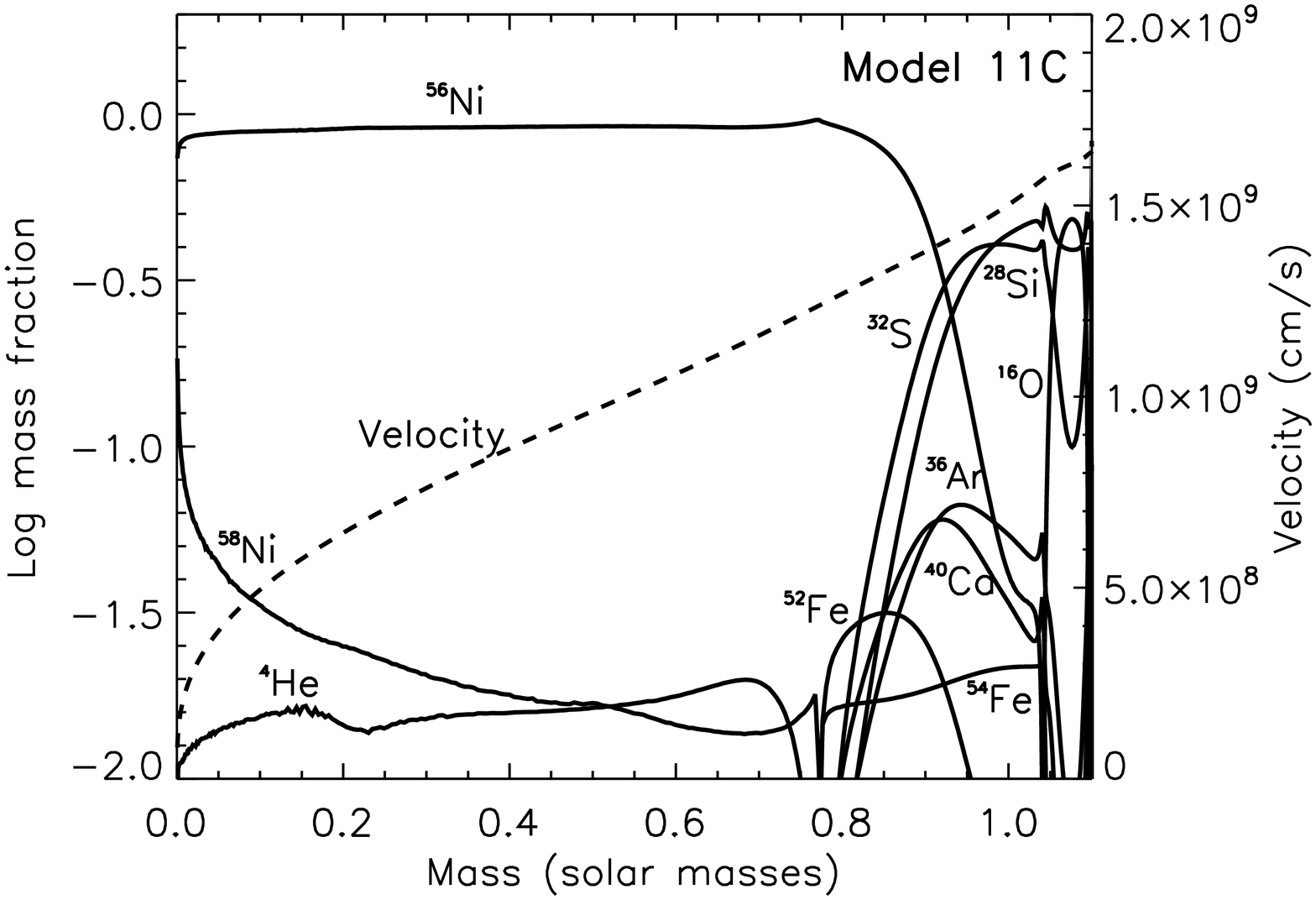}
\hfill
\includegraphics[width=0.475\textwidth]{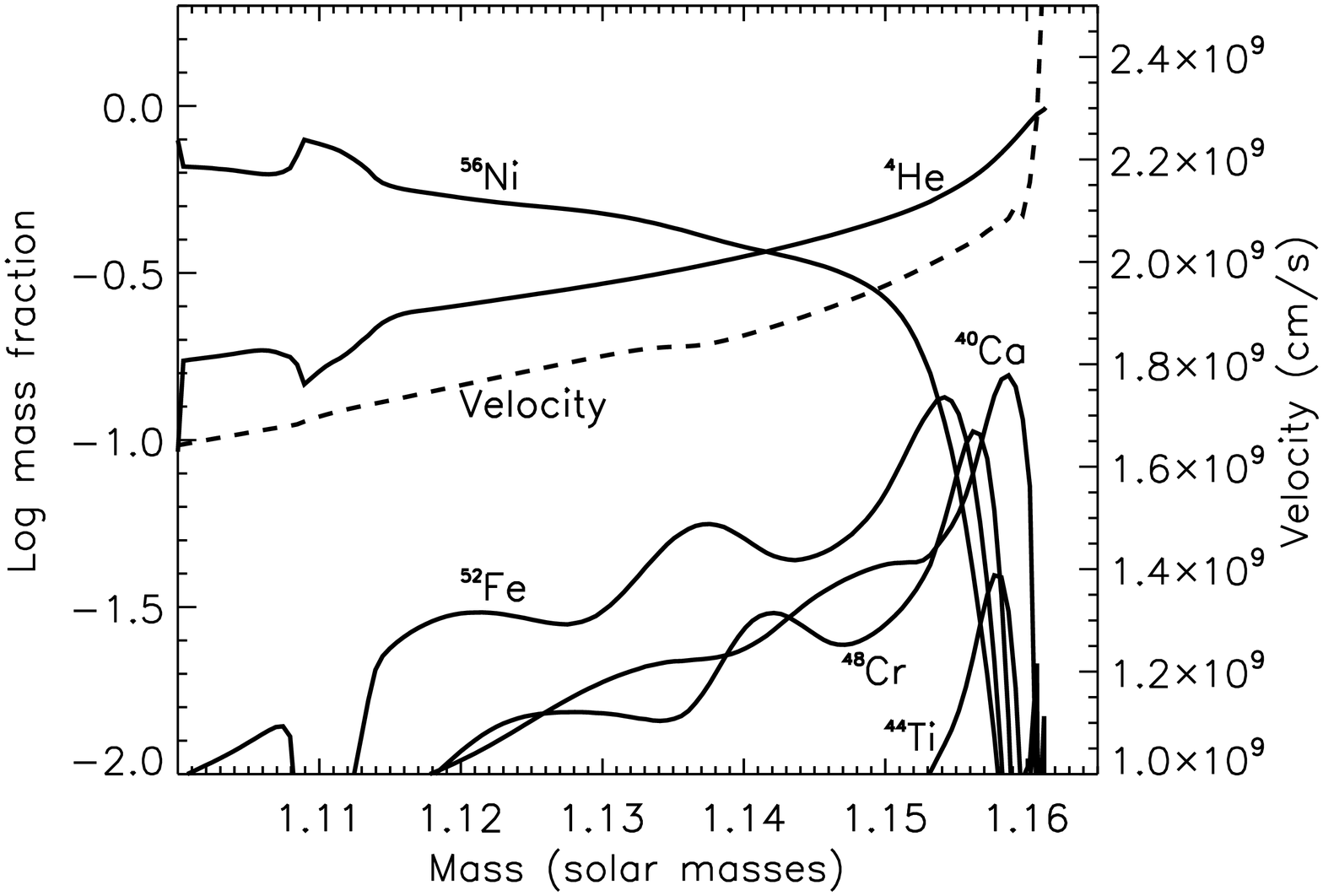}
\vskip 24pt
\includegraphics[width=0.475\textwidth]{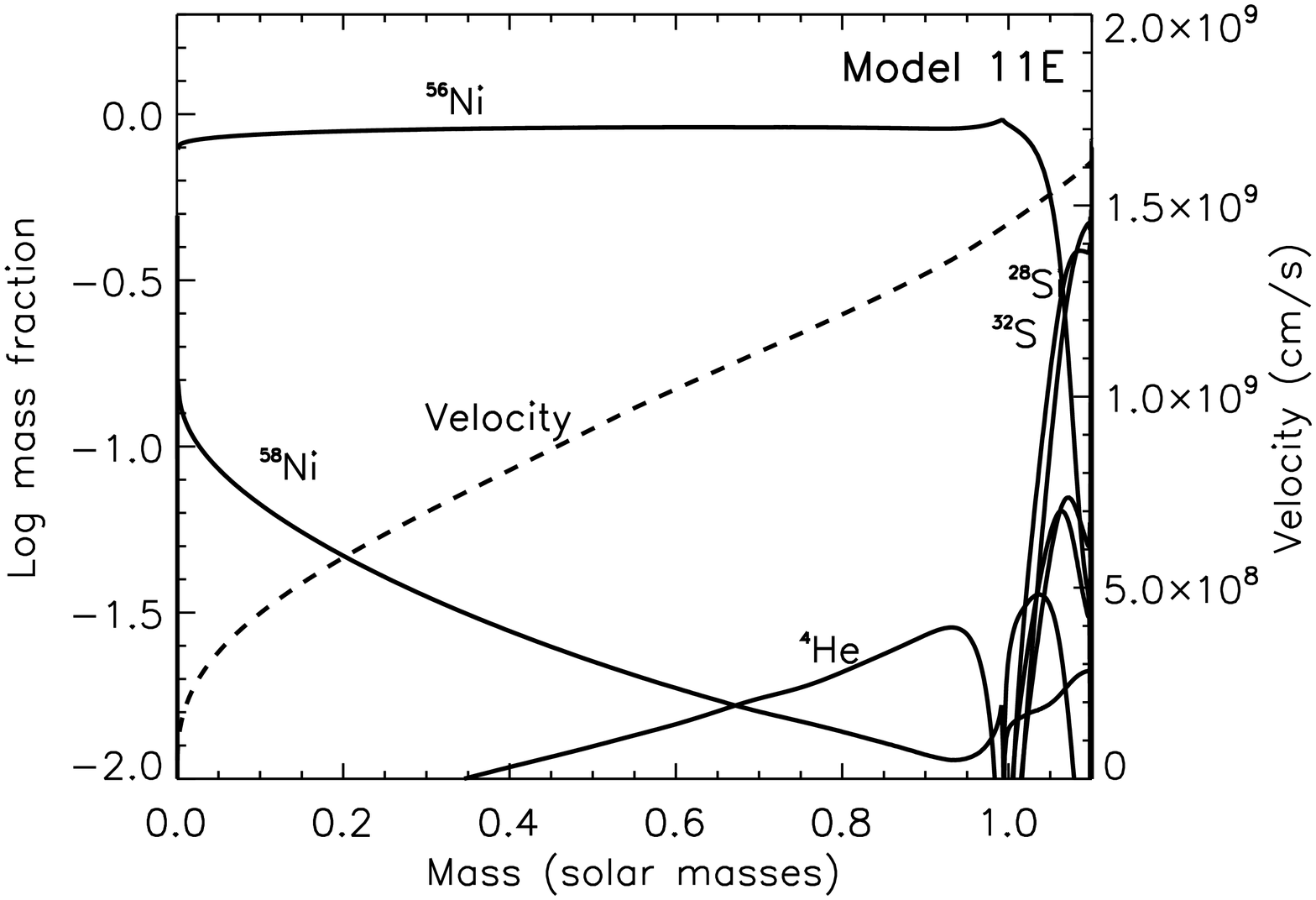}
\hfill
\includegraphics[width=0.475\textwidth]{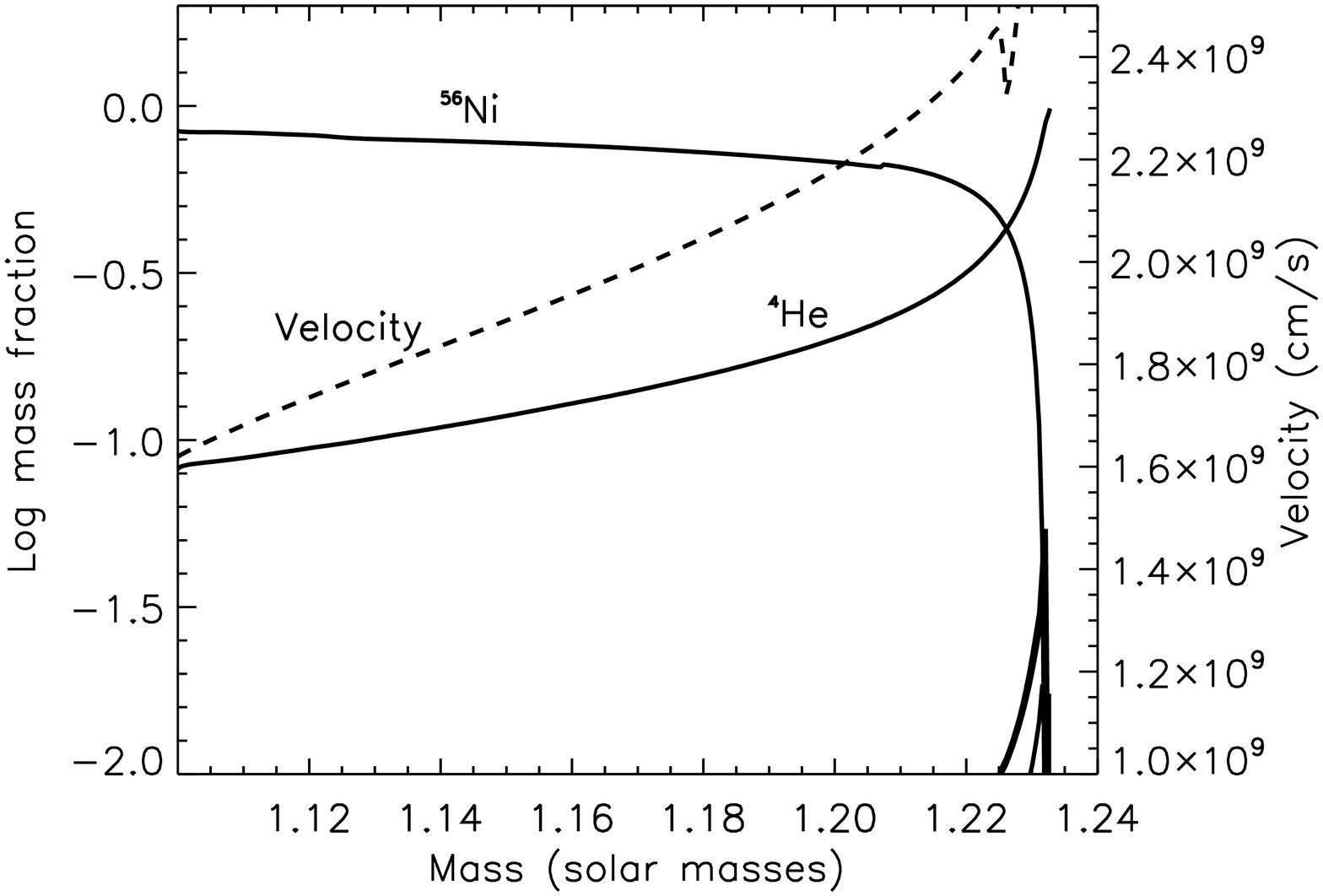}
\caption{Final abundances and velocities in two models with higher
  mass and density. These will make bright SN Ia with weaker lines of
  intermediate mass elements. Model 11C had a total mass of 1.162
  \Msun \ and ignited at 1.11 \Msun. Model 11E had a total mass of
  1.233 \Msun \ and ignited at the CO-He interface. Model 11C had a
  double detonation of the helium shell which propagated to the center
  of the star. Model 11E had an outwards moving helium detonation and
  a secondary central carbon detonation. \lFig{1p1a1}}
\end{center}
\end{figure}

\begin{figure}
\begin{center}
\includegraphics[width=0.9\textwidth]{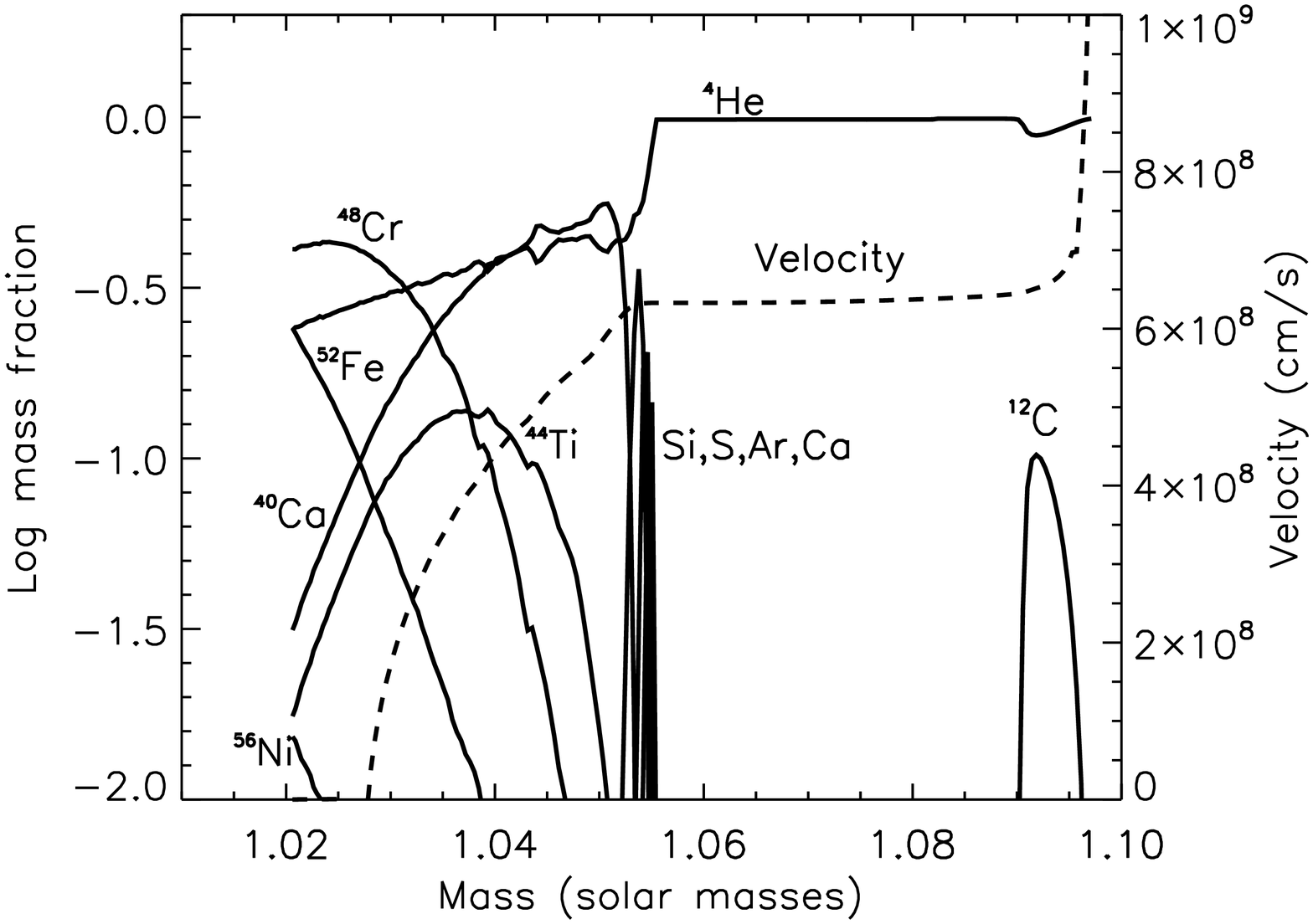}
\caption{Final abundances and velocity in Model 10DEFL. Burning is at
  a lower temperature and in a smaller fraction of the mass than in
  the models that detonate. Consequently the explosion speed is
  lower. The near constant speed in the unburned helium layer is a
  consequence of the one-dimensional nature of the simulation and
  would be different in 2D or 3D.  \lFig{deflg1c}}
\end{center}
\end{figure}

\begin{figure*}
\begin{center}
\includegraphics[width=0.48\textwidth]{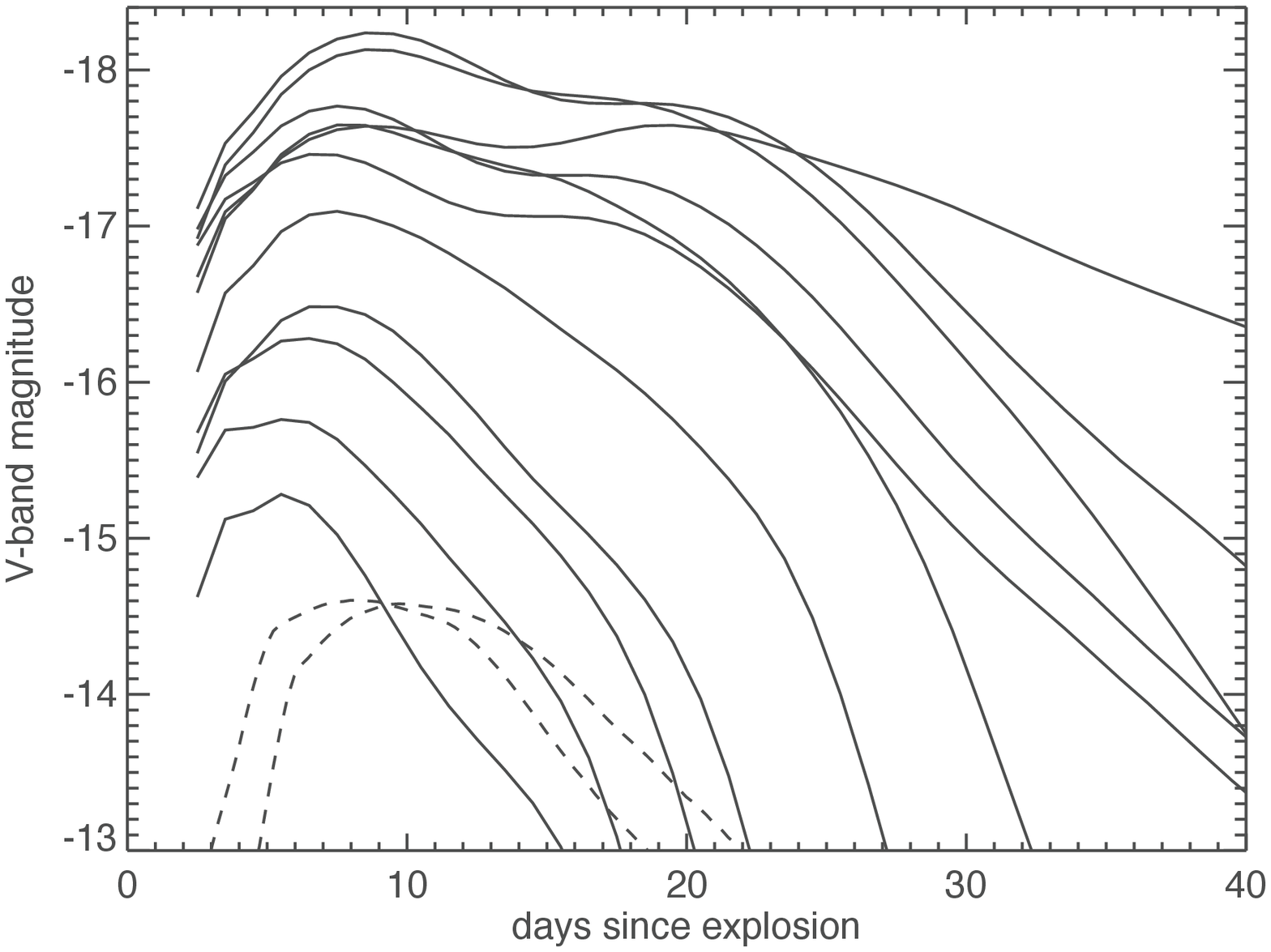}
\includegraphics[width=0.48\textwidth]{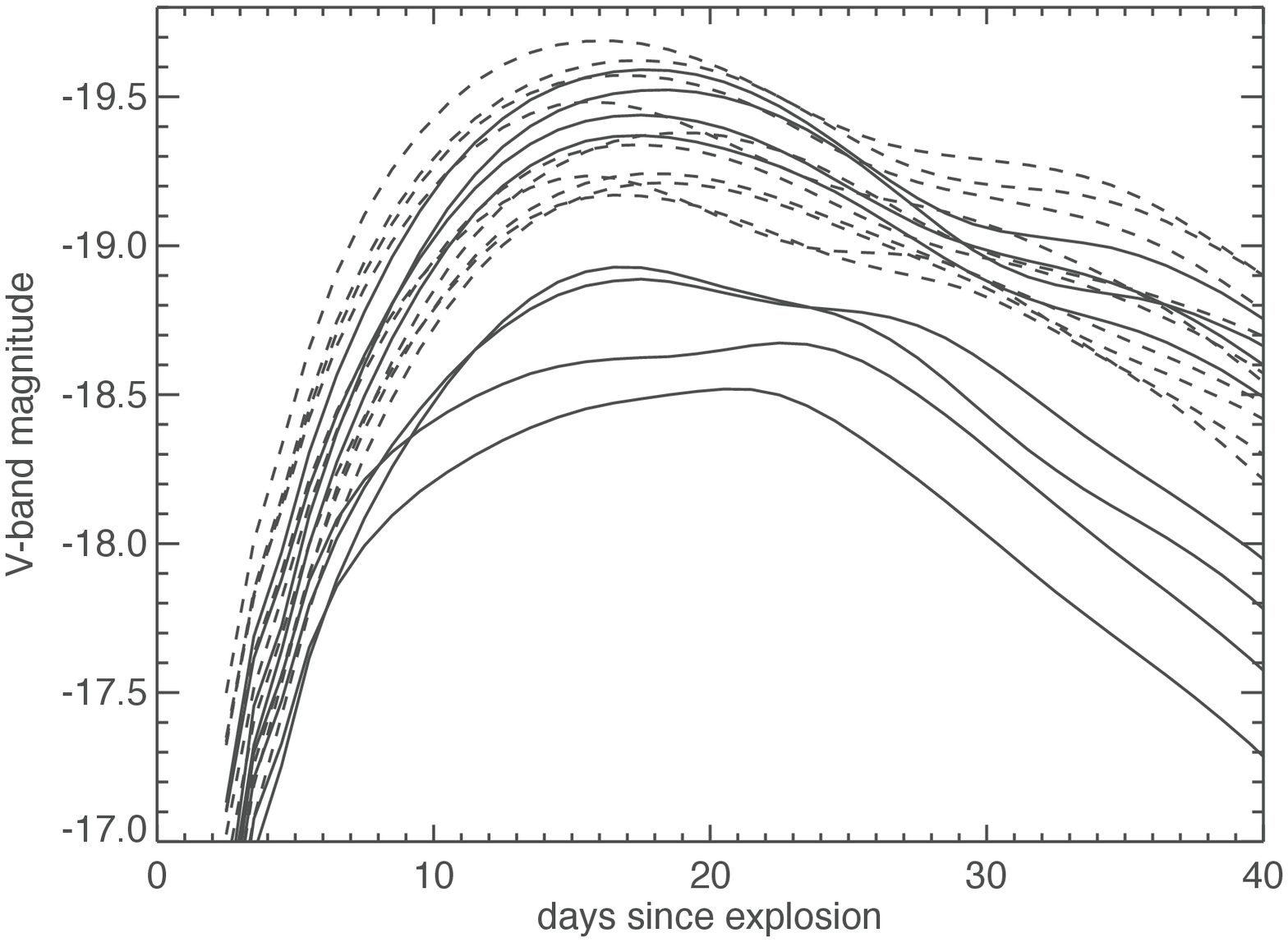}
\caption{A sample of V-band light curves illustrating the diversity of
  observable behaviors seen in the model set.  {\it Left:} Models in
  which only the helium shell exploded, either by detonation (case c,
  solid lines) or deflagration (case b, dashed lines).  {\it Right:}
  Models in which the entire star exploded, either by an inward
  detonation (case a, solid lines) or by compression of the CO-core
  triggering a secondary detonation (case d, dashed lines).
  \lFig{vband_lc}}
\end{center}
\end{figure*}

\begin{figure}
\begin{center}
\includegraphics[width=1.0\textwidth]{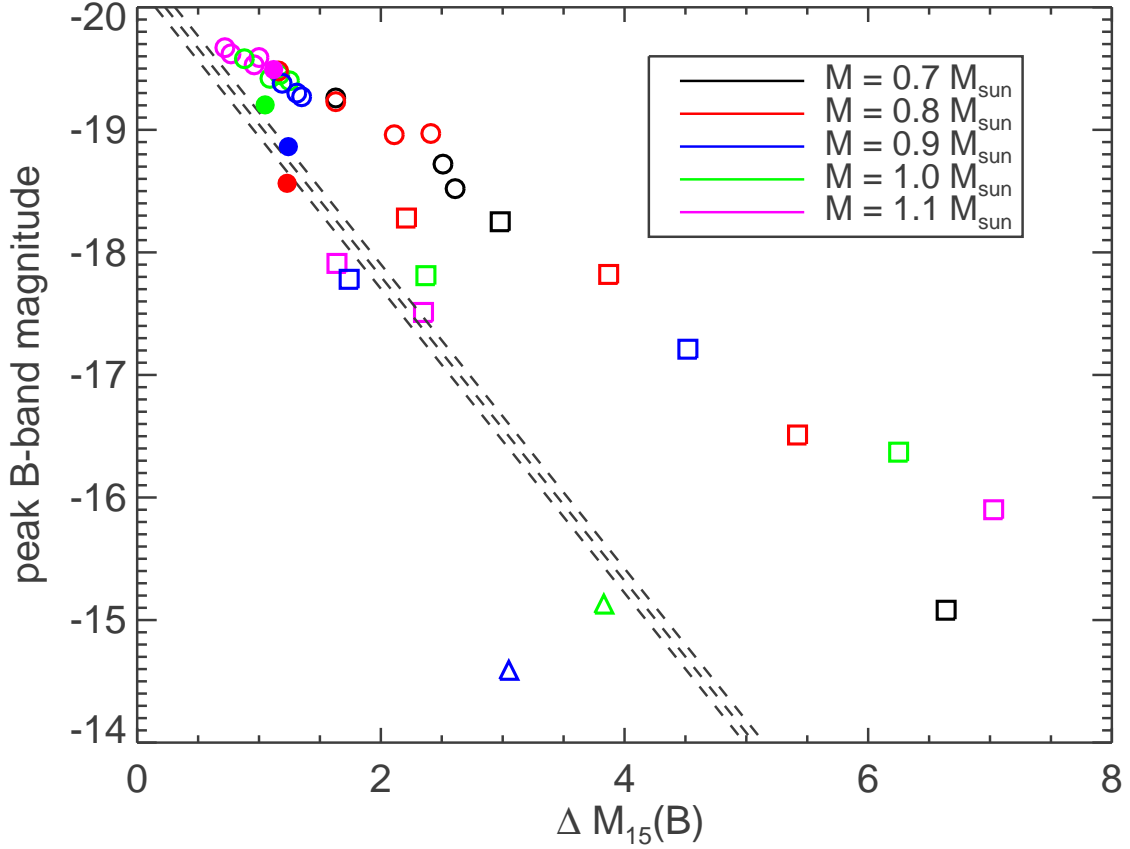}
\caption{Width-luminosity relation for the models.  Open
symbols represent ``cold'' models (i.e.,
  those where the white dwarf had an initial luminosity of 0.01
  \Lsun) while the closed circles represent four ``hot"
   models (initial luminosity $L = 1~\Lsun$; Models 8HBC, 9HC, 10HC, and 11HD). 
    Squares denote models where only the helium shell
  detonated, triangles, models where the shell deflagrated, and
  circles, models where both the helium shell and CO-core exploded.
  Color coding indicates the initial white dwarf mass. The hatched
  region shows the observed width-luminosity relation of
  \cite{Folatelli_2010}: $M_B = -19.12 + 1.241 (\dmb - 1.1)$, with a
  $\pm 0.1$~mag spread. This linear relation has been naively
  extrapolated into the region to $\dmb > 1.8$ where it actually no
  longer holds.  \lFig{mb_dm15}}
\end{center}
\end{figure}

\begin{figure}
\begin{center}
\includegraphics[width=1.0\textwidth]{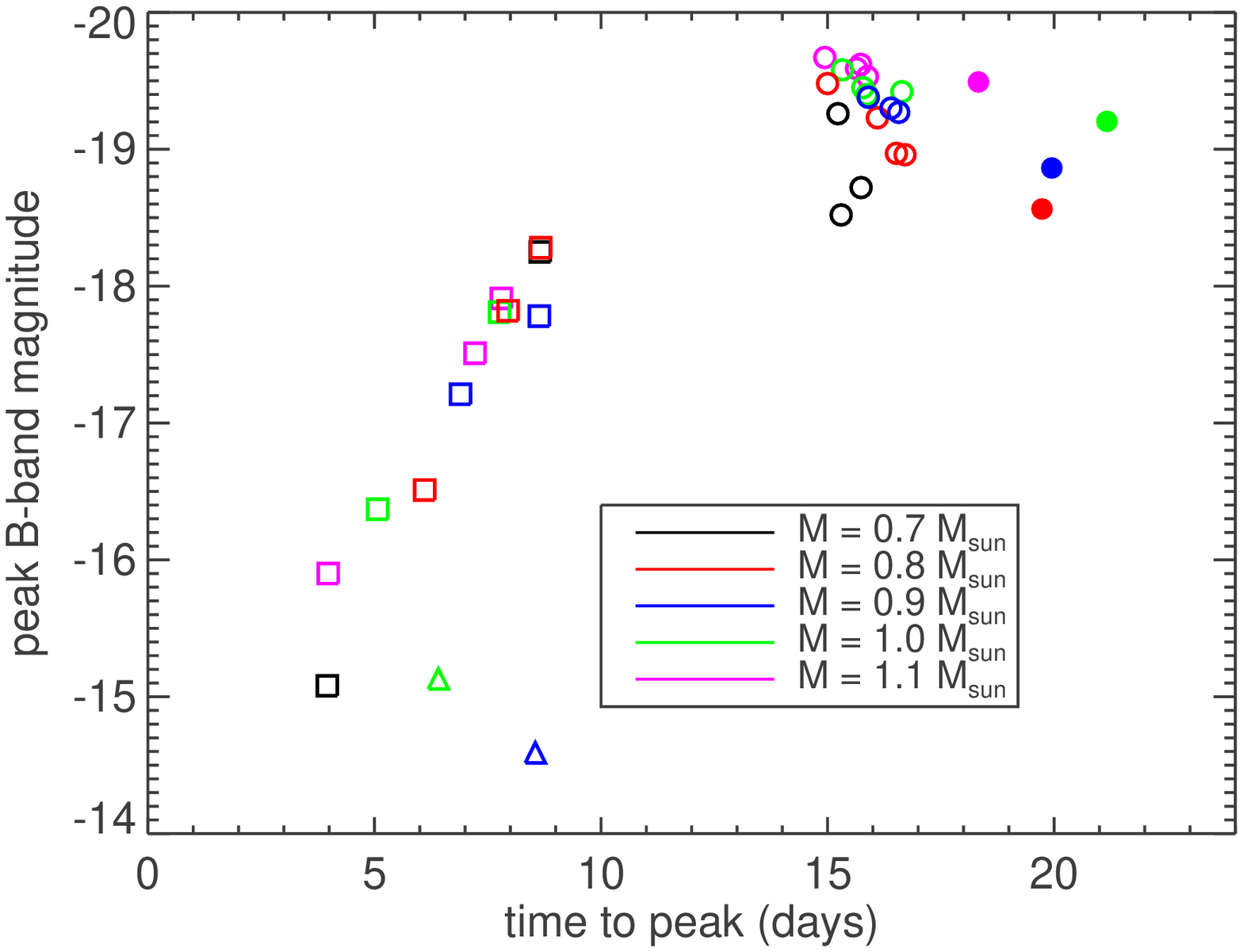}
\caption{Peak B-band luminosity versus the B-band rise time for the
  models described in \Fig{mb_dm15}.   Open (closed) symbols represent
  models with cold (hot) white dwarfs.  Squares denote models where only the
  helium shell detonated; triangles, models where the helium shell
  deflagrated, and circles, models where both the shell and core
  exploded.  Color coding indicates the initial white dwarf mass.
  \lFig{mb_tpeak}}
\end{center}
\end{figure}

\begin{figure}
\begin{center}
\includegraphics[width=1.0\textwidth]{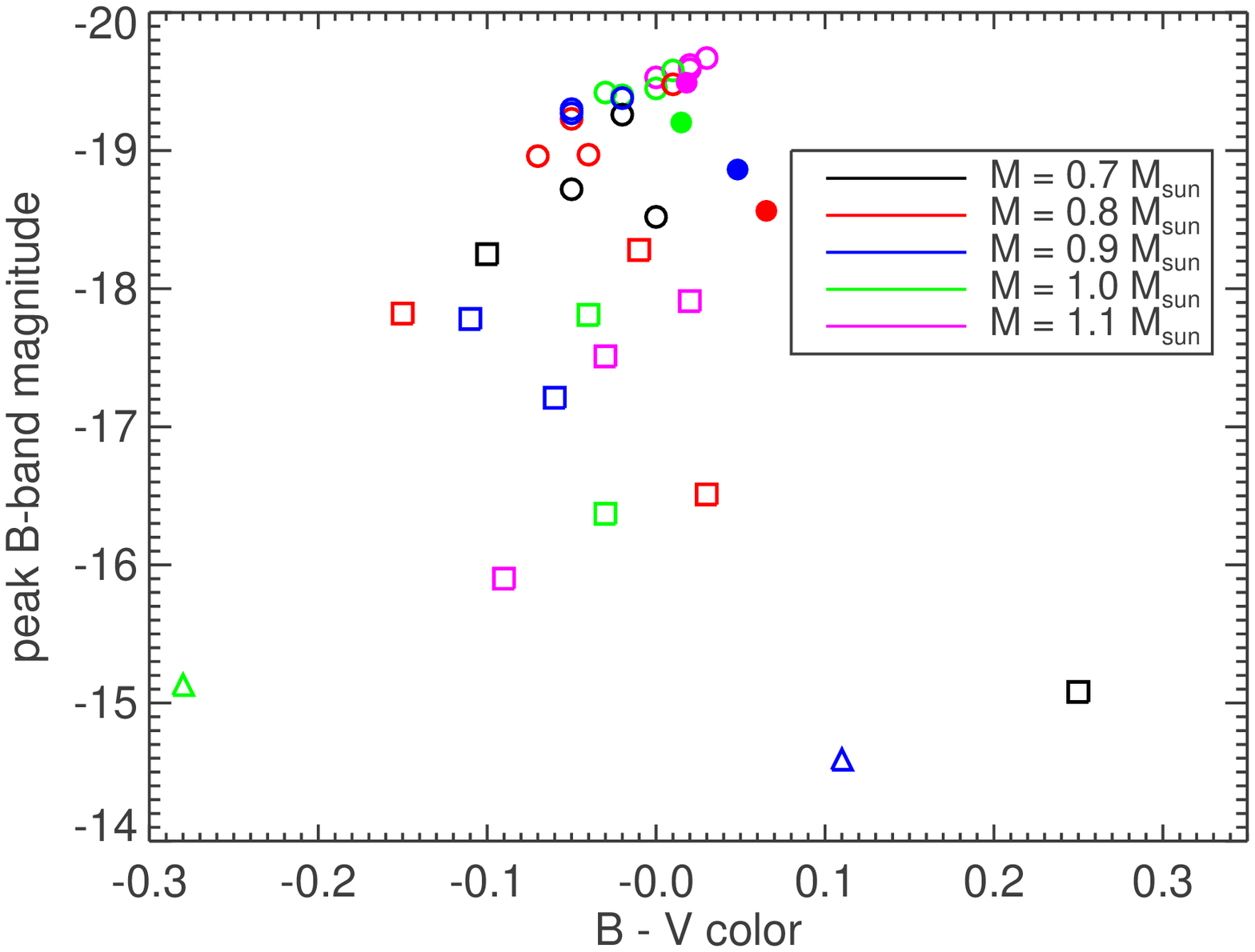}
\caption{Peak B-band luminosity versus the B-V color at B-maximum for
  the models described in \Fig{mb_dm15}.   Open (closed) symbols represent
  models with cold (hot) white dwarfs..  Squares denote models where only
  the helium shell detonated; triangles, models where the shell
  deflagrated; and circles, models where both shell and core exploded.
  The color coding indicates the initial white dwarf mass.
  \lFig{mb_color}}
\end{center}
\end{figure}

\begin{figure}
\begin{center}
\includegraphics[width=1.0\textwidth]{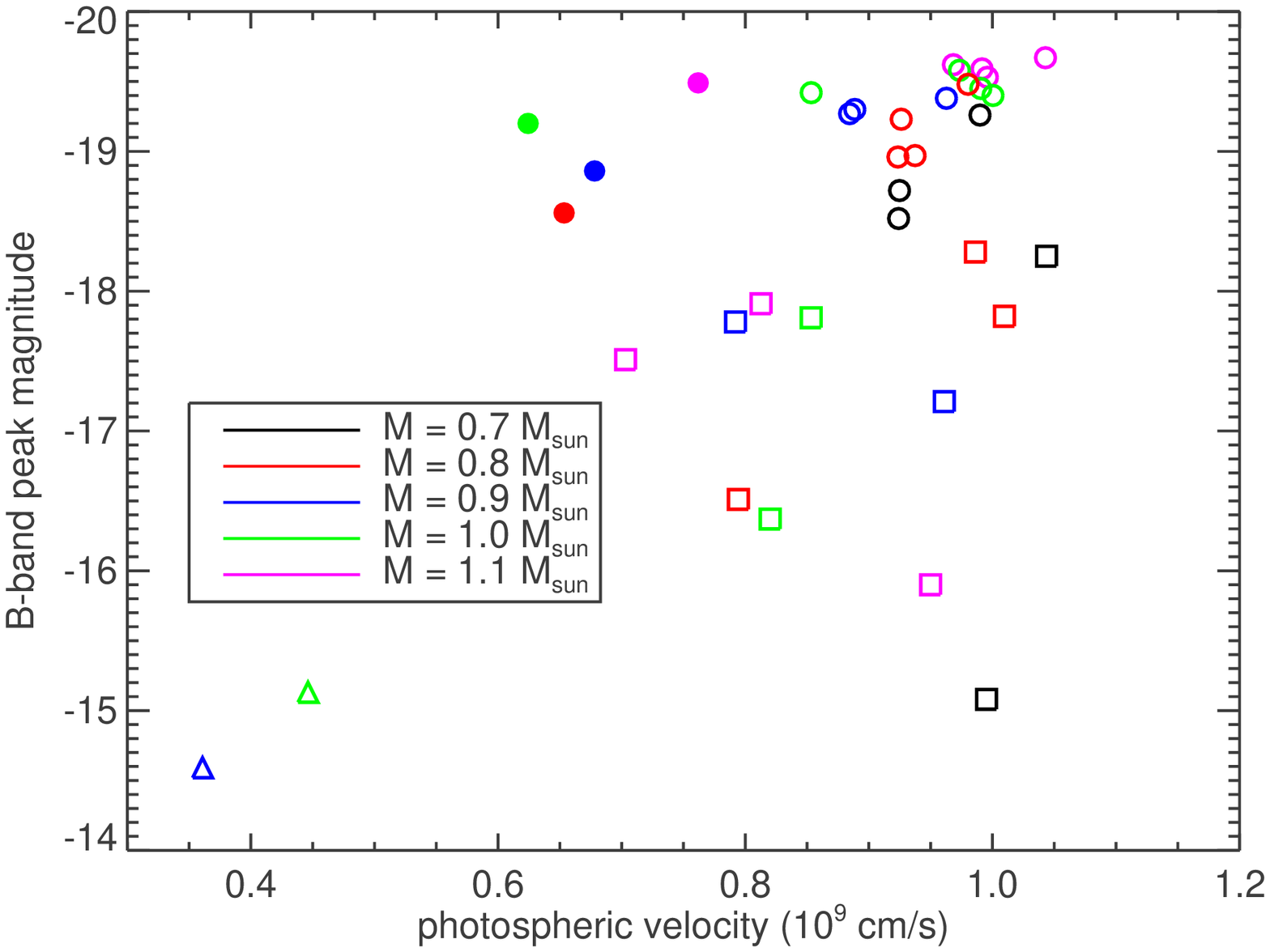}
\caption{Peak B-band luminosity versus the velocity of the electron
  scattering photosphere at B-maximum for the models described in
  \Fig{mb_dm15}.  Open (closed) symbols represent models with cold
  (hot) white dwarfs.  Squares denote models where only the shell
  detonated, triangles models where the shell deflagrated, and circles
  models where both shell and core exploded.  Color coding indicates
  the initial white dwarf mass.  \lFig{mb_vp}}
\end{center}
\end{figure}

\begin{figure}
\begin{center}
\includegraphics[width=1.0\textwidth]{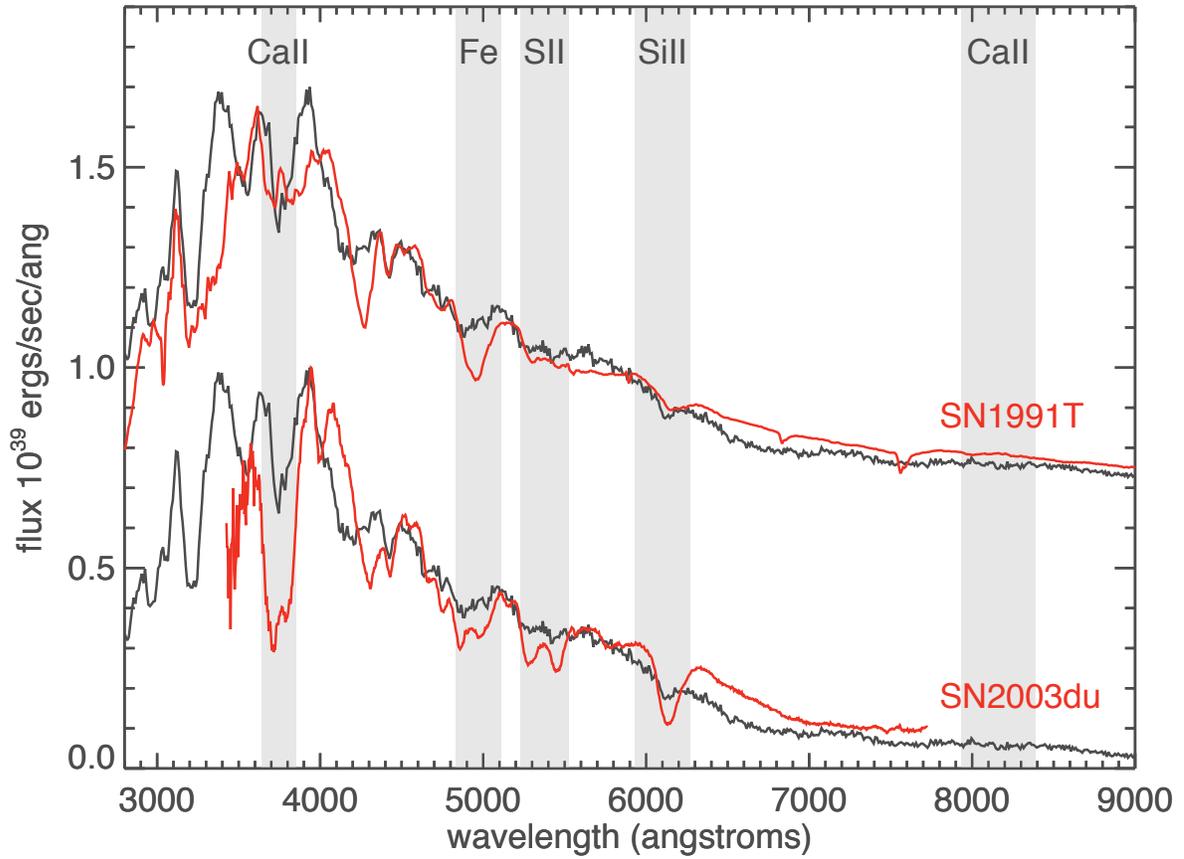}
\caption{Maximum light synthetic spectrum of the entire Model 9C
  (shell plus core; black lines) compared to observed spectra.  The
  model lacks the strong silicon, sulfur and calcium lines seen in the
  normal Type~Ia SN~2003du (bottom red line, \cite{Stanishev_2007}),
  and resembles more closely the peculiar Type~Ia SN~1991T (top red
  line, \cite{Filippenko_1992}).  \lFig{spec_compare}}
\end{center}
\end{figure}

\begin{figure}
\begin{center}
\includegraphics[width=1.0\textwidth]{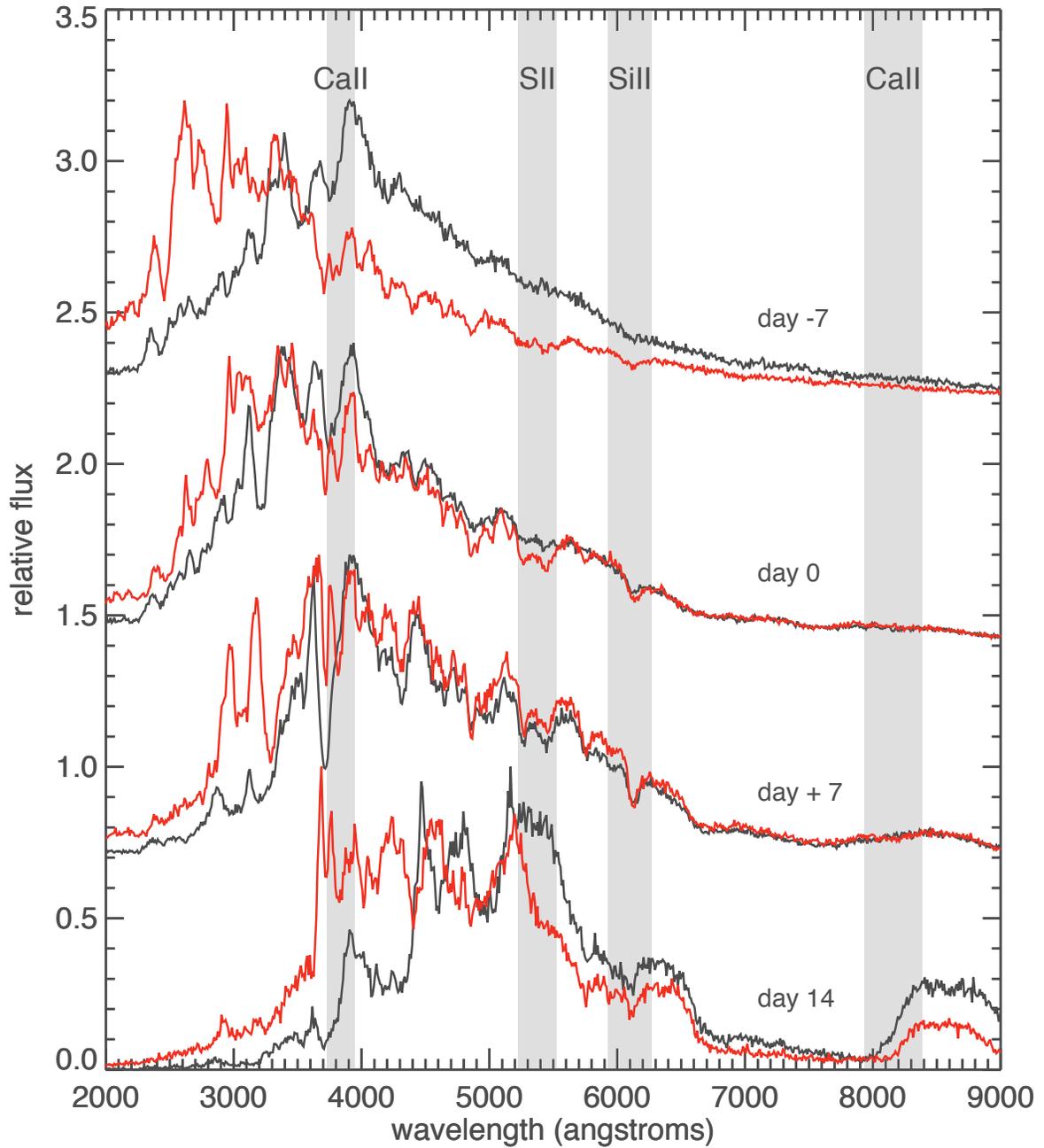}
\caption{Spectral time series of Model 9C, calculated with the
  external burned helium shell included (black lines) and with that
  shell removed (red lines).  Labels mark days since B-band maximum
  light.  The presence of the shell affects both the color and line
  features, especially before and after maximum light.
  \lFig{9C_series}}
\end{center}
\end{figure}

\begin{figure*}
\begin{center}
\includegraphics[width=1.0\textwidth]{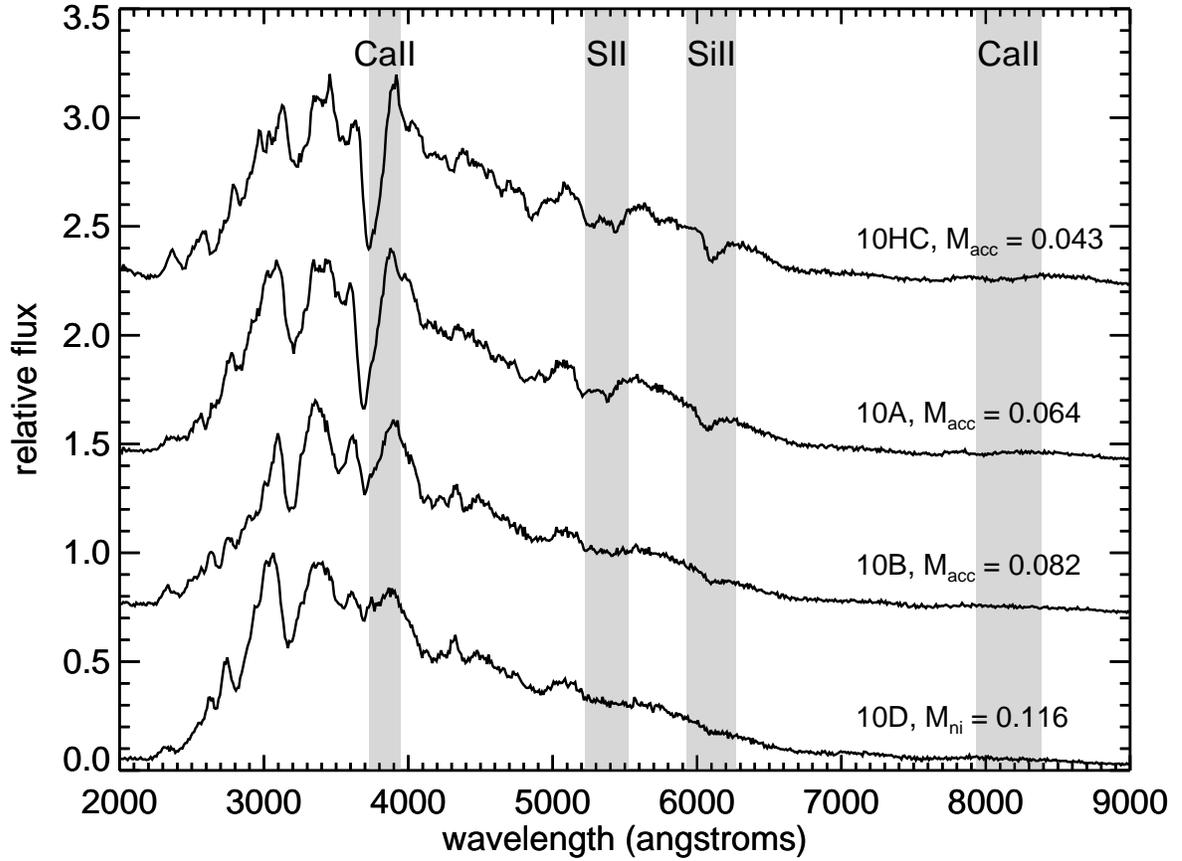}
\caption{Maximum light spectra of select full star (core plus shell)
  explosions.  These particular models all had the same core mass($ =
  1~\Msun$) but varied in the amount of helium accreted before a
  detonation took place.  Only the models with the lightest helium
  shells (Model 10HC and, to a lesser extent, Model 10A) show the IME
  absorption features typical of normal SNe~Ia \lFig{10_series}}
\end{center}
\end{figure*}

\begin{figure*}
\begin{center}
\includegraphics[width=1.0\textwidth]{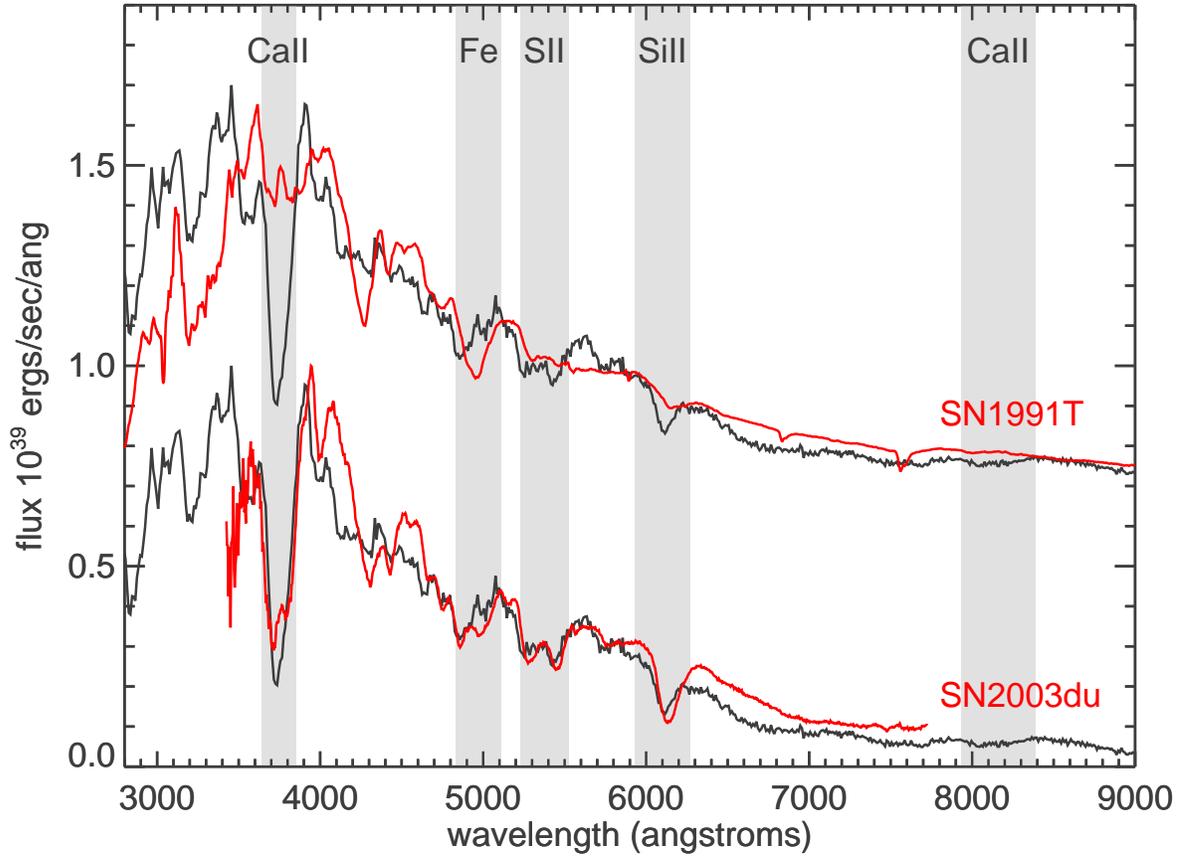}
\caption{Maximum light spectrum of Model~10HC (black lines) compared to
observed SNe~Ia.  The model shows reasonably strong absorption
features from IME which are in fair agreement with the normal
Type~Ia SN~2003du (bottom red line) but not the
peculiar SN~1991T (top red line).
\lFig{10HC_spec}}
\end{center}
\end{figure*}

\clearpage

\begin{figure*}
\begin{center}
\includegraphics[width=1.0\textwidth]{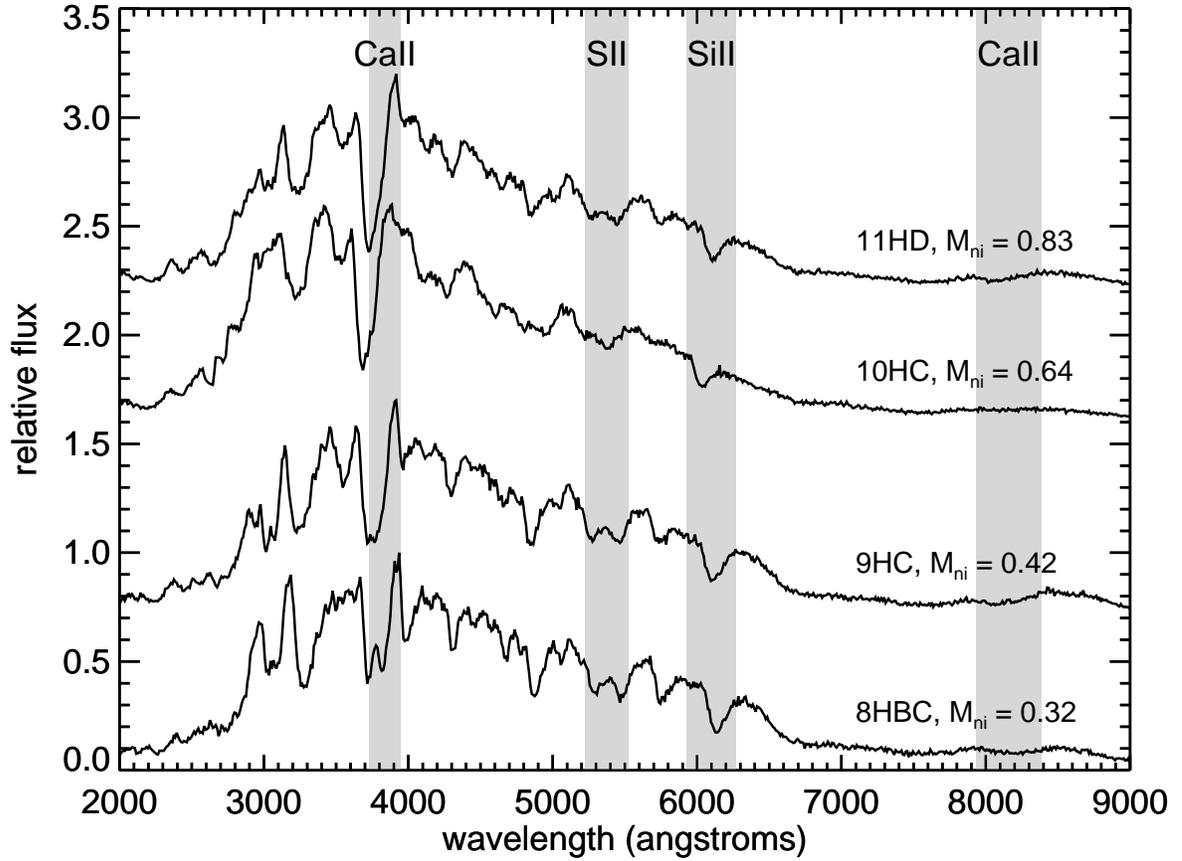}
\caption{ Maximum light spectra of four full star (core plus shell)
  explosions in which the white dwarf was ``hot" -- i.e., had an
  initial luminosity of $1~\Lsun$. The models vary in the initial mass
  of the white dwarf, and consequently the amount of \Nifs\ produced
  in the explosion. In all cases, the spectra resemble normal SNe~Ia,
  with noticeable line absorption features from IMEs (Si II, S II, and
  Ca II).  \lFig{hot_spectra}}
\end{center}
\end{figure*}

\begin{figure}
\begin{center}
\includegraphics[width=1.0\textwidth]{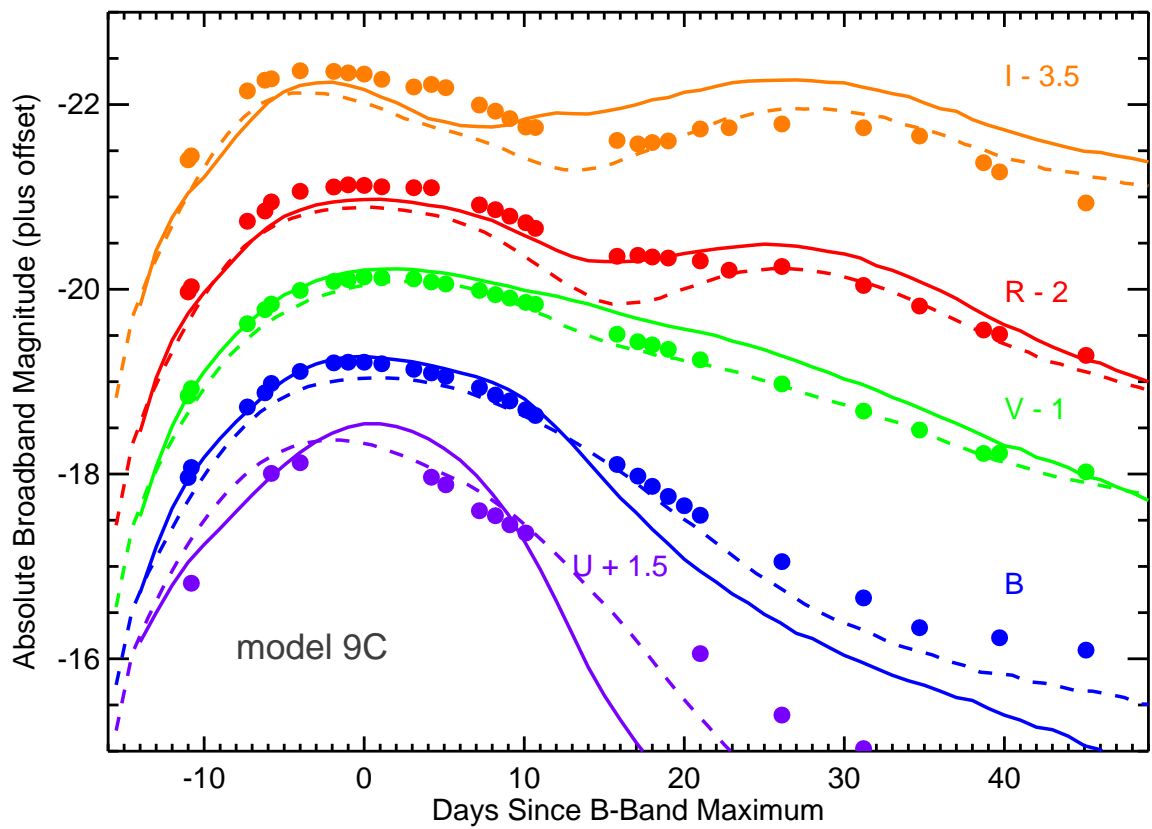}
\caption{Synthetic broadband UBVRI light curves of Model 9C (solid
  lines), in which the both the shell and star exploded producing
  $0.611$ \Msun \ of \Nifs.  To demonstrate the effect of the outer
  burned helium layer, we calculated the light curves with that shell
  included (solid lines) and with it removed (dashed lines).  The
  opacity of the shell causes the B-band light curves to decline after
  peak more rapidly than what is observed in the normal Type~Ia
  SN~2003du (circles, \cite{Stanishev_2007}).  \lFig{LC_9C}}
\end{center}
\end{figure}

\begin{figure}
\begin{center}
\includegraphics[width=1.00\textwidth]{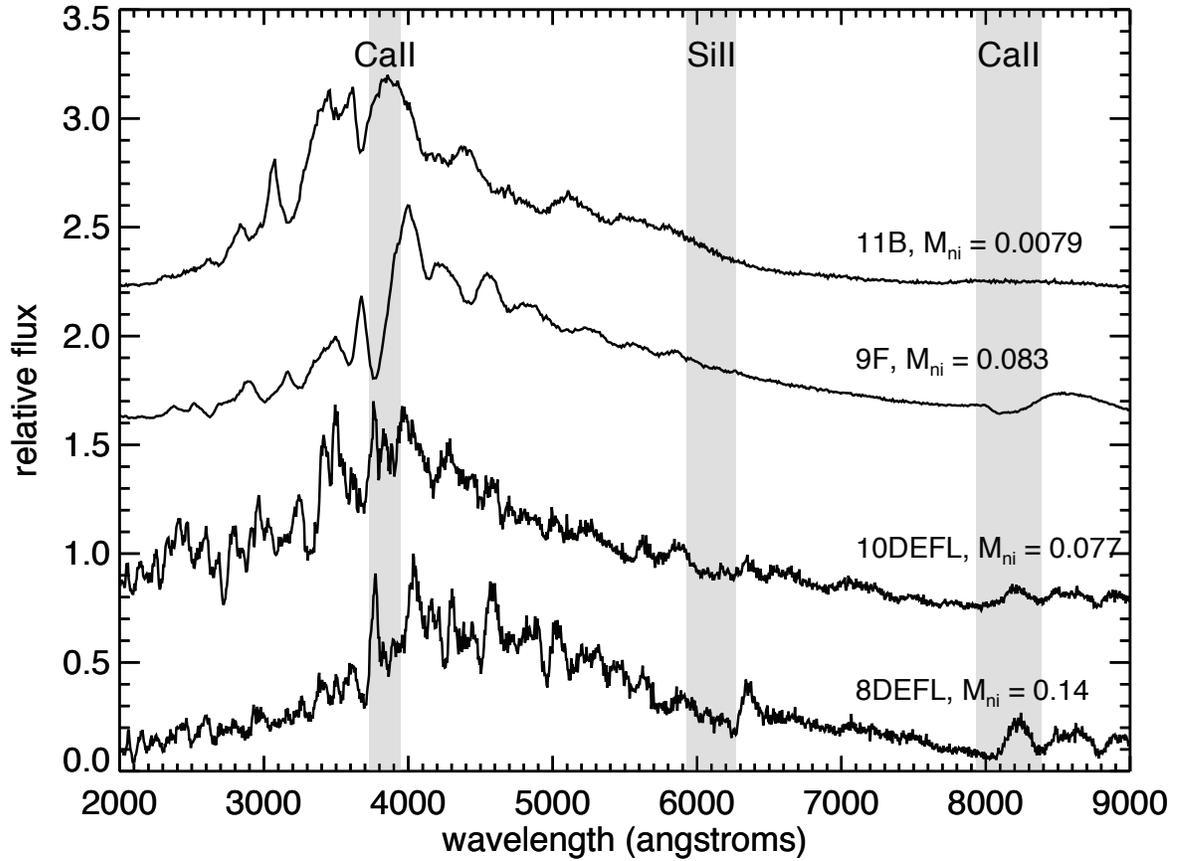}
\caption{Spectra at maximum light of select shell-only explosions.
  The top two spectra are those of helium detonation models, which
  show broad, high velocity absorptions.  The bottom two spectra are
  those of helium deflagrations, which show  narrower, lower
  velocity absorptions.  \lFig{shell_spec}}
\end{center}
\end{figure}

\begin{figure}
\begin{center}
\includegraphics[width=0.475\textwidth]{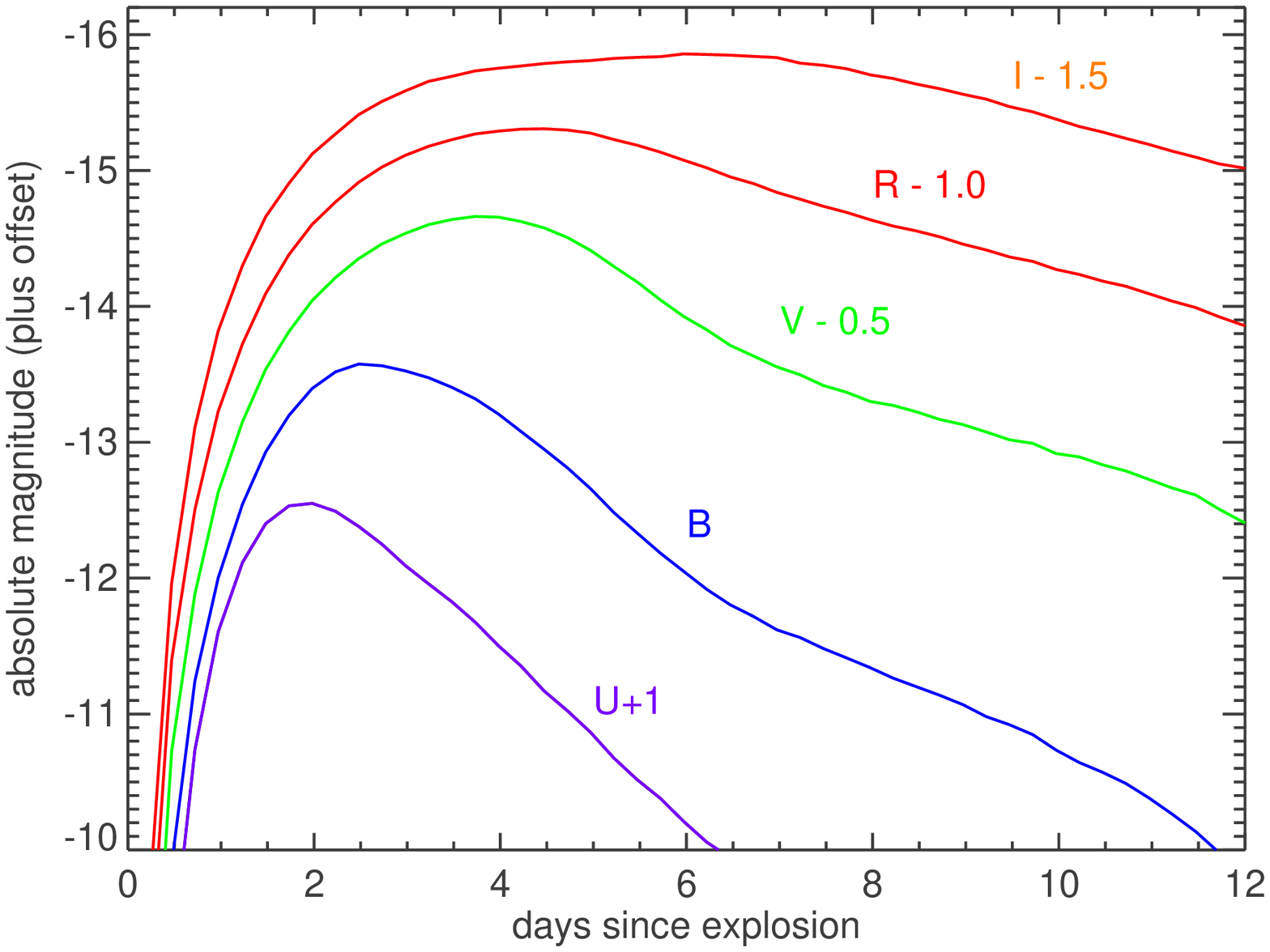}
\includegraphics[width=0.475\textwidth]{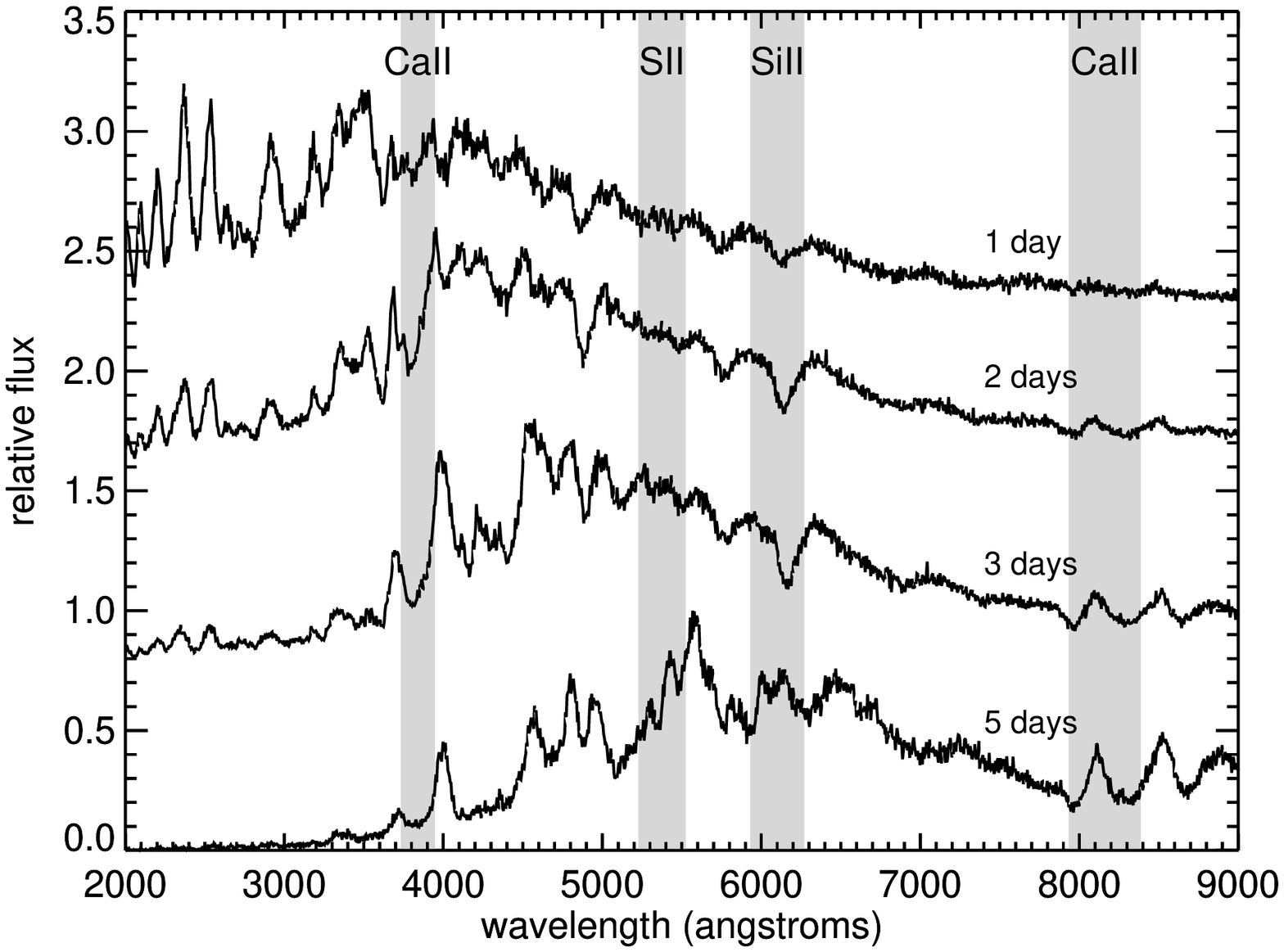}
\caption{Broadband light curves (left) and spectral time series
  (right, marked in time since explosion) of Model 8HBC1, which
  involved the explosion of a low mass helium shell.  The light curve
  is dim and evolves quickly; the spectra show notable features of
  intermediate mass elements moving at velocities near 9,000~\kms.
  The observational characteristics of Models 7B1, 9A1, and 10HCD1
  would probably be similar.
  \lFig{8HBC1}}
\end{center}
\end{figure}

\begin{figure}
\begin{center}
\includegraphics[width=0.475\textwidth]{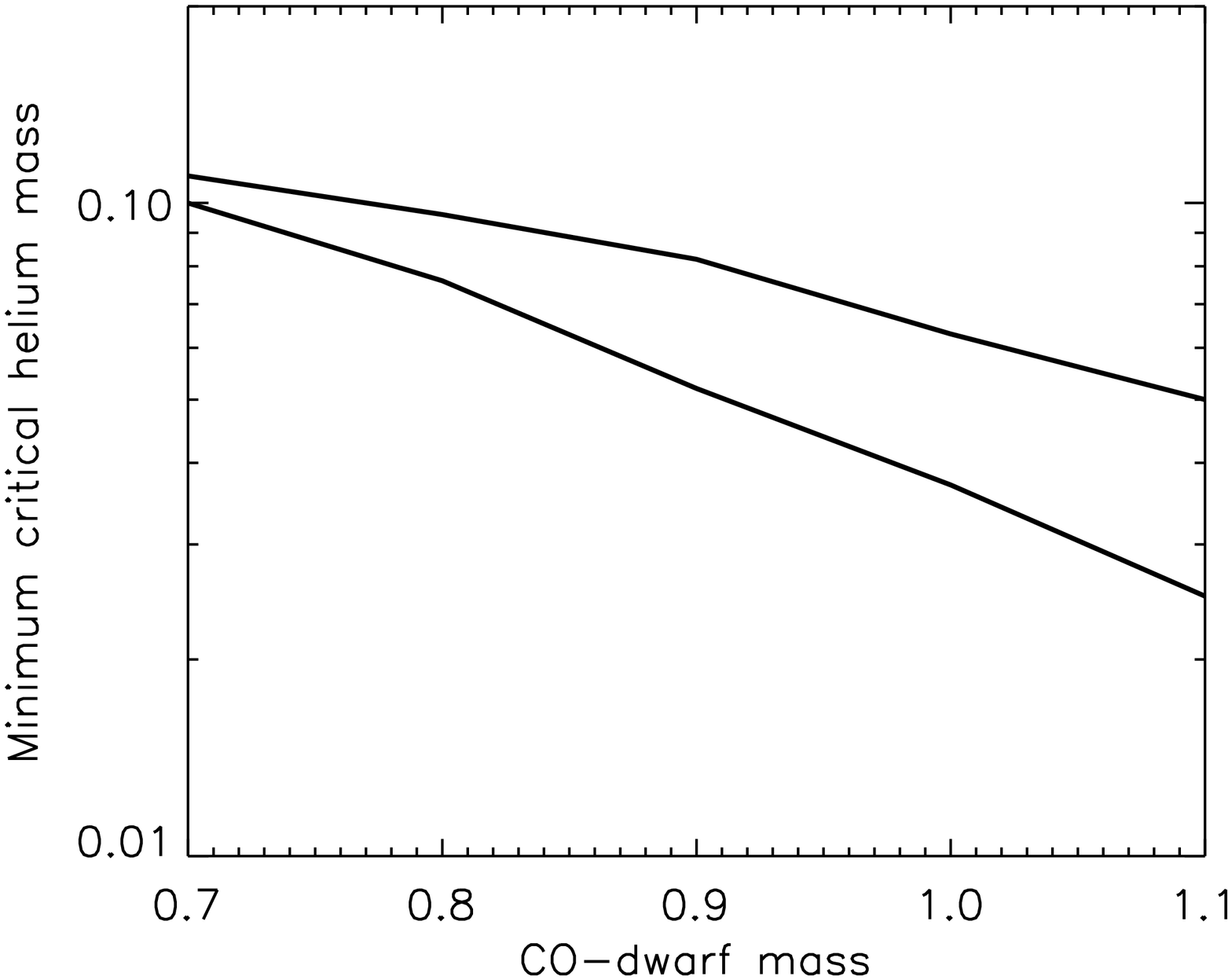}
\hfill
\includegraphics[width=0.475\textwidth]{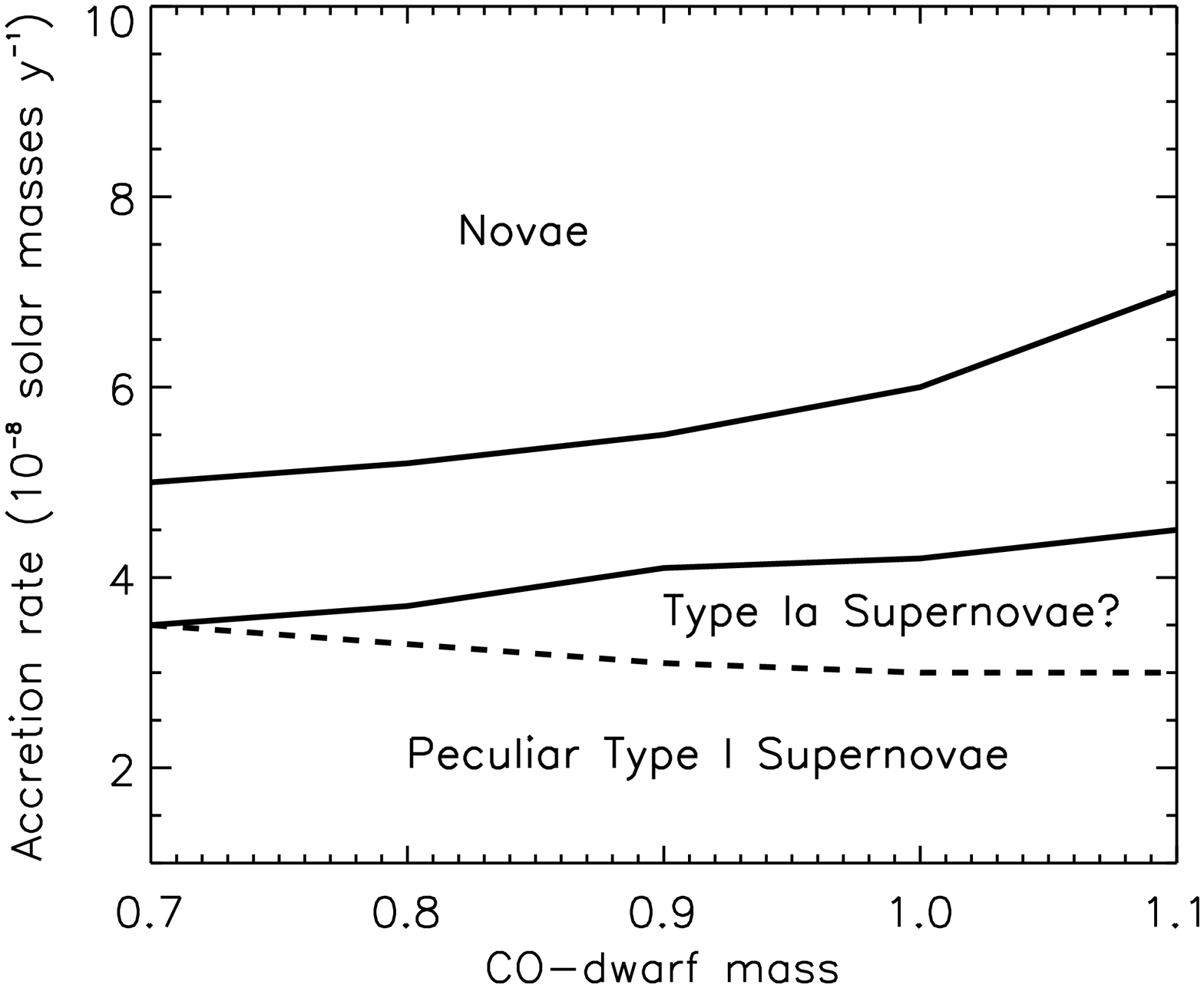}
\caption{The left frame shows the minimum mass of helium required to
  initiate a detonation or deflagration on a CO-dwarf of given
  mass. The top line is interpolated from the ``cold'' white dwarf
  models (Table 1) and the bottom from the ``hot'' models (Table
  2). Both curves lie within the range predicted by \citet{Bil07} (see
  their Fig. 2). The frame on the right shows the expected outcome for
  different accretion rates on CO-dwarfs of a given mass (again
  interpolated from Tables 1 and 2). The top solid line is for the
  ``cold'' models and the lower solid line is for the ``hot''
  models. Above the solid lines, nova-like outbursts are expected and
  not supernovae. Below the solid lines, one detonation or
  deflagration occurs.  It is assumed that a common SN Ia supernova
  requires an helium shell mass less than 0.05 \Msun \ or a $^{56}$Ni
  mass in that shell of less than 0.01 \Msun \ (see text). None of the
  ``cold'' models satisfy this criterion. The ``hot'' models that do
  satisfy these criteria are between the lower solid and dashed
  lines. However, it is also for this range of conditions that both
  helium and carbon detonation are most questionable (due to the low
  density and mass of the helium shell). For larger helium shell
  masses and larger $^{56}$Ni masses, one obtains peculiar Type I
  supernovae that have yet to be observed and classified.
  \lFig{mcrit}}
\end{center}
\end{figure}

\newpage

\tabletypesize{\scriptsize}


\begin{deluxetable}{ccccccccccc}
\tablecaption{Models - Cold White Dwarf Accretors}
\tablehead{
\colhead {Model} &
{Mass} &
{He}  &
{$\dot M_{-8}$} &
{M$_{\rm acc}$} &
{M$_{\rm ign}$} &
{$\rho_{\rm ign}$} &
{$\rho_{\rm run}$} &
{M($^{56}{\rm Ni}$)} &
{M$_{\rm out}(^{56}{\rm Ni}$)} &
{Note} \\
\colhead{}   &
{(\Msun)}    &
{zones$^e$}  &
{(\Msun \ y$^{-1}$)} &
{(\Msun)}    &
{(\Msun)}    &
{(g cm$^{-3}$)} &
{(g cm$^{-3}$)} &
{(\Msun)}    &
{(\Msun)}    &
\colhead{}
} 
\startdata
11AA & 1.1 &  86 & 8  & 0.043  &  0.018  & 1.5(6) & 4.9(5) &  0.    &  0.    & d \\    
11A  & 1.1 & 102 & 7  & 0.0510 &  0.0234 & 1.8(6) & 7.1(5) & 0.819  & 0.012  & a \\
11B  & 1.1 & 113 & 6  & 0.0568 &  0.0387 & 2.1(6) & 1.3(6) & 0.886  & 0.014  & b \\
11C  & 1.1 & 124 & 5  & 0.0620 &  0.0529 & 2.3(6) & 1.8(6) & 0.848  & 0.0268 & a \\
11D  & 1.1 & 176 & 4  & 0.0881 &  0.0816 & 3.8(6) & 3.3(6) & 0.927  & 0.0572 & a \\
11E  & 1.1 & 266 & 3  &  0.133 &  0.133  & 8.0(6) & 8.0(6) & 1.04   & 0.0949 & b \\
11F  & 1.1 & 398 & 2  &  0.200 &  0.200  & 2.4(7) & 2.4(7) & 1.11   & 0.162  & b 
\\[0.1 in]
11E1&  1.1 & 266 & 3  & 0.133  &  0.133  & 8.0(6) & 8.0(6) & 0.094  &  0.094 & c \\
11F1&  1.1 & 398 & 2  & 0.200  &  0.200  & 2.4(7) & 2.4(7) & 0.162  &  0.162 & c \\
    &      &     &    &        &         &        &        &        &        &   \\
10AA & 1.0 & 243 & 7  & 0.052  &  0.0262 & 1.1(6) & 3.8(5) &  0.    &  0.    & d \\
10A  & 1.0 & 377 & 6  & 0.064  &  0.0315 & 1.4(6) & 5.4(5) &  0.701 &  0.013 & b \\
10B &  1.0 & 163 & 5  & 0.0819 &  0.0618 & 1.8(6) & 1.2(6) &  0.745 & 0.0212 & b \\
10C &  1.0 & 180 & 4  & 0.0905 &  0.0744 & 2.1(6) & 1.6(6) &  0.709 & 0.0393 & a \\
10D &  1.0 & 231 & 3  & 0.116  &  0.114  & 3.1(6) & 2.8(6) &  0.785 & 0.0660 & a \\
10E &  1.0 & 354 & 2  & 0.178  &  0.178  & 6.6(6) & 6.6(6) &  0.969 & 0.125  & b  
\\[0.1 in]
10A1& 1.0  & 377 & 6  & 0.064  &  0.0315 & 1.4(6) & 5.4(5) &  0.    &  0.    & c \\
10B1& 1.0  & 163 & 5  & 0.0819 &  0.0618 & 1.8(6) & 1.2(6) &  0.0138& 0.0138 & c \\
10E1& 1.0  & 354 & 2  & 0.178  &  0.178  & 6.6(6) & 6.6(6) &  0.125 & 0.125  & c \\
    &      &     &    &        &         &        &        &        &        &   \\
9AA  & 0.9 & 159 & 6  & 0.0794 & 0.0418  & 1.1(6) & 4.1(5) &  0.    &  0.    & d \\
9A   & 0.9 & 199 & 5  & 0.099  & 0.0564  & 1.5(6) & 6.7(5) & 0.488  &  0.018 & a \\
9B   & 0.9 & 215 &4.5 & 0.109  & 0.0803  & 1.6(6) & 1.1(6) & 0.586  &  0.024 & b \\
9C  & 0.9  & 227 & 4  & 0.120  & 0.0960  & 1.8(6) & 1.3(6) & 0.611  & 0.0397 & b \\
9D  & 0.9  & 248 & 3  & 0.126  &  0.108  & 2.0(6) & 1.6(6) & 0.639  & 0.051  & b \\
9E  & 0.9  & 306 & 2  & 0.154  &  0.154  & 2.7(6) & 2.7(6) & 0.727  & 0.0833 & b  
\\[0.1 in]
9A1  & 0.9 & 198 & 5  & 0.099  & 0.0564  & 1.5(6) & 6.7(5) & 9.3(-5)& 9.3(-5)& c \\
9C1 &  0.9 & 227 & 4  & 0.120  & 0.0960  & 1.8(6) & 1.8(6) & 0.0397 & 0.0397 & c \\
9E1 &  0.9 & 306 & 2  & 0.154  & 0.154   & 2.7(6) & 2.7(6) & 0.0832 & 0.0832 & c \\
    &      &     &    &        &         &        &        &        &        &    \\
8AA  & 0.8 & 315 & 6  & 0.0839 & 0.0539  & 7.4(5) & 2.9(5) &  0.    &  0.    & d \\
8A   & 0.8 & 340 & 5  & 0.106  & 0.0650  & 1.0(6) & 4.5(5) & 0.331  &  0.011 & b \\
8B  & 0.8  & 142 & 4  & 0.142  &  0.114  & 1.6(6) & 1.1(6) & 0.388  & 0.0460 & a \\
8C  & 0.8  & 157 & 3  & 0.157  &  0.141  & 1.8(6) & 1.5(6) & 0.419  & 0.0644 & a \\
8D  & 0.8  & 175 & 2  & 0.175  &  0.175  & 2.1(6) & 2.1(6) & 0.579  & 0.0833 & b \\
8E  & 0.8  & 279 & 1  & 0.280  &  0.280  & 5.9(6) & 5.9(6) & 0.841  & 0.197  & b  
\\[0.1 in]
8A1  & 0.8 & 340 & 5  & 0.106  & 0.0650  & 1.0(6) & 4.5(5) & 0.0    &  0.0   & c \\
8D1 &  0.8 & 175 & 2  & 0.175  &  0.175  & 2.1(6) & 2.1(6) & 0.0833 & 0.0833 & c \\
8E1 &  0.8 & 279 & 1  & 0.280  &  0.280  & 5.9(6) & 5.9(6) & 0.197  & 0.197  & c \\
    &      &     &    &        &         &        &        &        &        &   \\
7AA  & 0.7 &  99 & 6  & 0.098  & 0.098   & 5.1(5) & 1.5(5) &  0.    &  0.    & d \\
7A   & 0.7 & 121 & 5  & 0.120  & 0.082   & 7.7(5) & 3.4(5) &  0.138 & 4.8(-3)& b \\
7B   & 0.7 & 153 & 4  & 0.153  & 0.105   & 1.1(6) & 6.0(5) &  0.256 & 0.015  & b \\
7C  & 0.7  & 185 & 3  & 0.185  & 0.172   & 1.6(6) & 1.3(6) &  0.262 & 0.068  & a \\
7D  & 0.7  & 208 & 2  & 0.207  & 0.182   & 2.0(6) & 1.5(6) &  0.327 & 0.0928 & a \\
7E  & 0.7  & 260 & 1  & 0.261  & 0.261   & 3.0(6) & 3.0(6) &  0.615 & 0.154  & b  
\\[0.1in]
7A1  & 0.7 & 121 & 5  & 0.120  & 0.082   & 7.7(5) & 3.4(5) &  0.    &  0.    & c \\
7B1  & 0.7 & 153 & 4  & 0.153  & 0.105   & 1.1(6) & 6.0(5) & 5.6(-4)& 5.6(-4)& c \\ 
7E1  & 0.7 & 260 & 1  & 0.261  & 0.261   & 3.0(6) & 3.0(6) &  0.154 & 0.154  & c  
\\[0.1 in]
8DEFL  & 0.8 & 157 & 3 & 0.157 & 0.141   & 1.8(6) &   ---  & 6.4(-5)& 6.4(-5)& d \\
10DEFL & 1.0 & 333 & 3 & 0.0972 & 0.0777 & 2.4(6) &   ---  & 9.0(-5)& 9.0(-5)& d \\
\enddata
\tablenotetext{a} {Inwards and outwards helium detonation; whole star
  explodes.}
\tablenotetext{b} {Outwards helium detonation, compresses CO-core and 
leads to secondary detonation.}
\tablenotetext{c} {Helium detonation only; no CO detonation.}
\tablenotetext{d} {No prompt detonation, deflagration or convective runaway.}
\tablenotetext{e} {All models had 375 zones in the CO core plus the number 
of helium zones indicated.}
\lTab{models1}
\end{deluxetable}

\newpage

\begin{deluxetable}{ccccccccccc}
\tablecaption{Models - Hot White Dwarf Accretor}
\tablehead{
\colhead {Model} &
{Mass} &
{He}  &
{$\dot M_{-8}$} &
{M$_{\rm acc}$} &
{M$_{\rm ign}$} &
{$\rho_{\rm ign}$} &
{$\rho_{\rm run}$} &
{M($^{56}{\rm Ni}$)} &
{M$_{\rm out}(^{56}{\rm Ni}$)} &
{Note} \\
\colhead{}   &
{(\Msun)}    &
{zones$^e$}  &
\colhead{}   &
{(\Msun)}    &
{(\Msun)}    &
{(g cm$^{-3}$)} &
{(g cm$^{-3}$)} &
{(\Msun)}    &
{(\Msun)}    &
\colhead{}
} 
\startdata
11HC & 1.1 & 149 & 5  & 0.015  &  0.015  & 2.6(5) & 2.6(5) &  0.    & 0.     & d \\
11HD & 1.1 & 227 & 4  & 0.027  &  0.026  & 8.7(5) & 6.6(5) & 0.826  & 2.9(-5)& b \\
     &     &     &    &        &         &        &        &        &        &   \\
11HD1& 1.1 & 227 & 4  & 0.027  &  0.026  & 8.7(5) & 6.6(5) & 3(-6)  & 3(-6)  & c \\
     &     &     &    &        &         &        &        &        &        &   \\
10HB & 1.0 & 129 & 5  & 0.0220 &  0.0220 & 6.0(4) & 6.0(4) &   0.   & 0.     & d \\
10HBC& 1.0 & 167 &4.5 & 0.0288 &  0.0288 & 3.6(5) & 3.6(5) &   0.   & 0.     & d \\
10HC & 1.0 & 143 & 4  & 0.0445 &  0.0426 & 9.2(5) & 6.6(5) & 0.636  & 8.1(-5)& b \\
10HCD& 1.0 & 263 &3.5 & 0.0638 &  0.0587 & 1.3(6) & 1.1(6) & 0.618  & 8.5(-3)& a \\
10HD & 1.0 & 277 & 3  & 0.0772 &  0.0735 & 1.7(6) & 1.4(6) & 0.663  & 0.025  & a \\
     &     &     &    &        &         &        &        &        &        &   \\
10HC1& 1.0 & 143 & 4  & 0.0445 &  0.0426 & 9.2(5) & 6.6(5) & 2.1(-5)& 2.1(-5)& c \\
10HCD1&1.0 & 263 &3.5 & 0.0638 &  0.0587 & 1.3(6) & 1.1(6) & 4.7(-3)& 4.7(-3)& c \\
     &     &     &    &        &         &        &        &        &        &   \\
9HB  & 0.9 & 113 & 5  & 0.028  & 0.028   & 1.0(5) & 1.0(5) &  0.    &  0.    & d \\
9HBC & 0.9 & 142 & 4.5& 0.037  & 0.037   & 2.0(5) & 2.0(5) &  0.    &  0.    & d \\
9HC  & 0.9 & 183 & 4  & 0.055  & 0.054   & 6.8(5) & 4.9(5) & 0.423  &  0.    & b \\
9HD  & 0.9 & 273 & 3  & 0.108  & 0.107   & 1.6(6) & 1.3(6) & 0.510  &  0.035 & a \\  
     &     &     &    &        &         &        &        &        &        &   \\
8HA  & 0.8 & 124 & 5  & 0.033  &  0.033  & 9.3(4) & 9.3(4) &  0.    &  0.    & d \\
8HB  & 0.8 & 187 & 4  & 0.059  &  0 059  & 2.7(5) & 2.7(5) &  0.    &  0.    & d \\
8HBC & 0.8 & 254 & 3.5& 0.097  &  0.0865 & 8.9(5) & 6.2(5) & 0.317  & 7(-4)  & b \\
8HC  & 0.8 & 319 & 3  & 0.139  &  0.127  & 1.5(6) & 1.2(6) & 0.364  & 0.043  & a \\
     &     &     &    &        &         &        &        &        &        &   \\
8HBC1 & 0.8 & 254 & 3.5& 0.097 &  0.0865 & 8.9(5) & 6.2(5) &1.5(-4) & 1.5(-4)& c \\
     &     &     &    &        &         &        &        &        &        &   \\
\enddata
\tablenotetext{a} {Inwards and outwards helium detonation; whole star
  explodes.}
\tablenotetext{b} {Outwards helium detonation, compresses CO-core and 
leads to secondary detonation.}
\tablenotetext{c} {Helium detonation only; no CO detonation.}
\tablenotetext{d} {Convective runaway - helium nova}
\tablenotetext{e} {All models had 375 zones in the CO core plus the number 
of helium zones indicated.}
\lTab{models2}
\end{deluxetable}

\newpage

\begin{deluxetable}{ccccc}
\tablecaption{Conditions at Ignition - 10$^{46}$ erg s$^{-1}$ and
  10$^{47}$ erg s$^{-1}$}
\tablehead{
\colhead {Model} &
{T$_{46}$} &
{$\rho_{46}$}  &
{T$_{46}$} &
{$\rho_{46}$}  \\
\colhead{}   &
{(10$^8$ K)}    &
{(10$^6$ g cm$^{-3}$)}  &
{(10$^8$ K)}    &
{(10$^6$ g cm$^{-3}$)}  
} 
\startdata
11HC & 3.37  &  0.378 & ...  &  ...   \\
11HD & 2.86  &  0.728 & 3.69 & 0.660  \\
     &       &        &      &        \\
10HB & 3.41  &  0.300 & ...  &  ...   \\
10HBC& 3.10  &  0.438 & ...  &  ...   \\
10HC & 2.81  &  0.717 & 3.51 & 0.655  \\
10HCD& 2.56  &  1.12  & 3.05 & 1.07   \\
10HD & 2.37  &  1.50  & 2.93 & 1.44   \\
     &       &        &      &        \\
9HB  & 3.60  &  0.204 & ...  &  ...   \\
9HBC & 3.17  &  0.323 & ...  &  ...   \\
9HC  & 2.80  &  0.558 & 3.62 & 0.487  \\
9HD  & 2.36  &  1.40  & 2.88 & 1.34   \\
     &       &        &      &        \\
8HA  & 4.42  &  0.0933& ...  &  ...   \\
8HB  & 3.00  &  0.362 & ...  &  ...   \\
8HBC & 2.63  &  0.690 & 3.28 & 0.625  \\
8HC  & 2.34  &  1.26  & 2.87 & 1.20   \\
     &       &        &      &        \\
11AA & 3.07  &  0.572 &  ... &  ...   \\
11A  & 2.81  &  0.789 & 3.72 & 0.713  \\
11B  & 2.48  &  1.35  & 3.14 & 1.29   \\
11C  & 2.34  &  1.82  & 2.81 & 1.77   \\
11D  & 2.09  &  3.35  & 2.46 & 3.29   \\
11E  & 1.84  &  8.10  & 2.10 & 8.04   \\
11F  & 1.58  &  23.6  & 1.82 & 23.5   \\
     &       &        &      &        \\
10AA & 3.09  &  0.460 &  ... &  ...   \\
10A  & 2.92  &  0.603 & 3.73 & 0.537  \\
10B  & 2.44  &  1.29  & 2.96 & 1.24   \\
10C  & 2.31  &  1.64  & 2.78 & 1.59   \\
10D  & 2.10  &  2.89  & 2.48 & 2.84   \\
10E  & 1.81  &  6.53  & 2.13 & 6.63   \\
     &       &        &      &        \\
9AA  & 2.95  &  0.487 & ...  &  ...   \\
9A   & 2.69  &  0.742 & 3.46 & 0.672  \\
9B   & 2.49  &  1.12  & 3.03 & 1.06   \\
9C   & 2.35  &  1.39  & 2.86 & 1.33   \\
9D   & 2.27  &  1.64  & 2.77 & 1.58   \\
9E   & 2.09  &  2.75  & 2.44 & 2.70   \\
     &       &        &      &        \\
8AA  & 3.00  &  0.376 & ...  &  ...   \\
8A   & 2.86  &  0.522 & 3.62 & 0.455  \\
8B   & 2.42  &  1.14  & 2.93 & 1.09   \\
8C   & 2.25  &  1.53  & 2.70 & 1.48   \\
8D   & 2.12  &  2.12  & 2.57 & 2.06   \\
8E   & 1.84  &  5.96  & 2.13 & 5.89   \\
     &       &        &      &        \\
7AA  & 3.10  &  0.291 & ...  &  ...   \\
7A   & 2.83  &  0.423 & 3.76 & 0.339  \\
7B   & 2.58  &  0.669 & 3.26 & 0.598  \\
7C   & 2.27  &  1.36  & 2.75 & 1.30   \\
7D   & 2.23  &  1.60  & 2.66 & 1.54   \\
7E   & 2.01  &  3.02  & 2.35 & 2.96   \\
\enddata
\lTab{ignpoint}
\end{deluxetable}

\newpage

\begin{deluxetable}{cccccccc}
\tablecaption{Explosion Characteristics}
\tablehead{
\colhead {Model} &
{KE}       &
{Remnant}  &
{Eject}    &
\colhead {Model} &
{KE}       &
{Remnant}  &
{Eject}    \\
\colhead {}   &
{(erg)}       &
{(\Msun)}     &
{(\Msun)}     &
\colhead{}    &
{(erg)}       &
{(\Msun)}     &
{(\Msun)}     
}
\startdata
11A &   1.4(51)  &   0    &  1.15  & 9HC &  1.1(51)  &  0    &   0.96 \\     
11B &   1.5(51)  &   0    &  1.16  & 9D  &  1.3(51)  &  0    &   1.02 \\   
11C &   1.5(51)  &   0    &  1.16  & 9HD &  1.2(51)  &  0    &   1.01 \\    
11D &   1.5(51)  &   0    &  1.19  & 9E  &  1.4(51)  &  0    &   1.05 \\  
11HD&   1.4(51)  &   0    &  1.13  & 9A1 &  3.4(49)  & 0.95  &  0.049 \\
11E &   1.7(51)  &   0    &  1.23  & 9C1 &  1.9(50)  & 0.90  &   0.12 \\  
11F &   1.9(51)  &   0    &  1.30  & 9E1 &  2.4(50)  & 0.90  &   0.15 \\
    &            &        &        &     &           & 	     &        \\
11HD1&  9.7(48)  &  1.11  &  0.017 & 8A  &  9.9(50)  &  0    &   0.91 \\  
11E1&   2.1(50)  &  1.11  &  0.12  & 8B  &  1.2(51)  &  0    &   0.94 \\
11F1&   3.3(50)  &  1.10  &  0.20  & 8HBC&  9.7(50)  &  0    &   0.90 \\  
    &            &        &        & 8C  &  1.2(51)  &  0    &   0.96 \\
    & 		 &	  & 	   & 8HC &  1.1(51)  &  0    &   0.94 \\   
10A &   1.3(51)  &   0    &  1.06  & 8D  &  1.3(51)  &  0    &   0.97 \\    
10B &   1.4(51)  &   0    &  1.08  & 8E  &  1.7(51)  &  0    &   1.08 \\
10C &   1.4(51)  &   0    &  1.09  &     &           &       &        \\  
10HC&   1.2(51)  &   0    &  1.05  & 8A1 &  3.6(49)  & 0.85  &   0.056\\ 
10HCD&  1.2(51)  &   0    &  1.07  & 8HBC1& 6.8(49)  & 0.81  &   0.087\\   
10D &   1.4(51)  &   0    &  1.12  & 8D1 &  2.7(50)  & 0.78  &   0.19 \\ 
10HD&   1.3(51)  &   0    &  1.08  & 8E1 &  5.3(50)  & 0.77  &   0.31 \\
10E &   1.7(51)  &   0    &  1.18  &     &           &	     &        \\ 
    &            &        &        & 7A  &  7.5(50)  &  0    &   0.82 \\
10A1&   1.1(49)  &  1.04  &  0.024 & 7B  &  9.4(50)  &  0    &   0.85 \\
10B1&   6.2(49)  &  1.03  &  0.05  & 7C  &  1.0(51)  &  0    &   0.89 \\  
10HC1&  2.3(49)  &  1.01  &  0.035 & 7D  &  1.1(51)  &  0    &   0.91 \\
10HCD1& 5.4(49)  &  1.01  &  0.057 & 7E  &  1.4(51)  &  0    &   0.96 \\
    &            &        &        & 7A1 &  3.7(49)  & 0.74  &   0.081\\ 
10E1&   3.1(50)  &  0.99  &  0.19  & 7B1 &  1.0(50)  & 0.75  &   0.10 \\  
    & 		 &	  &        & 7E1 &  4.6(50)  & 0.69  &   0.27 \\
9A  &   1.2(51)  &   0    &  1.00  &     &           &       &        \\
9B  &   1.2(51)  &   0    &  1.01  & 10DEFL& 2.3(49)  & 1.03  &  0.069\\ 
9C  &   1.3(51)  &   0    &  1.02  & 8DEFL & 3.0(49)  & 0.82  &  0.14 \\
\enddata
\lTab{explosions}
\end{deluxetable}

\newpage

\begin{deluxetable}{cccccccc}
\tablecaption{Radioactivities Produced in the Explosions (Solar Masses)}
\tablehead{
\colhead {Model} &
{$^{44}$Ti} &
{$^{48}$Cr} &
{$^{52}$Fe} &
{$^{55}$Co} &
{$^{56}$Ni} &
{$^{57}$Ni} &
{$^{59}$Cu + $^{59}$Ni}
} 
\startdata
11A & 1.2(-3) & 3.1(-3) & 6.8(-3) & 1.2(-3) & 8.2(-1) & 1.9(-2) & 6.5(-4) \\
11B & 6.2(-4) & 2.8(-3) & 9.1(-3) & 1.4(-3) & 8.9(-1) & 2.0(-2) & 6.2(-4) \\ 
11C &   2.7(-4) & 1.5(-3) & 7.4(-3) & 1.3(-3) & 8.5(-1) & 2.1(-2) & 9.1(-4) \\
11D &   1.6(-4) & 8.3(-4) & 5.6(-3) & 1.1(-3) & 9.3(-1) & 2.4(-2) & 9.0(-4) \\
11HD&  4.6(-4) & 4.1(-4) & 6.1(-3) & 1.2(-3) & 8.3(-1) & 1.7(-2) & 4.7(-4) \\
11E &   9.6(-5) & 4.4(-4) & 3.1(-3) & 5.0(-4) & 1.0(00) & 2.9(-2) & 1.2(-3) \\
11F &   4.4(-5) & 1.1(-4) & 5.8(-4) & 9.0(-4) & 1.1(00) & 3.8(-2) & 1.3(-3)  
\\[0.1 in]
11HD1& 4.3(-4) & 5.8(-5) &  ...    &   ...   &   ...   &  ...    &  ...    \\
11E1&   8.4(-5) & 2.9(-4) & 5.2(-4) & 3.0(-5) & 9.4(-2) & 4.0(-3) & 4.3(-4) \\
11F1&   3.8(-5) & 1.3(-4) & 2.2(-4) & 1.2(-5) & 1.6(-1) & 5.7(-3) & 3.1(-4) \\
    &           &         &         &         &         &         &         \\
10A &  2.6(-4) & 5.8(-4) & 7.7(-3) & 1.5(-3) & 7.0(-1) & 1.4(-2) & 7.2(-4) \\
10B &   8.4(-4) & 3.8(-3) & 1.3(-2) & 1.7(-3) & 7.4(-1) & 1.7(-2) & 5.3(-4) \\
10C &   4.5(-4) & 2.3(-3) & 1.0(-2) & 1.7(-3) & 7.1(-1) & 1.8(-2) & 8.3(-4) \\
10HC&  1.1(-3) & 5.8(-4) & 7.4(-3) & 1.5(-3) & 6.4(-1) & 1.2(-2) & 3.0(-4) \\
10HCD& 1.3(-3) & 5.1(-3) & 1.7(-2) & 2.2(-3) & 6.2(-1) & 1.3(-2) & 3.8(-4) \\
10D &   2.6(-4) & 1.3(-3) & 7.7(-3) & 1.4(-3) & 7.9(-1) & 2.0(-2) & 8.0(-4) \\
10HD&  6.4(-4) & 3.0(-3) & 1.2(-2) & 1.8(-3) & 6.6(-1) & 1.6(-2) & 5.3(-4) \\
10E &   1.4(-4) & 6.5(-4) & 4.2(-3) & 7.2(-4) & 9.7(-1) & 2.6(-2) & 1.2(-3) 
\\[0.1 in]
10A1  & 1.9(-4)&1.2(-5)  &  ...    &  ...    &  ...    &  ...    &  ...    \\
10B1  & 7.7(-4)& 3.3(-3) & 6.5(-3) & 4.2(-4) & 1.4(-2) & 1.7(-3) & 4.2(-5) \\
10HC1 & 1.1(-3)& 1.7(-4) & 4.1(-5) &  ...    & 2.1(-5) &  ...    &  ...    \\
10HCD1&1.7(-3) & 7.6(-3) & 7.1(-3) & 4.9(-4) & 4.7(-3) & 6.1(-4) & 3.4(-5) \\
10E1  & 1.2(-4)& 4.5(-4) & 8.3(-4) & 4.9(-5) & 1.2(-1) & 5.4(-3) & 5.0(-4) \\
               &         &         &         &         &         &         \\
9A  & 3.4(-3) & 6.5(-3) & 1.0(-2) & 1.7(-3) & 4.9(-1) & 1.0(-2) & 2.8(-4) \\
9B  & 1.5(-3) & 7.0(-3) & 1.6(-2) & 2.1(-3) & 5.9(-1) & 1.3(-2) & 5.3(-4) \\  
9C  &  9.6(-4) & 4.4(-3) & 1.3(-2) & 1.8(-3) & 6.1(-1) & 1.5(-2) & 4.8(-5) \\
9HC&  3.6(-4) & 4.1(-4) & 8.0(-3) & 1.7(-3) & 4.2(-1) & 6.3(-3) & 1.4(-4) \\     
9D  &  6.5(-4) & 3.3(-3) & 1.1(-2) & 1.6(-3) & 6.4(-1) & 1.7(-2) & 7.3(-4) \\
9HD&  9.3(-4  & 4.2(-3) & 1.5(-2) & 2.1(-3) & 5.1(-1) & 1.3(-2) & 4.1(-4) \\
9E  &  4.1(-4) & 2.0(-3) & 8.5(-3) & 1.4(-3) & 7.3(-1) & 2.0(-2) & 7.1(-4)  
\\[0.1 in]
9A1&  2.2(-3) & 5.9(-4) & 1.5(-4) & 1.0(-5) & 9.3(-5) & 1.3(-5) &  ...   \\
9C1 &  9.3(-4) & 4.1(-3) & 7.0(-3) & 4.9(-4) & 4.0(-2) & 4.2(-3) & 2.0(-4) \\
9E1 &  3.9(-4) & 1.7(-3) & 3.1(-3) & 2.1(-4) & 8.3(-2) & 7.2(-3) & 4.3(-4) \\
    &          &         &         &         &         &         &        \\
8A  & 5.1(-4) & 1.4(-3) & 1.0(-2) & 2.0(-3) & 3.3(-1) & 5.3(-3) & 3.8(-4) \\
8B  & 1.7(-3) & 6.4(-3) & 2.0(-2) & 2.3(-3) & 3.9(-1) & 9.4(-3) & 2.6(-4) \\
8HBC&3.7(-3) & 1.9(-3) & 8.5(-3) & 1.7(-3) & 3.2(-1) & 4.4(-3) & 9.4(-5) \\
8C  & 8.9(-4) & 4.4(-3) & 1.4(-2) & 2.0(-3) & 4.2(-1) & 1.3(-2) & 6.6(-4) \\
8HC& 1.5(-3) & 6.1(-3) & 1.8(-2) & 2.2(-3) & 3.6(-1) & 1.0(-2) & 2.5(-4) \\
8D  & 6.4(-4) & 3.3(-3) & 1.2(-2) & 1.6(-3) & 5.8(-1) & 1.7(-2) & 7.4(-4) \\
8E  & 2.2(-4) & 1.0(-3) & 5.9(-3) & 9.6(-4) & 8.4(-1) & 2.3(-2) & 1.1(-3)  
\\[0.1 in]
8A1& 3.1(-4) & 1.2(-5) &  ...    &  ...    &  ...    &  ...    &   ...   \\
8HBC1& 3.6(-3) & 9.6(-4) & 2.5(-4) & 2.0(-5) & 1.5(-4)& 2.1(-5) & ...    \\
8D1 & 6.1(-4) & 3.0(-3) & 5.6(-3) & 3.7(-4) & 8.4(-2) & 8.2(-3) & 5.9(-4) \\
8E1 & 2.1(-4) & 8.1(-4) & 1.5(-3) & 9.4(-5) & 2.0(-1) & 8.5(-3) & 7.5(-4) \\
    &         &         &         &         &         &         &         \\
7A  & 6.5(-4) & 2.7(-3) & 9.9(-3) & 1.8(-3) & 1.4(-1) & 1.6(-3) & 2.7(-5) \\
7B  & 5.0(-3) & 5.3(-3) & 1.1(-2) & 1.9(-3) & 2.6(-1) & 4.5(-3) & 4.2(-4) \\ 
7C  & 1.4(-3) & 6.2(-3) & 1.7(-2) & 2.0(-3) & 2.6(-1) & 9.9(-3) & 4.5(-4)  \\
7D  & 1.0(-3) & 4.9(-3) & 1.5(-2) & 2.0(-3) & 3.3(-1) & 1.3(-2) & 8.4(-4) \\
7E  & 6.0(-4) & 2.6(-3) & 9.4(-3) & 1.4(-3) & 6.2(-1) & 2.0(-2) & 6.4(-4)   
\\[0.1in]
7A1& 5.6(-5) &   ...   &  ...    &   ...   &  ...    &   ...   &   ...   \\
7B1& 4.8(-3) & 3.9(-3) & 1.9(-3) & 1.2(-4) & 5.6(-4) & 5.7(-5) & 1.0(-5) \\ 
7E1 & 5.8(-4) & 2.3(-3) & 4.2(-3) & 2.8(-4) & 1.5(-1) & 1.1(-2) & 4.2(-4)   
\\[0.1 in]
10DEFL & 2.5(-3) & 6.3(-3) & 1.6(-3) & 1.3(-3) & 9.0(-5) & 7.3(-6) & 4.9(-8) \\
8DEFL  & 3.5(-3) & 3.7(-3) & 4.6(-4) & 1.1(-5) & 6.4(-5) & 6.7(-6) & 3.0(-7) \\
\enddata
\lTab{radio}
\end{deluxetable}

\clearpage

\begin{deluxetable}{cccccccccccc}
\tablecaption{Major Production Factors - Double Detonation - High Mass Dwarfs}
\tablehead{
\colhead {Species} &
{11A}   & 
{11B}   & 
{11C}   &
{11D}   &
{11E}   &
{11F}   &
{10A}   & 
{10B}   &
{10C}   &
{10D}   &
{10E}
} 
\startdata
$^{28}$Si & 100 & 70  & 91  & 60  & 18  & ... & 130 & 110 & 140 & 99  & 32  \\
$^{32}$S  & 178 & 130 & 160 & 120 & 39  & ... & 240 & 190 & 250 & 180 & 65  \\
$^{34}$S  & 18  & 9.2 & 22  & ... & ... & ... & 23  & 22  & 24  & 22  & ... \\
$^{36}$Ar & 130 & 100 & 110 & 90  & 36  & ... & 180 & 140 & 170 & 130 & 57  \\
$^{38}$Ar & 12  & 5.9 & 14  & ... & ... & ... & 14  & 12  & 18  & ... & ... \\
$^{39}$K  & 27  & 19  & 11  & ... & ... & ... & 62  & 27  & 19  & ... & ... \\
$^{40}$Ca & 220 & 160 & 140 & 110 & 51  & ... & 290 & 210 & 220 & 160 & 77  \\
$^{43}$Ca & 110 & 160 & 85  & 68  & 53  & 27  & 120 & 230 & 140 & 99  & 78  \\
$^{44}$Ca & 630 & 320 & 140 & 81  & 46  & 20  & 140 & 460 & 240 & 140 & 69  \\
$^{45}$Sc & 6.3 & 2.3 & 1.3 & ... & ... & ... & 3.7 & 2.5 & 2.1 & 1.0 &1.2  \\
$^{46}$Ti & 230 & 190 & 290 & 82  & 36  & ... & 82  & 350 & 400 & 160 & 61  \\
$^{47}$Ti & 300 & 310 & 240 & 96  & 54  & 23  & 67  & 530 & 380 & 170 & 84  \\
$^{48}$Ti & 1100& 980 & 510 & 280 & 140 & 44  & 220 & 1400 & 830 & 480 & 220 \\
$^{49}$Ti & 130 & 76  & 65  & 54  & 43  & 11  & 90  & 90  & 93  & 75  & 59  \\
$^{51}$V  & 320 & 360 & 250 & 130 & 55  & 15  & 110 & 540 & 390 & 210 & 87  \\
$^{50}$Cr & 110 & 74  & 81  & 51  & 20  & ... & 110 & 110 & 130 & 75  & 34  \\
$^{52}$Cr & 360 & 480 & 390 & 290 & 150 & 27  & 440 & 720 & 560 & 420 & 220 \\
$^{53}$Cr & 170 & 170 & 140 & 120 & 71  & 17  & 210 & 210 & 190 & 160 & 98  \\
$^{55}$Mn & 70  & 83  & 76  & 60  & 28  & 47  & 99  & 110 & 100 & 85  & 42  \\
$^{54}$Fe & 67  & 56  & 60  & 49  & 19  & 46  & 92  & 75  & 87  & 70  & 30  \\
$^{56}$Fe & 570 & 610 & 580 & 620 & 670 & 680 & 520 & 550 & 520 & 560 & 650 \\
$^{57}$Fe & 570 & 590 & 620 & 690 & 810 & 100 & 440 & 530 & 540 & 620 & 740 \\
$^{59}$Co & 140 & 130 & 210 & 190 & 250 & 250 & 170 & 120 & 190 & 180 & 250 \\
$^{58}$Ni & 280 & 250 & 300 & 350 & 500 & 100 & 170 & 190 & 210 & 230 & 350 \\
$^{60}$Ni & 730 & 790 & 810 & 920 & 1000& 950 & 570 & 690 & 690 & 810 & 1000\\
$^{61}$Ni & 810 & 860 & 1100& 1000& 970 & 930 & 510 & 1000& 1200& 1000& 1000\\
$^{62}$Ni & 840 & 750 & 930 & 1100& 1400& 160 & 500 & 600 & 670 & 760 & 1300\\
$^{63}$Cu & 12  & 18  & 37  & 35  & 20  & 27  & 21  & 19  & 50  & 41  & 24  \\
$^{65}$Cu & 47  & 67  & 64  & 46  & 32  & 24  & 23  & 99  & 87  & 65  & 39  \\
$^{64}$Zn & 210 & 300 & 300 & 340 & 190 & 120 & 190 & 420 & 410 & 550 & 240 \\
$^{66}$Zn & 86  & 88  & 100 & 150 & 160 & 160 & 56  & 79  & 76  & 120 & 160 \\
$^{68}$Zn & 31  & 35  & 41  &  20 & ... & ... & 14  & 63  & 59  & 41  & 12  \\
$^{69}$Ga & 17  & 44  & 19  &  13 & ... & ... & 12  & 69  & 29  & 21  & ... \\
$^{70}$Ge & 10  & 18  & 21  &  18 & 13  & ... & 12  & 44  & 32  & 26  & 16  \\
$^{72}$Ge & 4.2 & 2.9 & 10  & ... & ... & ... & 2.2 & 11  & 17  & ... & ... \\
$^{74}$Se & 21  & 45  & 72  & 35  & 12  & ... & 69  & 220 & 130 & 73  & 23  \\
$^{78}$Kr & 9.0 & 23  & 51  & 42  & 10  & ... & 200 & 190 & 88  & 120 & 25  \\
$^{80}$Kr & 4.2 & 1.6 & 17  & ... & ... & ... & 11  & 26  & 31  & ... & ... \\
\enddata
\lTab{himddyield}
\end{deluxetable}

\clearpage

\begin{deluxetable}{cccccccccccccccc}
\tablecaption{Major Production Factors - Double Detonation - Low Mass Dwarfs}
\tablehead{
\colhead {Species} &
{9A}  & 
{9B}  & 
{9C}   &
{9D}   &
{9E}   &
{8A}  & 
{8B}   &
{8C}   &
{8D}   &
{8E}   &
{7A}  & 
{7B}  & 
{7C}   &
{7D}   &
{7E}
} 
\startdata
$^{28}$Si & 210 & 160 & 150 & 140 & 100 & 270 & 260 & 220 & 150 & 46  & 350 & 280 & 300 & 260 & 110 \\
$^{32}$S  & 350 & 260 & 230 & 210 & 180 & 410 & 440 & 330 & 230 & 94  & 480 & 410 & 420 & 380 & 200 \\
$^{34}$S  & 47  & 39  & 40  & 40  & 22  & 81  & 40  & 84  & 40  & ... & 120 & 81  & 120 & 97  & 21  \\
$^{36}$Ar & 250 & 190 & 170 & 150 & 130 & 300 & 290 & 220 & 170 & 81  & 370 & 280 & 270 & 250 & 140 \\
$^{38}$Ar & 25  & 18  & 18  & 16  & 11  & 33  & 26  & 29  & 17  & ... & 47  & 33  & 53  & 37  & 11  \\
$^{39}$K  & 74  & 56  & 32  & 25  & 11  & 140 & 73  & 37  & 50  & ... & 290 & 190 & 65  & 46  & 15  \\
$^{40}$Ca & 480 & 290 & 250 & 220 & 170 & 510 & 400 & 310 & 230 & 110 & 450 & 790 & 380 & 340 & 200 \\
$^{43}$Ca & 290 & 360 & 270 & 200 & 160 & 290 & 400 & 260 & 230 & 130 & 270 & 940 & 390 & 320 & 220 \\
$^{44}$Ca & 2000& 870 & 560 & 370 & 230 & 330 & 1000& 550 & 390 & 120 & 470 & 3500& 930 & 670 & 370 \\
$^{45}$Sc &  18 & 5.3 & 2.6 & 2.2 & 1.8 & 6.2 & 5.5 & 2.8 & 2.4 & 1.4 & 5.1 & 140 & 4.8 & 3.3 & 2.3 \\
$^{46}$Ti & 370 & 280 & 530 & 580 & 220 & 170 & 310 & 770 & 510 & 120 & 290 & 260 & 970 & 900 & 260 \\
$^{47}$Ti & 680 & 590 & 800 & 670 & 290 & 180 & 680 & 950 & 690 & 150 & 210 & 600 & 1400& 1200& 350 \\
$^{48}$Ti & 2600& 2800& 1700& 1300& 750 & 630 & 2700& 1800& 1400& 390 & 1300& 2500& 2800& 2200& 1100 \\
$^{49}$Ti & 280 & 170 & 100 & 94  & 83  & 120 & 170 & 130 & 100 & 82  & 130 & 210 & 150 & 140 & 95  \\
$^{51}$V  & 770 & 920 & 670 & 540 & 280 & 380 & 1100& 810 & 520 & 150 & 830 & 1200& 1200& 950 & 390 \\
$^{50}$Cr & 200 & 140 & 140 & 130 & 92  & 180 & 210 & 170 & 130 & 50  & 190 & 210 & 210 & 190 & 110 \\
$^{52}$Cr & 620 & 940 & 800 & 690 & 490 & 680 & 1300& 930 & 720 & 330 & 740 & 790 & 1200& 990 & 600 \\
$^{53}$Cr & 280 & 280 & 210 & 200 & 170 & 270 & 320 & 250 & 210 & 130 & 280 & 280 & 270 & 250 & 180 \\
$^{55}$Mn & 120 & 140 & 120 & 110 & 89  & 150 & 170 & 140 & 110 & 61  & 150 & 150 & 150 & 150 & 100 \\
$^{54}$Fe & 120 & 99  & 88  & 81  & 68  & 140 & 150 & 120 & 88  & 44  & 140 & 130 & 120 & 120 & 75  \\
$^{56}$Fe & 390 & 460 & 480 & 500 & 550 & 290 & 330 & 350 & 470 & 620 & 130 & 240 & 240 & 290 & 510 \\
$^{57}$Fe & 350 & 430 & 500 & 550 & 650 & 200 & 340 & 450 & 610 & 710 & 68  & 180 & 380 & 490 & 710 \\
$^{59}$Co & 72  & 130 & 120 & 180 & 170 & 110 & 69  & 170 & 190 & 270 & 8.7 & 120 & 130 & 230 & 170 \\
$^{58}$Ni & 120 & 140 & 150 & 170 & 190 & 68  & 87  & 100 & 150 & 250 & 21  & 56  & 63  & 85  & 140 \\
$^{60}$Ni & 400 & 920 & 650 & 730 & 860 & 200 & 410 & 530 & 780 & 1000& 32  & 180 & 460 & 560 & 950 \\
$^{61}$Ni & 740 & 920 & 1400& 1600& 1300& 390 & 1100& 1800& 1800& 1100& 170 & 450 & 2000& 2300& 1700\\
$^{62}$Ni & 340 & 440 & 520 & 610 & 640 & 180 & 270 & 380 & 540 & 1200& 38  & 160 & 290 & 380 & 650 \\
$^{63}$Cu & 16  & 20  & 35  & 63  & 51  & 28  & 30  & 93  & 71  & 35  & 3.2 & 33  & 87  & 140 & 110 \\
$^{65}$Cu & 66  & 100 & 150 & 140 & 100 & 44  & 150 & 170 & 160 & 55  & 23  & 48  & 230 & 220 & 130 \\
$^{64}$Zn & 300 & 410 & 690 & 660 & 850 & 170 & 620 & 810 & 980 & 350 & 8.4 & 210 & 1000& 1100& 1100\\
$^{66}$Zn & 34  & 65  & 73  & 97  & 120 & 30  & 61  & 51  & 80  & 190 & 11  & 26  & 50  & 58  & 190 \\
$^{68}$Zn & 52  & 51  & 120 & 92  & 68  & 24  & 96  & 130 & 140 & 220 & ... & 30  & 190 & 150 & 63  \\
$^{69}$Ga & 28  & 63  & 81  & 54  & 34  & 27  & 78  & 54  & 56  & 150 & 11  & 23  & 82  & 66  & 43  \\
$^{70}$Ge & 15  & 35  & 81  & 52  & 40  & 28  & 36  & 58  & 56  & 23  & 16  & 20  & 80  & 75  & 52  \\
$^{72}$Ge & 8.6 & 7.5 & 32  & 24  & ... & 7.6 & ... & 35  & 17  & ... & 3.2 & 7.2 & 46  & 38  & ... \\
$^{74}$Se & 52  & 130 & 500 & 230 & 130 & 210 & 94  & 210 & 170 & 36  & 100 & 92  & 320 & 290 & 130 \\
$^{78}$Kr & 23  & 170 & 450 & 200 & 230 & 330 & 26  & 120 & 180 & 37  & 89  & 79  & 160 & 220 & 180 \\
$^{80}$Kr & 13  & 16  & 92  & 61  & 14  & 30  & ... & 45  & 22  & ... & 6.1 & 7.4 & 53  & 69  & 13  \\
$^{84}$Sr & 1.3 & 5.9 & 79  & 14  & ... & 9.7 & ... & ... & ... & ... & 6.7 & 11  & 1.3 & 11  & ... \\
\enddata
\lTab{lomddyield}
\end{deluxetable}

\clearpage

\begin{deluxetable}{ccccccccccccccc}
\tablecaption{Major Production Factors - Helium Shell Detonations}
\tablehead{
\colhead {Species} &
{11E1}  &
{11F1}  &
{10A1}  & 
{10B1}  &
{10E1}  &
{9A1}   & 
{9C1}   &
{9E1}   &
{8A1}   & 
{8D1}   &
{8E1}   &
{7A1}   & 
{7B1}   & 
{7E1} 
} 
\startdata
$^{28}$Si & ... & ... & ... & ... & ... & ... & ... & ... & 1.0 & ... & 2.6 & 2.7 & ... & 1.5 \\
$^{32}$S  & ... & ... & 2.4 & ... & ... & 2.5 & ... & ... & 6.5 & 1.0 & 1.2 & 17  & 4.2 & 1.0 \\
$^{34}$S  & 1.3 & ... & 1.5 & 2.3 & 2.1 & ... & 6.0 & 5.5 & 3.1 & 9.8 & 3.9 & 7.7 & 1.8 & 6.1 \\
$^{36}$Ar & ... & ... & 13  & 2.7 & ... & 12  & 3.6 & 1.5 & 37  & 2.9 & 1.2 & 83  & 21  & 2.7 \\
$^{38}$Ar & ... & ... & ... & ... & ... & ... & 1.7 & 1.0 & ... & 2.4 & 1.0 & 1.1 & 1.2 & 1.2 \\
$^{39}$K  & 3.8 & 2.0 & 52  & 18  & 5.6 & 56  & 23  & 5.6 & 130 & 17  & 9.4 & 250 & 170 & 9.0 \\
$^{40}$Ca & 5.7 & 2.6 & 100 & 51  & 8.9 & 230 & 72  & 30  & 230 & 55  & 17  & 130 & 510 & 46  \\
$^{43}$Ca & 52  & 23  & 120 & 200 & 78  & 270 & 260 & 140 & 230 & 210 & 130 & 110 & 830 & 220 \\
$^{44}$Ca & 40  & 17  & 110 & 420 & 61  & 1300& 550 & 220 & 200 & 370 & 120 & 40  & 3300& 360 \\
$^{45}$Sc & ... & ... & 1.2 & 1.1 & ... & 8.8 & 1.2 & ... & 2.2 & ... & ... & ... & 130 & ... \\
$^{46}$Ti & 35  & 8.4 & ... & 320 & 60  & 4.8 & 590 & 240 & ... & 570 & 120 & ... & 44  & 250 \\
$^{47}$Ti & 53  & 21  & 6.0 & 480 & 84  & 97  & 830 & 300 & 9.2 & 710 & 150 & ... & 390 & 350 \\
$^{48}$Ti & 93  & 38  & 4.6 & 1200& 150 & 1600& 240 & 630 & 5.6 & 1200& 300 & ... & 1800& 950 \\
$^{49}$Ti & 13  & 11  & ... & 20  & 18  & 24  & 11  & 19  & ... & 25  & 27  & ... & 88  & 27  \\
$^{51}$V  & 36  & 15  & 1.4 & 460 & 59  & 620 & 100 & 230 & 1.3 & 470 & 110 & ... & 770 & 340 \\
$^{50}$Cr & ... & ... & ... & 16  & ... & 31  & 2.4 & 7.4 & ... & 21  & ... & ... & 18  & 8.9 \\
$^{52}$Cr & 26  & 10  & ... & 370 & 43  & 420 & 9.5 & 180 & ... & 350 & 85  & ... & 130 & 270 \\
$^{53}$Cr & ... & ... & ... & 27  & ... & 14  & 2.1 & ... & ... & 9.5 & ... & ... & 30  & ... \\
$^{55}$Mn & ... & ... & ... & 26  & ... & 33  & ... & 14  & ... & 26  & 6.0 & ... & 9.6 & 20  \\
$^{56}$Fe & 61  & 99  & ... & 10  & 84  & 31  & ... & 63  & ... & 68  & 140 & ... & ... & 130 \\
$^{57}$Fe & 110 & 150 & ... & 52  & 160 & 140 & ... & 230 & ... & 280 & 270 & ... & 2.3 & 400 \\
$^{59}$Co & 75  & 59  & ... & 9.6 & 110 & 51  & ... & 100 & ... & 150 & 180 & ... & 2.9 & 110 \\
$^{58}$Ni & 25  & 53  & ... & ... & 33  & 11  & ... & 22  & ... & 30  & 56  & ... & ... & 26  \\
$^{60}$Ni & 200 & 220 & ... & 73  & 280 & 200 & ... & 310 & ... & 380 & 480 & ... & ... & 570 \\
$^{61}$Ni & 270 & 220 & ... & 460 & 400 & 1100& 3.3 & 940 & ... & 1500& 700 & ... & 5.7 & 1400\\
$^{62}$Ni & 260 & 290 & ... & 36  & 370 & 110 & 1.4 & 110 & ... & 180 & 620 & ... & 2.2 & 300 \\
$^{63}$Cu & 18  & 22  & ... & 7.1 & 22  & 38  & ... & 51  & ... & 79  & 34  & ... & ... & 110 \\
$^{65}$Cu & 18  & 8.9 & ... & 78  & 27  & 150 & 1.4 & 92  & ... & 150 & 47  & ... & 1.8 & 120 \\
$^{64}$Zn & 140 & 70  & ... & 260 & 190 & 660 & ... & 810 & ... & 950 & 320 & ... & ... & 1100\\
$^{66}$Zn & 57  & 41  & ... & 16  & 81  & 33  & ... & 66  & ... & 45  & 130 & ... & ... & 150 \\
$^{68}$Zn & 7.6 & ... & ... & 55  & 12  & 120 & ... & 70  & ... & 140 & 22  & ... & ... & 64  \\
$^{69}$Ga & 5.8 & ... & ... & 59  & 9.3 & 80  & 1.8 & 33  & ... & 54  & 15  & ... & 1.6 & 43  \\
$^{70}$Ge & 8.0 & ... & ... & 36  & 12  & 78  & 1.6 & 37  & ... & 51  & 21  & ... & 1.6 & 51  \\
$^{72}$Ge & ... & ... & ... & 10  & ... & 33  & ... & 6.8 & ... & 20  & ... & ... & ... & 6.6 \\
$^{75}$As & ... & ... & ... & 5.3 & ... & 13  & ... & ... & ... & ... & ... & ... & ... & ... \\
$^{74}$Se & 12  & ... & ... & 200 & 23  & 490 & 4.8 & 130 & ... & 170 & 35  & ... & 3.2 & 130 \\
$^{76}$Se & ... & ... & ... & 12  & ... & 35  & ... & ... & ... & 5.2 & ... & ... & ... & ... \\
$^{77}$Se & ... & ... & ... & 7.9 & ... & 26  & ... & 6.9 & ... & 9.0 & ... & ... & ... & 6.4 \\
$^{78}$Kr & 10  & ... & ... & 130 7 25  & 450 & 1.0 & 230 & ... & 170 & 37  & ... & ... & 180 \\
$^{80}$Kr & ... & ... & ... & 23  7 ... & 92  & ... & 14  & ... & 22  & ... & ... & ... & 13  \\
\enddata
\lTab{hedetyield}
\end{deluxetable}

\begin{deluxetable}{ccccccccc}
\tablecaption{Major Production Factors - Double Detonations - Hot WD Models}
\tablehead{
\colhead {Species} &
{11HD}  & 
{10HC}  & 
{10HCD} & 
{10HD}  & 
{9HC}   & 
{9HD}   &  
{8HBC}  & 
{8HC}     
} 
\startdata
$^{28}$Si &   96 &  160 &  160 &  140 &  210 &  190 &  270 &  250  \\
$^{32}$S  &  170 &  260 &  240 &  230 &  350 &  290 &  390 &  350  \\
$^{34}$S  &   18 &   36 &   52 &   45 &   47 &   65 &   75 &   88  \\
$^{36}$Ar &  130 &  190 &  170 &  160 &  250 &  200 &  270 &  240  \\
$^{38}$Ar &  8.5 &   16 &   18 &   15 &   25 &   21 &   30 &   27  \\
$^{39}$K  &   28 &   50 &   61 &   20 &   74 &   32 &  120 &   55  \\
$^{40}$Ca &  230 &  370 &  250 &  220 &  480 &  280 &  690 &  350  \\
$^{43}$Ca &   67 &  140 &  250 &  170 &  290 &  250 &  440 &  380  \\
$^{44}$Ca &  240 &  630 &  710 &  350 & 2000 &  550 & 2500 &  940  \\
$^{45}$Sc &  4.2 &  7.2 &  4.9 &  1.8 &   19 &  2.6 &  19  &  3.9  \\
$^{46}$Ti &  8.1 &   14 &   26 &  390 &  370 &  510 &   25 &  470  \\
$^{47}$Ti &   18 &   41 &  250 &  460 &  680 &  680 &  180 &  830  \\
$^{48}$Ti &  150 &  220 & 1900 & 1100 & 2600 & 1700 &  840 & 2600  \\
$^{49}$Ti &   67 &   89 &  120 &   91 &  280 &  120 &  130 &  140  \\
$^{51}$V  &   66 &  110 &  910 &  500 &  770 &  720 &  440 & 1000  \\
$^{50}$Cr &   79 &  120 &  110 &  110 &  200 &  140 &  170 &  160  \\
$^{52}$Cr &  330 &  430 & 1000 &  670 &  620 &  890 &  580 & 1200  \\
$^{53}$Cr &  170 &  220 &  310 &  200 &  280 &  240 &  270 &  280  \\
$^{55}$Mn &   75 &  100 &  140 &  110 &  120 &  140 &  130 &  160  \\
$^{54}$Fe &   67 &   97 &   92 &   86 &  120 &  110 &  130 &  120  \\
$^{56}$Fe &  580 &  480 &  460 &  490 &  390 &  400 &  280 &  310  \\
$^{57}$Fe &  510 &  380 &  420 &  500 &  350 &  430 &  170 &  370  \\
$^{59}$Co &  110 &   72 &   90 &  120 &   72 &  100 &   27 &   70  \\
$^{58}$Ni &  220 &  150 &  170 &  180 &  120 &  130 &   62 &   81  \\
$^{60}$Ni &  660 &  480 &  490 &  630 &  400 &  520 &  160 &  430  \\
$^{61}$Ni &  490 &  340 &  560 & 1100 &  740 & 1300 &  130 & 1400  \\
$^{62}$Ni &  630 &  420 &  520 &  590 &  340 &  410 &  140 &  270  \\
$^{63}$Cu &  1.4 & ...  &  6.7 &   25 &   16 &   34 &  1.8 &   30  \\
$^{65}$Cu &   10 &  7.4 &   40 &  100 &   66 &  140 &   10 &  180  \\
$^{64}$Zn &   42 &   31 &   59 &  430 &  300 &  590 &   11 &  740  \\
$^{68}$Zn & ...  & ...  &  8.4 &   89 &   52 &  120 & ...  &  140  \\
$^{69}$Ga & ...  & ...  &   17 &   46 &   28 &   64 &  8.3 &   91  \\
$^{70}$Ge &  2.5 &  1.9 &   30 &   45 &   15 &   55 &   17 &   70  \\
$^{72}$Ge & ...  & ...  &  4.6 &   17 &  8.6 &   23 &  2.6 &   21  \\
$^{74}$Se & ...  & ...  &  180 &  200 &   52 &  240 &  160 &  300  \\
$^{78}$Kr & ...  & ...  &   99 &   91 &   23 &  100 &  200 &  150  \\
$^{80}$Kr & ...  & ...  &   10 &   16 &   13 &   21 &   14 &   26  \\
\enddata
\lTab{hotwdyield}
\end{deluxetable}

\clearpage

\begin{deluxetable}{ccccccc}
\tablecaption{Major Production Factors - Single Detonations and Deflagrations}
\tablehead{
\colhead {Species} &
{11HD1} & 
{10HC1} & 
{10HCD1}&  
{8HBC1} &   
{8DEFL} &  
{10DEFL}  
} 
\startdata
$^{28}$Si &   ...  &  ...  & ...  & ...  & ...  & ...   \\
$^{32}$S  &    1.3 &   2.0 & ...  &  4.0 &  1.4 &  1.0  \\
$^{34}$S  &   ...  &  ...  & ...  &  1.3 & ...  & ...   \\
$^{35}$Cl &    4.9 &   8.3 &  9.6 &   19 &   10 &  6.0  \\
$^{36}$Ar &    7.1 &    11 &  3.5 &   22 &  9.3 &  4.6  \\
$^{38}$Ar &   ...  &  ...  & ...  & ...  &  1.0 & ...   \\
$^{39}$K  &    23  &    40 &   70 & 100  &   82 &   47  \\
$^{41}$K  &    3.5 &   6.2 &   12 &   15 &   28 &   20  \\
$^{40}$Ca &    89  &   170 &   85 &  420 &  220 &  110  \\
$^{42}$Ca &   ...  &   1.6 &   29 &  5.1 &   37 &   35  \\
$^{43}$Ca &    84  &   160 &  320 &  440 &  240 &  160  \\
$^{44}$Ca &   230  &   610 &  970 & 2400 & 2200 & 1300  \\
$^{45}$Sc &    2.1 &   4.6 &  7.0 &   15 &   38 &   33  \\
$^{46}$Ti &    1.4 &   2.9 &  8.4 &  7.0 &   23 &   23  \\
$^{47}$Ti &   17   &   43  &  270 &  160 &  330 &  280  \\
$^{48}$Ti &   20   &   66  & 2900 &  430 & 1500 & 2400  \\
$^{49}$Ti &    1.3 &   3.5 &  170 &   18 &  280 &  490  \\
$^{51}$V  &    8.6 &   28  & 1200 &  170 &  410 &  540  \\
$^{50}$Cr &   ...  &  ...  &   35 &  3.9 &   44 &   65  \\
$^{52}$Cr &   ...  &   2.4 &  410 &   17 &   30 &  100  \\
$^{53}$Cr &   ...  &  ...  &  140 &  3.5 &  27  &   68  \\
$^{55}$Mn &   ...  &  ...  &   42 &  1.5 & ...  & ...   \\
$^{54}$Fe &   ...  &  ...  &  2.9 & ...  & ...  & ...   \\
$^{56}$Fe &   ...  &  ...  &  3.5 & ...  & ...  & ...   \\
$^{57}$Fe &   ...  &  ...  &   19 & ...  & ...  & ...   \\
$^{59}$Fe &   ...  &  ...  &  8.4 &  1.2 & ...  & ...   \\
$^{58}$Ni &   ...  &  ...  &  5.1 & ...  & ...  & ...   \\
$^{60}$Ni &   ...  &  ...  &  9.4 & ...  & ...  & ...   \\
$^{61}$Ni &   ...  &  ...  &  110 &  5.7 & ...  & ...   \\
$^{62}$Ni &   ...  &  ...  &   26 &  2.7 & ...  & ...   \\
$^{63}$Cu &   ...  &  ...  &  4.6 & ...  & ...  & ...   \\
$^{65}$Cu &   ...  &  ...  &   17 &  2.8 & ...  & ...   \\
$^{64}$Zn &   ...  &  ...  &  1.7 & ...  & ...  & ...   \\
$^{66}$Zn &   ...  &  ...  &  8.0 &  1.5 & ...  & ...   \\
$^{68}$Zn &   ...  &  ...  &  1.2 & ...  & ...  & ...   \\
$^{69}$Ga &   ...  &  ...  &  6.2 &  4.1 & ...  & ...   \\
$^{70}$Ge &   ...  &  ...  &   16 &  4.1 & ...  & ...   \\
$^{72}$Ge &   ...  &  ...  &  4.3 & ...  & ...  & ...   \\
$^{74}$Se &   ...  &  ...  &   58 &   13 & ...  & ...   \\
$^{78}$Kr &   ...  &  ...  &   37 &  3.6 &  3.6 &  3.6  \\
$^{80}$Kr &   ...  &  ...  &  4.4 & ...  & ...  & ...   \\
\enddata
\lTab{deflyield}
\end{deluxetable}

\clearpage

\begin{deluxetable}{cccccc}
\tablecaption{Radioactive Progenitors}
\tablehead{
\colhead {Species} &
{Made As} &
{Species} &
{Made As} &
{Species} &
{Made As}  
} 
\startdata
$^{43}$Ca & $^{43}$Sc,$^{43}$Ti & $^{56}$Fe & $^{56}$Ni & $^{66}$Zn & $^{66}$Ge \\
$^{44}$Ca & $^{44}$Ti      & $^{57}$Fe & $^{57}$Ni & $^{68}$Zn & $^{68}$Se \\
$^{47}$Ti & $^{47}$V,$^{47}$Cr  & $^{59}$Co & $^{59}$Cu & $^{69}$Ga & $^{69}$Se \\
$^{48}$Ti & $^{48}$Cr      & $^{60}$Ni & $^{60}$Zn & $^{70}$Ge & $^{70}$Se \\
$^{49}$Ti & $^{49}$Cr      & $^{61}$Ni & $^{61}$Zn & $^{72}$Ge & $^{72}$Kr \\
$^{51}$V  & $^{51}$Mn      & $^{62}$Ni & $^{62}$Zn & $^{74}$Se & $^{74}$Kr \\
$^{52}$Cr & $^{52}$Fe      & $^{63}$Cu & $^{63}$Ga & $^{78}$Kr & $^{78}$Sr \\
$^{53}$Cr & $^{53}$Fe      & $^{65}$Cu & $^{65}$Ge & $^{80}$Kr & $^{80}$Zr \\
$^{55}$Mn & $^{55}$Co      & $^{64}$Zn & $^{64}$Ge & $^{45}$Sc & $^{45}$Ti \\
\enddata
\lTab{prog}
\end{deluxetable}

\end{document}